# Universal Analytical Modeling of Coated Plasmonic Particles

Nikolai G. Khlebtsov[1,2,*], Sergey V. Zarkov[1,3]

[1]Institute of Biochemistry and Physiology of Plants and Microorganisms, "Saratov Scientific Centre of the Russian Academy of Sciences," 13 Prospekt Entuziastov, Saratov 410049, Russia

[2]Saratov State University, 83 Ulitsa Astrakhanskaya, Saratov 410012, Russia

[3]Institute of Precision Mechanics and Control, "Saratov Scientific Centre of the Russian Academy of Sciences," 24 Ulitsa Rabochaya, Saratov 410028, Russia

[*]To whom correspondence should be addressed. E-mail: (NGK) khlebtsov@ibppm.ru

**Abstract.** From a structural point of view, plasmonic nanoparticles are always at least two-layer structures with a dielectric layer of stabilizing, targeting, fluorescent, Raman, or other functional molecules, often embedded in a matrix such as porous silica. To optimize the optical properties of such bioconjugates, one needs efficient analytical models based on simple physical ideas and with reasonable accuracy comparable to rigorous numerical methods requiring significant computer resources. Recently, we suggested an analytical approach (*J. Phys. Chem. C* **2024**, *128*, 15029−15040) based on a combination of the modal expansion method (MEM) and the dipole equivalence method (DEM). The MEM+DEM method was exemplified by coated gold nanorods excited along the symmetry axis. Here, we extend the MEM+DEM approach for particles of various shapes and orientations, including random ones. To illustrate the possibilities of our method, we calculate extinction and scattering spectra of gold and silver nanorods, nanodiscs, triangle nanoprisms, bicones, and bipyramids with a dielectric coating thickness from 0 to 30 nm. For the main plasmonic peaks, we found excellent agreement between our analytical method and numerical simulations performed with COMSOL, except for some disagreement between MEM and COMSOL solutions for bicones with small rounding radii. This drawback of MEM for coated particles is explained by the unusual local field distribution with strong localization near the particle tips. To fix the point, we propose a factorized version of MEM that brings analytical spectra of coated bicones and bipyramids in good agreement with numerical counterparts. To illustrate the application of our method to real experimental colloids, we prepared 12 AuNR samples using a chemical etching with fine tuning of the LPR over the 1020–580 nm range. By including the CTAB shell in the simulation model, we achieved excellent agreement between the calculated and measured dependence of the LPR peak position on the AuNR aspect ratio. In summary, our method makes it possible to use the MEM parameters of the original metal particles. It does not require additional numerical calculations to build analytical models for bilayered or multilayered conjugates. The presented analytical solutions depend on the particle shape, the shell thickness, and the shell refractive index, regardless of the particle size and the metal's nature. Due to the simplicity and reasonable accuracy of the MEM+DEM analytical models, they can help apply machine learning to predict various plasmonic responses such as light absorption, scattering, surface enhanced Raman scattering (SERS), and metal-enhanced fluorescence (MEF).

## 1. Introduction

Modern methods of chemical synthesis[1] make it possible to obtain gold and silver nanoparticles (NPs) with precise control of their size, shape, and structure. The resulting NPs should be stabilized and functionalized with suitable surface ligands for biomedical applications.[2-4] NPs obtained by laser ablation[5] are chemically pure, but in biological environments, they will immediately aggregate if their surface is not coated with a stabilizer.

Functionalization of NPs provides desirable properties for many promising applications, such as biosensing,[6] cardiovascular[7] therapy, photodynamic and photothermal[8] cancer therapy,[9,10] gene[11] and drug[12] delivery, enhancing adjuvant properties and protection of immune system cells,[13] etc. In the case of "green synthesis,"[14] the resulting NPs are always stabilized by the components of the reaction mixture after the synthesis is completed. The most well-known stabilizers are hexadecyltrimethylammonium bromide (CTAB) bilayer on the surface of gold nanorods[15] and steals-immune coating by m-PEF-SH.[13] Thus, in the real world, stabilized and functionalized NPs are always at least bilayer particles with a plasmonic core and a dielectric shell. Therefore, their plasmonic properties differ from those of the original bare NPs.

Currently, there are several powerful numerical techniques to simulate the classical electromagnetic response from individual or interacting particles of an arbitrary structure and geometry in a complex environment. In particular, most popular are the finite element method (FEM)[16], the boundary element method (BEM),[17,18] the finite difference time domain method (FDTDM)[19], and the discrete dipole approximation (DDA).[20] Applications of these methods to specific plasmonic problems have been discussed in many focused reviews.[21-24] To compare the advantages and disadvantages of different methods, readers are referred to the works.[25-27]

Although the above numerical methods can simulate plasmonic properties of NPs for almost arbitrary geometry, such simulations may require significant computer resources for real polydisperse and polymorphic colloids. From the experimental point of view, it would be desirable to have simple yet accurate analytical approximations based on clear physical ideas capable of predicting layered NP colloids' plasmonic properties with a common desktop PC just *in situ*. To meet this goal, we proposed[28] a combination of the improved electrostatic approximation (IEA)[29] with the DEM[30] for coated gold and silver spheroids. However, the spheroid is a convenient but simplified model just to account for basic shape effects. To go beyond the spheroid shape, we proposed[28] an analytical model using a combination of the MEM developed recently by Yu *et al*.[31] with DEM. However, our previous consideration[28] was restricted by the coated gold nanorods excited along the symmetry axis. Moreover, only the first expansion mode was considered, thus making it impossible to make any calculations for real colloid samples.

Here, we extend our previous treatment in several ways. First, we included all expansion modes excited at arbitrary polarization of the incident field. This allows for the treatment of real colloids with random particle orientations. Second, to illustrate the possibilities of our method, we calculate extinction and scattering spectra of gold and silver nanorods (NRs), nanodiscs (NDs), triangle nanoprisms (NTs), bicones (BCs), and bipyramids (BPs). The accuracy of the MEM+DEM approach is examined with benchmark COMSOL simulations. We generally found excellent agreement between numerical and analytical methods in predicting the amplitude and spectral position of main localized plasmon resonance (LPR) peaks for rods, disks, and prisms. However, we discovered notable disagreement between MEM and COMSOL solutions for coated bicones and bipyramids. This disagreement is explained by the unusual local field distribution inside such NPs with strong localization near the particle tips. Such localization violates the dipole physical picture behind the DEM method. We introduce the third improvement of the previous MEM+DEM treatment to fix the point.[28] Namely, we introduce a factorized version of MEM that brings analytical spectra of coated bicones and bipyramids in excellent agreement with numerical counterparts.

To illustrate the application of the MEM+DEM approach to real experimental colloids, we prepared 12 AuNR samples using a chemical etching method that enables fine tuning of the LPR over the 1020–580 nm range. We demonstrate that the TEM-based simulation model of bare AuNRs does not accurately describe the experimental dependence of the LPR peak wavelength on the average aspect ratio. In contrast, by including the CTAB shell in the simulation model, we achieved excellent agreement between the calculated and measured dependence of the LPR peak position on the AuNR aspect ratio.

It should be emphasized that our analytical approach does not require any additional numerical calculations except for those for bare metal NPs. Our solution is described by four simple functions depending on a generalized aspect ratio, the shell thickness, and its refractive index for a specific particle shape. Moreover, the proposed analytical model can be easily extended to multilayer coating.

## 2. Methods

### *2.1. Experimental*

AuNRs were synthesized by a seed-mediated protocol as described previously.[32] In short, Au seeds were obtained by adding 0.6 mL of 10 mM sodium borohydride to a mixture of 10 mL of 0.1 M CTAB and 0.25 mL of 10 mM $HAuCl_4$. Then, 3.5 mL of 0.1 M silver nitrate was added to a solution of $HAuCl_4$ (500 mL, 0.5 mM) in 0.1 M CTAB, followed by the addition of 25 mL of 0.1 M hydroquinone, with mixing by hand until the growing mixture became clear. Finally, 8 mL of the seed solution was added to the growing mixture and left overnight at 30°C without stirring. To eliminate impurity particles, the nanorod colloid was centrifuged at 10,000 g for 20 minutes, and the residue was redissolved in 30 mL of 200 mM CTAC. This product was left unstirred to form a brown film of aggregated AuNRs. The supernatant containing impurities was removed, and the aggregated AuNRs were redissolved in 50 mM CTAB. The purification process was repeated twice to minimize the percentage of impurity particles. The final concentration of AuNRs was adjusted to produce an absorption peak of 2 at 1020 nm in a 2 mm cuvette.

AuNRs with various LPR wavelengths were obtained by chemical etching of the as-prepared AuNR sample.[32] To this end, 50 mL of the initial AuNR colloid in 50 mM CTAB was titrated by adding 20 μL of 2 mM $HAuCl_4$ solution every 10 minutes. In total, 12 samples with desired LPR wavelengths were obtained, ranging from 1020 nm to 580 nm.

Extinction spectra were recorded using Specord 250 and Specord S300 spectrophotometers (Analytik Jena, Germany). Transmission electron microscopy (TEM) images were obtained using a Libra-120 transmission electron microscope (Carl Zeiss, Germany) at the Simbioz Center for the Collective Use of Research Equipment in the Field of Physico-Chemical Biology and Nanobiotechnology (IBPPM RAS, Saratov).

*2.2. Theoretical methods.*

We used a commercial package COMSOL Multiphysics 5.1 (Wave Optics module) for benchmark electrodynamic simulations in standard 3D and new 2.5D[33] options. The latter ensures considerable computational savings regarding memory and processing time due to expanding all fields in the cylindrical harmonics and treating each harmonic independently.[34] This accelerates the computation speed considerably, thus making the random orientation averaging available with acceptable CPU time.

In addition to COMSOL simulations, we also used T-matrix method (TMM) for axially symmetrical metallic NRs and NDs with random orientations. Compared to COMSOL, TMM codes[35] provide the fastest calculations for randomly oriented ensembles of axially symmetrical homogeneous metallic particles without coatings. In summary, we used 3D and 2.5D COMSOL calculations as benchmarks for MEM+DEM simulations and TMM codes to verify the accuracy of COMSOL simulations for bare NRs and NDs.

## 3. Modal expansion method for coated particles

*3.1. Models*

Figure 1 shows five NP models that are described by a characteristic size $L$, a generalized aspect ratio $AR$, and the complex metal permittivity $\varepsilon_1$. A homogeneous coating (Figure 1B) is characterized by a constant shell thickness $s$ and permittivity $\varepsilon_2$; a homogeneous host medium is characterized by a dielectric function $\varepsilon_m$ (water in this work). The dielectric functions of gold and silver were calculated by tabulated data[36] and by a spline described in the Supporting Information file[28], respectively. We aim to examine the MEM+DEM model *vs* COMSOL simulations, so we used bulk optical constants without any size correction. Regarding the Drude model[37,38] the bulk dielectric functions can be easily corrected to account for known contributions from radiation damping,[39] surface electron scattering[40,41] and chemical interface damping.[42-44] We also note that our homogeneous coating model enables the evaluation of the primary effects resulting from the functionalization of plasmonic particles. As a type of mean-field theory approach, MEM+DEM is not designed to assess the more subtle effects associated with coating heterogeneity in real hybrid nanostructures.

The SI file gives details of the particle shape geometries and explicit formulas for the core and total particle volumes (Section S1). For example, the core and total volumes of NRs with semispherical ends are equal to $V_i = L_i^3 \pi (3AR_i - 1)/(12AR_i^3)$ and are characterized by the radii of equivolume spheres $R_{iev} = \sqrt[3]{3V_i / 4\pi}$, where indices $i = 1,2$ stand for core and shell, respectively, $L_1 \equiv L$, $AR_1 = AR$, $L_2 = L + 2s$, $AR_2 = (L + 2s)/(d + 2s)$.

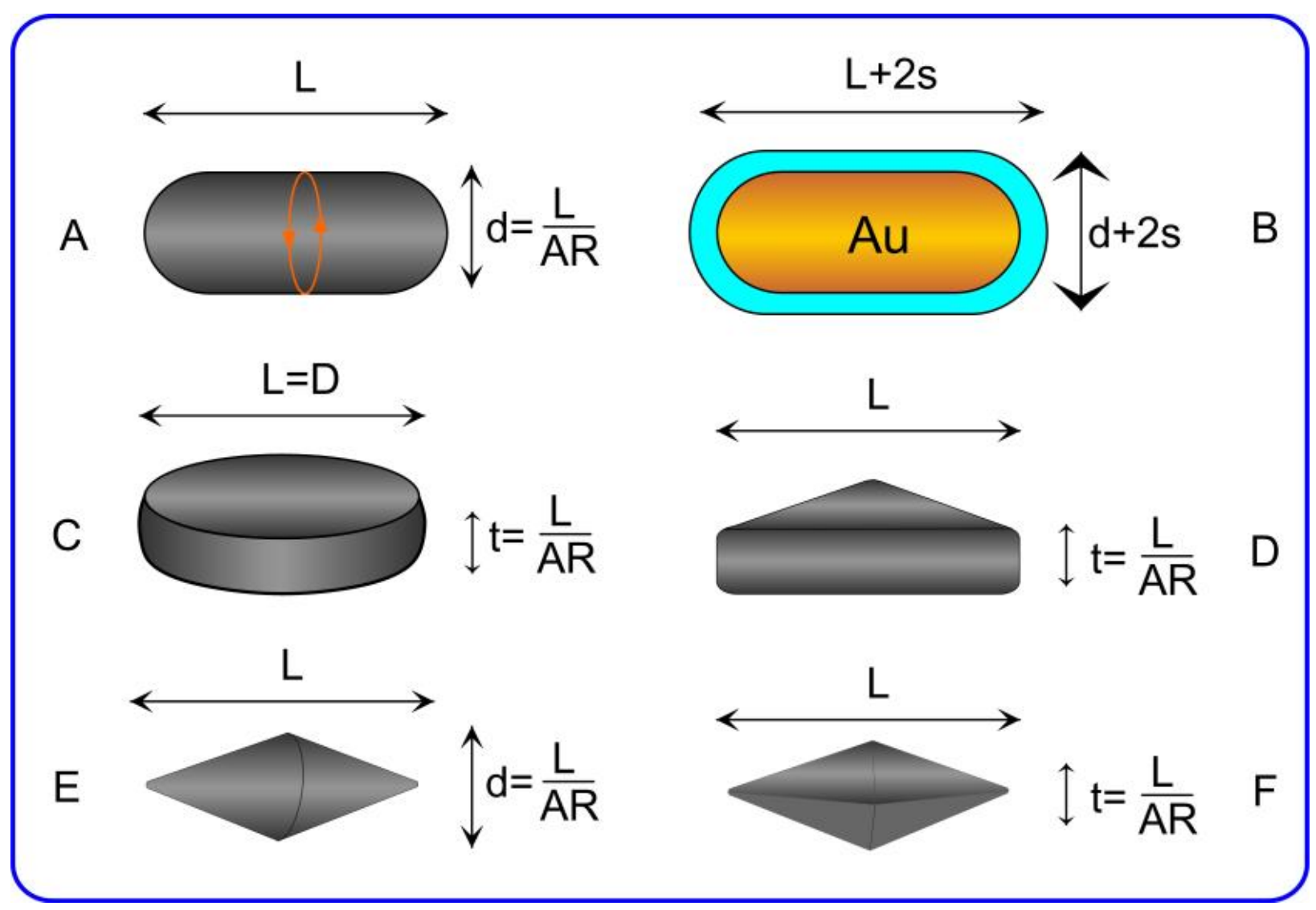

**Figure 1**. Geometrical models for five particle morphologies are characterized by the characteristic length $L$ and a generalized aspect ratio $AR$. Panel B illustrates NP coating with a dielectric shell of a constant thickness $s$. NRs have hemispherical caps at the tips (A); NDs have semicircular edge profile (C); NTs have smooth edges with rounding radius $(L+t)/40$ (D); BCs

have rounded tips and circular base edge with rounding radii $d/40$ and $d/100$, respectively, (E); BPs have pentagonal smooth base and rounded tips as indicated in SI file (F).

*3.2 Polarizability of coated metal particles in MEM*

In MEM, the particle polarizability is expanded in terms of electrostatic eigenvalues

$$\alpha(\lambda) = \frac{1}{4\pi}\sum_j \bar{V}_j \left[ \frac{\varepsilon_m}{\varepsilon_1 - \varepsilon_m} - \frac{1}{\eta_j - 1} - A_j \right]^{-1}, \quad (1)$$

$$A_j = a_{j2}x^2 + i\frac{4\pi^2 \bar{V}_j}{3L^3}x^3 + a_{j4}x^4, \quad (2)$$

where $x = \sqrt{\varepsilon_m} L/\lambda$, $L$ is a characteristic particle size (for example, the sphere diameter or rod length), $\bar{V}_j$ is the modal volume that obeys the sum rule $\sum_j \bar{V}_j = V$, $V$ is the total particle volume, $\eta_j$ is the eigenvalue of *j*-mode ($j = 1$ corresponds to the dipole eigenmode), $a_{j2}$ and $a_{j4}$ are MEM parameters depending on the generalized aspect ratio. Our definition of polarizability follows Bohren and Huffman's book.[38] It differs from Eq. (3) of Ref.[31] by the absence of host dielectric function in front of Eq. (1). This does not affect the resulting physical quantities as the induced dipole moments coincide in both definitions.

In Eq. (2), the first term ($\sim x^2$) describes the retardation correction of polarizability (or dynamic depolarization[45,46]) due to finite particle size, the second term describes the radiative reaction correction ($\sim x^3$),[47] and the last term ($\sim x^4$) corresponds to an approximate high-order correction with neglecting the interactions between different electrostatic modes (cross terms). It should be noted that there is some close similarity between the MEM, quasi-static approximation (QSA)[48,49] and previously reported analytical approximation by Kuwata *et al.*[50] However, the important difference between MEM and Kuwata *et al.* approximation[50] is that the latter was developed as an extension of Mie theory for small spheres and then applied empirically for scattering spectra of plasmonic rods. By contrast, the MEM provides a universal description of the extinction spectra for a representative library of the NP shapes and morphologies. Besides,

MEM is formulated using electrostatic eigenmode expansion accomplished by size- and shape-dependent corrections for dynamic depolarization and radiation damping.[51,52]

Four parameters $\eta_j$, $\bar{V}_j$, $a_{j2}$, $a_{j4}$ are calculated by minimizing the deviations of the MEM solution from benchmark numerical simulations. A remarkable advantage of MEM is that the fitting parameters for metal particles depend only on the generalized aspect ratio $h \equiv AR$ and do not depend on the particle size, composition, or environment. In summary, MEM provides straightforward, accurate analytical models to simulate extinction, scattering spectra, and other electromagnetic properties of arbitrarily shaped plasmonic NPs. Owing to their analytical simplicity, MEM models can be easily implemented on any desk PC to simulate plasmonic response from realistic experimental models involving size and shape polydisperse complicated by random orientations.

In principle, the MEM approach can also be applied to the coated particles; however, it would require multiple benchmark simulations to find MEM parameters for a particular NP coating. Fortunately, this obstacle can be avoided, and the problem can be easily solved using the previously developed DEM. In our previous report,[28] we applied this approach to coated nanorods excited along the major axis. Here, we generalize the MEM+DEM method for variously shaped particles with possible random orientations.

To combine MEM with DEM, we recast first Eq. (1) as

$$\alpha_1^p = R_{1ev}^3 \sum_j q_j^p(h_1)\beta_{1j}^p , \tag{3}$$

$$\beta_{1j}^p = \frac{\varepsilon_1 - \varepsilon_m}{3\varepsilon_m + 3L_{1mj}^p(\varepsilon_1 - \varepsilon_m)} , \tag{4}$$

$$q_j^p(h_1) = \bar{V}_j^p / V_1 , \tag{5}$$

where the superscript $p$ stands for a specific polarization of the incident electric field $\mathbf{E}_0^p$; the modal volume fraction $q_j^p(h_1)$ depends on the core aspect ratio $h_1 = AR$. The generalized polarizability (3) is calculated as a weighted sum of the normalized modal polarizabilities (4) that have the same form as common normalized electrostatic polaizabilities[28] calculated with the effective modal depolarization factors $L^p_{1mj}$. For arbitrarily shaped particles and a specific polarization of the incident field, these factors account for retardation and radiation-damping effects in the following MEM form

$$L^p_{1mj} = \frac{1}{1-\eta^p_j(h_1)} - A^p_j(h_1,\varepsilon_m)\,, \tag{6}$$

$$A^p_j(h_1,\varepsilon_m) = a^p_{j2}(h_1)x^2_{1m} + i\frac{2}{9}q^p_j(h_1)X^3_{1m} + a^p_{j4}(h_1)x^4_{1m}\,, \tag{7}$$

$$x_{1m} = \frac{L_1 k_m}{2\pi}\,,\ X_{1m} = R_{1ev}k_m\,,\ k_m = k\sqrt{\varepsilon_m} = \frac{2\pi}{\lambda}\sqrt{\varepsilon_m}\,, \tag{8}$$

where $R_{1ev}$ and $L_1 = L$ are the equivolume sphere radius of the core and its characteristic size, respectively. The first term in Eq. (7) corresponds to the shape-dependent retardation correction, whereas the second one describes the radiation damping in terms of the shape-independent equivolume radius parameter $X_{1m} = k_m R_{1ev}$ multiplied by the modal volume fractions. The last term in Eq. (7) accounts for small contributions $\sim x^4_{1m}$ multiplied by shape-dependent modal coefficients $a_{j4}(h_1)$. If we consider the first dipole mode only ($j=1$), the radiation damping term in Eq. (4) denominator reduces to well-known[45-47] shape-independent form $i\frac{2}{3}X^3_{1m}$.

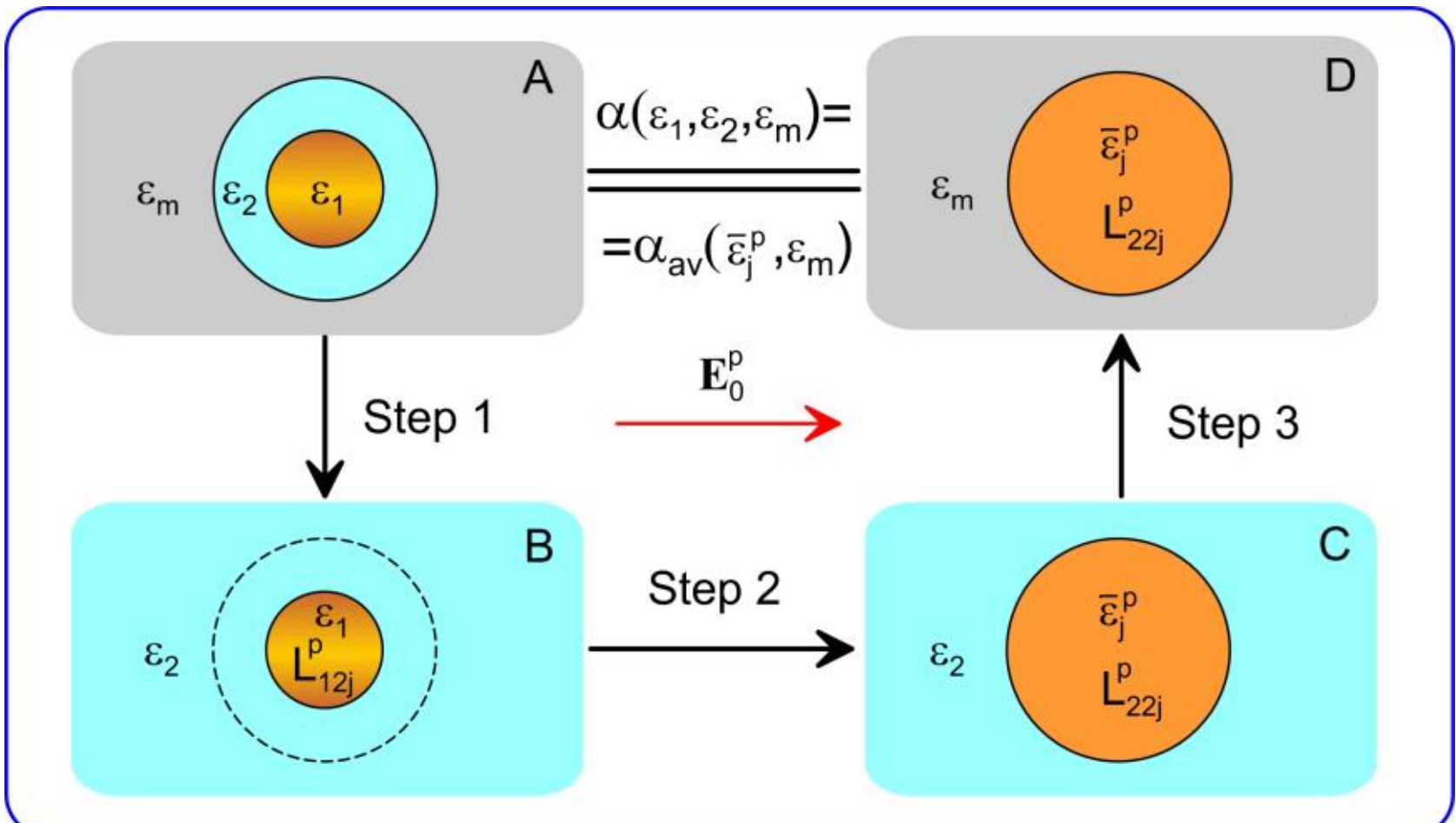

**Figure 2**. Scheme for calculating the MEM polarizability of coated plasmonic nanoparticles using DEM. In the first step, the initial coated particle (A) is placed in a homogeneous shell medium (B). In the second step, the average dielectric function and generalized depolarization factors are calculated using DEM principle (C). In the final third step, the averaged particle is plased in the external medium (D). According to DEM+MEM method, the generalized polarizabilities in panels A and D are equivalent.

Our goal is to obtain an analytical expression for the generalized MEM polarizability of coated particles by combining MEM Eqs. (3-4) with DEM[30] (Figure 2). In the first step, we place the plasmonic core in a homogeneous host medium with the shell dielectric permittivity $\varepsilon_2$. Suppose we surround the core with an imaginary surface coinciding with the shell of the original two-layer particle (dashed line in panel B). In that case, the core polarizability will not change. In the second step, we replace the imaginary particle with an averaged particle with equivalent MEM polarizability (panel C). By equating both polarizabilities, we arrive at the following DEM equation

$$R_{1ev}^3 \sum_j q_{1j}^p \frac{\varepsilon_1 - \varepsilon_2}{\varepsilon_2 + L_{12j}^p(\varepsilon_1 - \varepsilon_2)} = R_{2ev}^3 \sum_j q_{2j}^p \frac{\bar{\varepsilon}_j^p - \varepsilon_2}{\varepsilon_2 + L_{22j}^p(\bar{\varepsilon}_j^p - \varepsilon_2)} . \quad (9)$$

Unfortunately, Eq. (9) cannot be resolved with respect to $\bar{\varepsilon}_j^p$. To avoid this difficulty, we propose the following modal extension of DEM: we apply Eq. (9) to each mode $j$ separately. The validity of this assumption will be shown below when compared with rigorous calculations of cross sections. For independent modes, Eq. (9) gives the following simple solution

$$\bar{\varepsilon}_j^p = \varepsilon_2 \left( 1 + \frac{B_{12j}^p}{1 - L_{22j}^p B_{12j}^p} \right), \quad (10)$$

$$B_{12j}^p = f_{12} \frac{q_j^p(h_1)}{q_j^p(h_2)} \frac{\varepsilon_1 - \varepsilon_2}{\varepsilon_2 + L_{12j}^p(\varepsilon_1 - \varepsilon_2)}, \quad (11)$$

where $f_{12} = (R_{1ev} / R_{2ev})^3$ is the core volume fraction. Explicit expressions for effective depolarization factors and related quantities are as follows:

$$L_{i2j}^p = \frac{1}{1 - \eta_j^p(h_i)} - A_j^p(h_i, \varepsilon_2), \quad i = 1, 2, \quad (12)$$

$$A_j^p(h_i, \varepsilon_2) = a_{j2}^p(h_i) x_{i2}^2 + i \frac{2}{9} q_j^p(h_i) X_{i2}^3 + a_{j4}^p(h_i) x_{i2}^4, \quad i = 1, 2, \quad (13)$$

$$x_{i2} = \frac{L_i k_2}{2\pi}, \; X_{i2} = R_{iev} k_2, \; i = 1, 2, \quad (14)$$

where $k_2 = k\sqrt{\varepsilon_2} = (2\pi/\lambda)\sqrt{\varepsilon_2}$, $L_1$ and $L_2$ are the characteristic sizes of core and shell, respectively; the modal volume fractions $q_j^p(h_i)$ depend on the aspect ratio of core ($h_1$) and shell ($h_2$), respectively.

In the third step, we place the averaged particle in panel C in the host medium $\varepsilon_m$ (panel D) to calculate the averaged polarizability

$$\alpha_{av}^p = R_{2ev}^3 \sum_j q_{2j}^p \frac{\bar{\varepsilon}_j^p - \varepsilon_m}{3\varepsilon_m + 3L_{2mj}^p(\bar{\varepsilon}_j^p - \varepsilon_m)}, \tag{15}$$

$$L_{2mj}^p = \frac{1}{1-\eta_j^p(h_2)} - A_j^p(h_2, \varepsilon_m), \tag{16}$$

$$A_j^p(h_2, \varepsilon_m) = a_{j2}^p(h_2) x_{2m}^2 + i\frac{2}{9} q_j^p(h_2) X_{2m}^3 + a_{j4}^p(h_2) x_{2m}^4, \tag{17}$$

$$x_{2m} = \frac{L_2 k_m}{2\pi},\ X_{2m} = R_{2ev} k_m,\ k_m = k\sqrt{\varepsilon_m}. \tag{18}$$

By comparing physical pictures in panels A and D, we conclude that Eq. (15) gives the desired expression for the MEM polarizability of coated particles. Remarkably, Eq. (15) includes the core MEM parameters only, while the coating is described by its external size, shell thickness, and dielectric function, and these shell quantities are used with the same MEM functions of the core.

The above MEM+DEM scheme can easily be generalized for an arbitrary multilayered coating by recursive application of MEM to the coated averaged particle in panel D. A simplest case of a plasmonic nanorod with two-layered coating has been considered in our previous work.[28]

The averaged over random particle orientations cross sections read[35]

$$\langle C_{ext} \rangle = \pi R_2^2 \langle Q_{ext} \rangle = 4\pi \frac{1}{k_m'} \mathrm{Im}\left( k_m^2 \frac{\alpha_{av}^x + \alpha_{av}^y + \alpha_{av}^z}{3} \right), \tag{19}$$

$$\langle C_{sca} \rangle = \pi R_2^2 \langle Q_{sca} \rangle = \frac{8\pi}{3} |k_m|^4 \frac{|\alpha_{av}^x|^2 + |\alpha_{av}^y|^2 + |\alpha_{av}^z|^2}{3}, \tag{20}$$

where the angular brackets designate orientation averaging, $k_m' = \mathrm{Re}(k_m) = (2\pi/\lambda)\mathrm{Re}(n_m)$, and superscripts *x, y, z* stand for polarizations of the incident light along the corresponding frame axes related to the particle. Equations (19-20) are applicable for an arbitrary absorbing host and for a dielectric environment, they reduce to known results.[38] A simple form of Eqs. (19-20) implies that the optical cross sections depend on the incident light polarization but not on the direction of the incident wave vector $\mathbf{k}$. Clearly, this property holds only for small particles—less than 100–150 nm. Otherwise, orientation averaging should be performed, considering random $\mathbf{k}$ directions. For example, in the case of triangular nanoprisms, such numerical averaging is described in Ref.[53] (Supporting Information, Section S3). In the TMM, the orientation-averaged extinction and scattering cross sections are proportional to the trace and the norm of the particle T-matrix, respectively.[54] Thus, no numerical integration is needed.

## 4. Results and discussion.

### 4.1. Gold and silver nanorods

Before calculating the extinction and scattering spectra of coated particles, we performed several accuracy tests by comparing the extinction and scattering spectra obtained with COMSOL and TMM codes for metal nanorods. In all simulations, the particle diameter was 15 nm, typical for most experimental samples used in various applications.[55] Figure S6 demonstrates excellent agreement between both methods for a most challenging test with long nanorods ($AR = 6$). Similarly, excellent agreement was obtained for other particle models. Finally, we performed extensive simulations for AuNRs with fixed and random orientations using MEM codes in comparison with benchmark TMM calculations (Figure S7). MEM parameters for AuNRs were taken from Ref.[31] and are listed in SI. We found excellent agreement between the major extinction and scattering plasmonic peaks at longitudinal excitation and small differences for a minor perpendicular peak.

Figure 3 shows the extinction spectra of randomly oriented gold nanorods with a diameter of 15 nm, aspect ratios from 2 to 6, and the shell thicknesses of 0, 3, and 30 nm. From a practical point of view, the region of the main spectral peak associated with the longitudinal dipole resonance is of the most significant interest. With an increase in the coating thickness, the extinction peak moves to the red[56,57] and is strongly (for short nanorods) or slightly (for longer ones) increased in the peak amplitude.

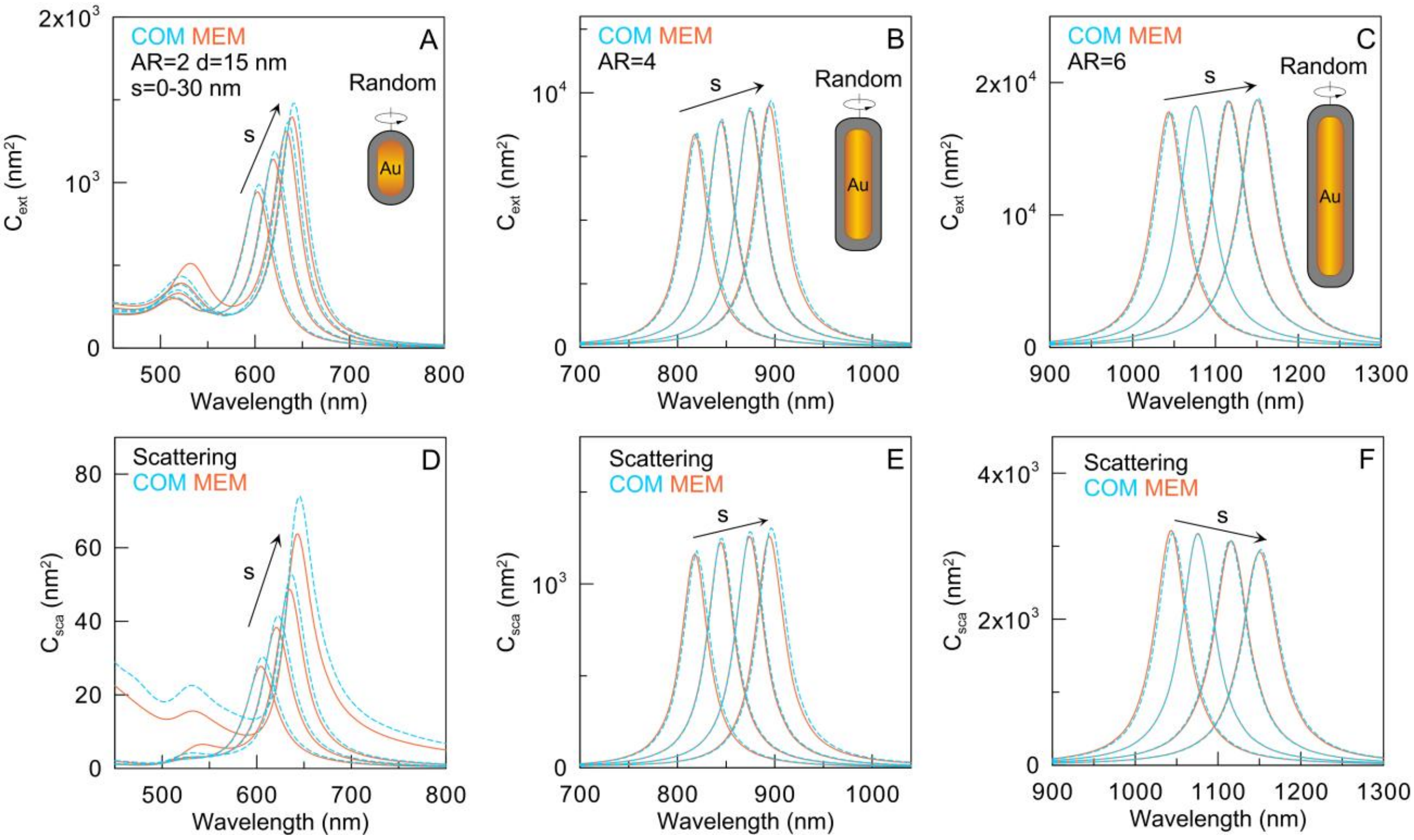

**Figure 3**. Extinction (A-C) and scattering (D-F) spectra of randomly oriented Au bare and coated randomly oriented nanorods were calculated using COM (blue) and MEM (red). The particle diameter is 15 nm, and their aspect ratio is 2 (A, D), 4 (B, E), and 6 (C, F). The coating thicknesses are 0, 3, 10, and 30 nm; the coating refractive index is 1.5.

Interestingly, for long rods, the scattering peak does not increase but decreases with increasing thickness of the coating (panel F). Also, for small AuNRs, an increase in the shell thickness from 3 to 30 nm strongly increases the short-wavelength scattering (panel D).

Extinction and scattering MEM spectra of randomly oriented silver nanorods reproduce the main longitudinal plasmonic TMM peak quite accurately (Figure S8). In contrast to data for gold nanorods, the short wavelength portion of extinction and scattering spectra demonstrates a complicated structure. MEM reproduces two minor peaks near 355 and 375 nm with acceptable accuracy. In addition, spectra in Figure S8 exhibit multipole resonances that MEM cannot reproduce.

The physical origin of multipole peaks is elucidated by Figure 4, where panels A and B show extinction and scattering spectra for a randomly oriented ensemble of AgNRs with $AR = 4$, whereas panels C and D show similar spectra for two specific fixed orientations. A 475-nm peak in panel A is related to quadrupole[58] absorption because it is absent in scattering spectra (panel B). This quadrupole resonance is excited most strongly by TM plane wave (electric field lies in (**a,k**) plane; **a** is the rod symmetry axis) whose wave vector is directed at 54 degrees with respect to the symmetry axis **a** (dashed spectrum in panel C). It should be emphasized that the 475-nm quadrupole peak disappears at longitudinal excitation due to symmetry constraints. Similarly to the scattering spectrum for random ensemble (panel B), this peak is not seen in scattering spectrum in panel D.

The second multipole peak in panel A is located near 410 nm, which is also caused by absorption rather than scattering. However, in contrast to the 475-nm quadrupole, this 410-nm peak is most effectively excited by the longitudinal electric field (solid blue line in panel C). Although we observe a small similar scattering peak in panel D, its contribution is negligible compared to the absorption.

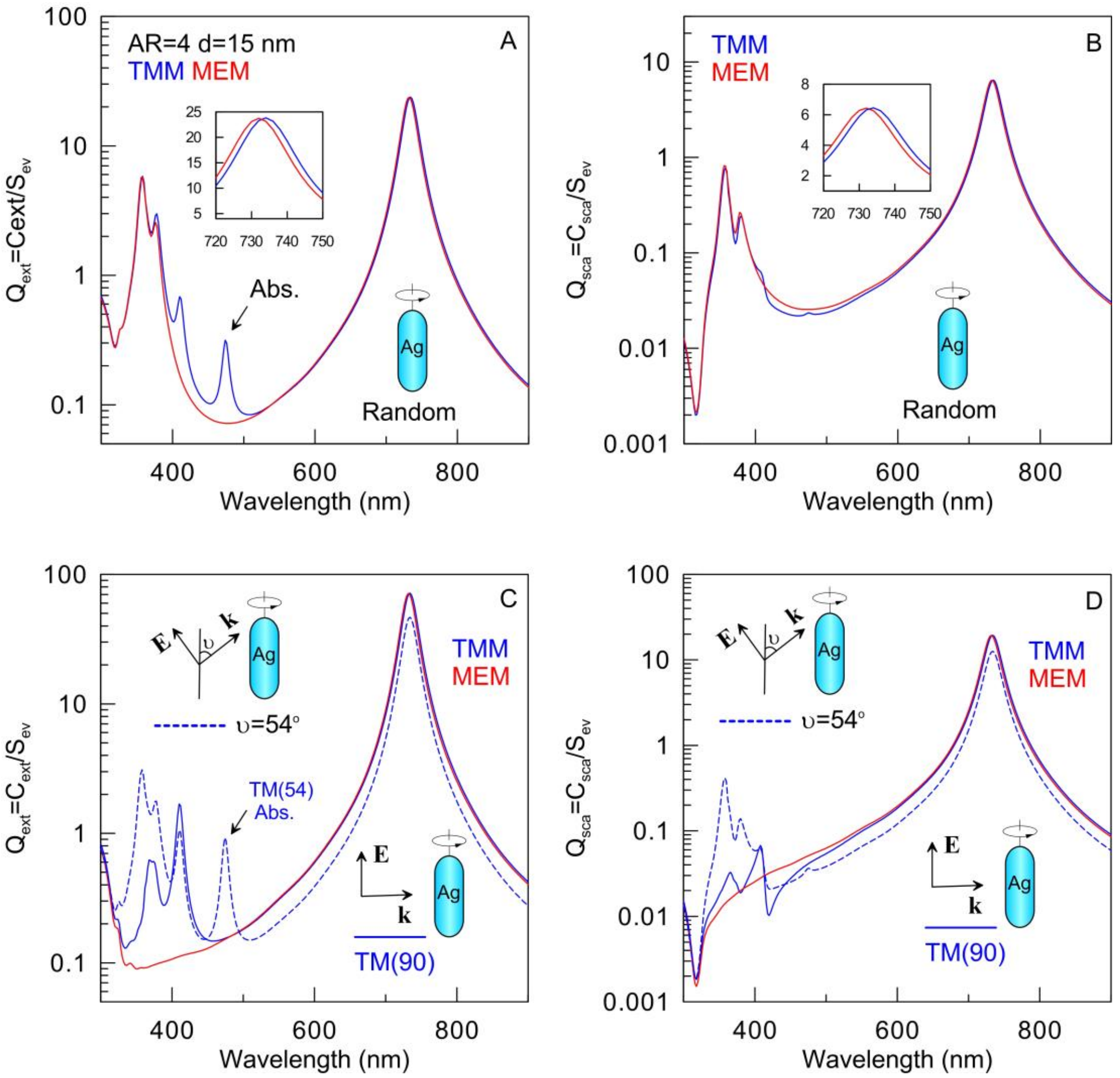

**Figure 4**. Extinction and scattering spectra calculated by TMM (blue) and MEM (red) for random particle orientations (A, B) and excitation by the TM incident wave with wave vector directed perpendicularly (solid blue line) and at 54 degrees (dashed blue line) to the rod symmetry axis (C, D). The arrow in panel C indicates the quadrupole absorption peak. Note that there is excellent agreement between TMM and MEM in the description of the main plasmonic peak (solid blue and red curves).

Figure 5 illustrates the quality of MEM approximation in describing the main plasmonic peaks of coated silver nanorods. Some blue deviation of MEM spectra is observed for a minimal aspect ratio $AR = 2$, while the agreement becomes almost perfect for longer nanorods. We emphasize that we used the same universal MEM parameters as those for gold particles in this case. If one makes an independent MEM fitting for silver nanorods, the quality of modeling the spectra of coated particles will be higher.

For short AgNRs ($AR_1 \le 2$) with a dielectric coating, the scattering, absorption, and extinction cross sections increase. This behavior is expected based on the limiting case of coated silver nanospheres. In Section S2 of the SI file, we derive a simple yet accurate approximation for the LPR scattering cross section of small coated spheres:

$$C_{sca,2} = \pi R_2^2 Q_{sca,2} = \pi R_2^2 \frac{8}{3}\left(k_m R_2\right)^4 \left|\frac{\varepsilon_{av} - \varepsilon_m}{\varepsilon_{av} + 2\varepsilon_m}\right|^2, \quad (21)$$

$$\varepsilon_{av} = \varepsilon_2 \left(1 + 3f + i\frac{9\varepsilon_2 f}{\varepsilon_1''}\right), \quad (22)$$

$f \equiv f_{12} = (R_1 / R_2)^3$ is the core volume fraction and $\varepsilon_1''$ is the imaginary part of silver dielectric function at LPR wavelength. As Eqs. (21–22) predict, increasing the thickness of the dielectric shell leads to an increase in scattering, in agreement with numerical and MEM+DEM simulations. Regarding long AuNRs, panels E and F demonstrate an unexpected decrease in scattering amplitude for coated silver nanorods. To gain insight into the possible physical mechanisms behind this unusual behavior, we performed separate calculations of the absorption, scattering, and extinction spectra (Figure S9). The absorption spectra show an increasing trend for short ($AR_1 = 2$), moderate ($AR_1 = 4$), and long ($AR_1 = 6$) nanorods, whereas the scattering increases with shell thickness for short nanorods but decreases for long ones. Since absorption dominates, the extinction spectra follow the absorption trend.

The different dependence of absorption and scattering on shell thickness can be qualitatively explained as follows: the absorption cross-section efficiency is primarily determined by the metal core losses and is weakly affected by the outer dielectric shell. This accounts for the moderate increase in absorption cross sections across all aspect ratios. In contrast, the scattering cross section is determined by the squared modulus of the polarizability (see Eq. (20)). It is known that the polarizability of coated particles can become zero under certain conditions (see, e.g., Eq. (5.36) in Ref.[38] and the discussion on p. 150). Therefore, the decrease in scattering observed in panels E and F of Figure 5 can be qualitatively explained by similar reasons, where the coating acts as a clearing agent.

An instructive example is the scattering from elongated silver nanorods with a confocal dielectric coating. For this model, there exist analytical solution in the electrostatic approximation (EA)[28]

$$\left\langle C_{sca}^{EA} \right\rangle = \pi R_{2ev}^2 \left\langle Q_{sca}^{EA} \right\rangle = \frac{8\pi}{3} |k|^4 \frac{\left|\bar{\alpha}_c^{EA}\right|^2 + 2\left|\bar{\alpha}_a^{EA}\right|^2}{3}, \tag{23}$$

where the average polarizabilities along the spheroid semiaxes $a = b$ and $c \geq a$ are calculated by usual electrostatic formulae for homogeneous spheroidal particles[38]

$$\bar{\alpha}_a^{EA} = R_{2ev}^3 \frac{\bar{\varepsilon}_a - \varepsilon_m}{3\varepsilon_m + 3L_{2a}(\bar{\varepsilon}_a - \varepsilon_m)}, \ \bar{\alpha}_c^{IEA} = R_{2ev}^3 \frac{\bar{\varepsilon}_c - \varepsilon_m}{3\varepsilon_m + 3L_{2c}(\bar{\varepsilon}_c - \varepsilon_m)}. \tag{24}$$

However, in the case of coated particles, the average dielectric functions $\bar{\varepsilon}_a$ and $\bar{\varepsilon}_b$ are calculated using the DEM method. [28] Note that, even for isotropic metal particles and isotropic coatings, the average DEM dielectric function is anisotropic with the principal components $\bar{\varepsilon}_a$ and $\bar{\varepsilon}_b$. To eliminate the influence of the quadratic dependence of the scattering cross section on particle volume and to isolate shape effects, we consider particles with a fixed volume (corresponding to the diameter of an equivalent sphere of 15 nm) and a variable aspect ratio $AR_1$ from 1 to 6. Figure S10 shows that, with increasing aspect ratio, the absorption of the silver spheroid increases. Since absorption dominates over scattering, the extinction also increases. However, the scattering cross section initially increases rapidly and then begins to decrease. Covering the spheroid with a dielectric shell has little effect on the overall dependence of the cross sections on the core aspect ratio. While for spheres and short spheroids the coating increases scattering, for long particles, the coating, on the contrary, decreases scattering. Thus, we conclude that the increase or decrease in the scattering cross section with increasing shell thickness results from a delicate balance between the scattering influenced by the plasmonic core and the scattering contributions from the shell.

In most sensing, imaging, and other biological and medical applications, the maximal sizes of silver and gold nanorods are less than $20\times120$ nm. Therefore, we conclude that a simple MEM+DEM approach can be used to simulate the plasmonic properties of dielectric-coated nanorods with reasonable accuracy.

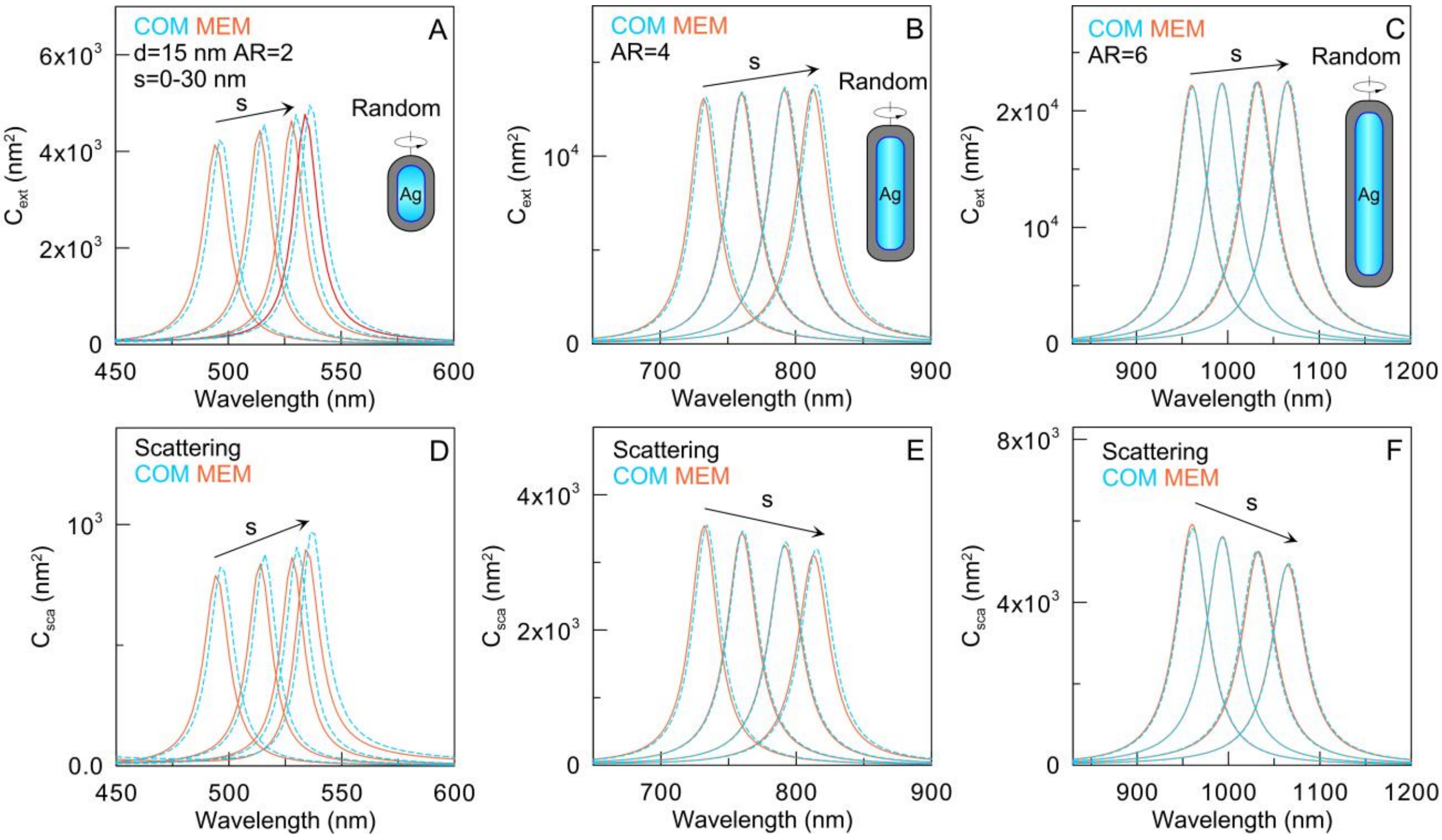

**Figure 5**. Extinction (A-C) and scattering (D-F) spectra of randomly oriented Ag bare and coated nanorods were calculated using COM (blue) and MEM (red). The particle diameter is 15 nm, and their aspect ratio is 2 (A), 4 (B), and 6 (C). The coating thickness is 0, 3, 10, and 30 nm; the coating refractive index is 1.5. Note decreased scattering peaks with increased shell thickness (panels E and F).

### 4.2. Gold and silver nanodisks

To verify the accuracy of MEM for disks, we compared benchmark TMM spectra with those calculated by MEM. Figure S11 illustrates excellent agreement between TMM and MEM extinction and scattering spectra for thin bare Au disks with 10 and 20 nm thicknesses. At the same time, some minor differences are observed for thicker 30-nm disks. Similarly, for coated gold nanodisks, we observed excellent agreement between MEM and COMSOL extinction and scattering spectra of thin 10-nm (data not shown) and good agreement for 20-nm particles with the coating thickness ranging from 3 to 30 nm (Figure 6 A,B).

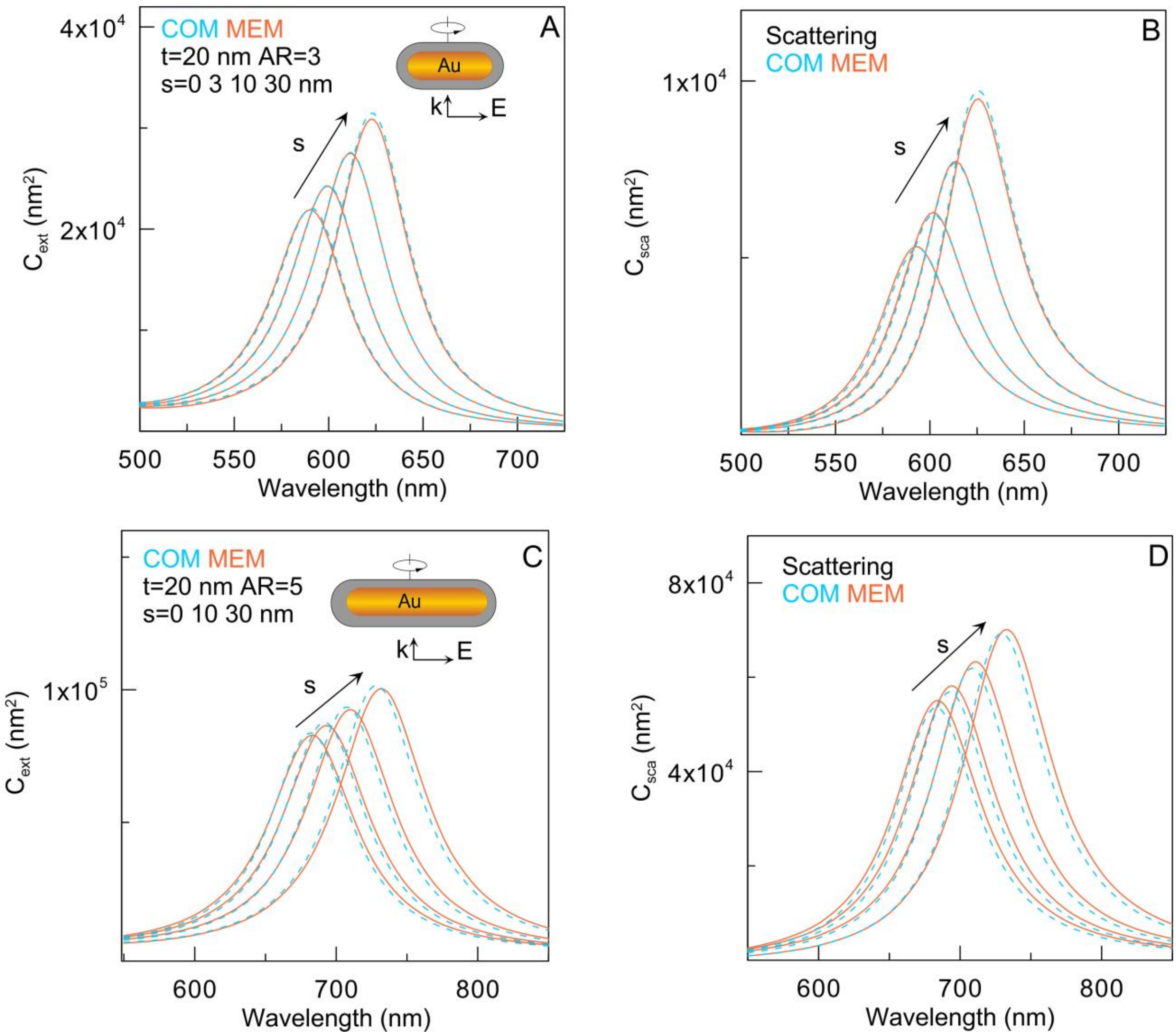

**Figure 6**. Extinction (A, C) and scattering (B, D) spectra of gold bare and coated nanodisks were calculated using MEM (red lines) and COMSOL (blue dashed lines). The particle thickness is $L = 20$ nm, and their aspect ratios are 3 (A, B) and 5 (C, D). The coating thickness varies from 0 to 30 nm; the coating refractive index is 1.5. The wave vector is directed along the disk axis.

However, the MEM approximation accuracy becomes lower with increased gold disk thickness and diameter (Figure 6 C, D). In particular, Figure 7A shows an evident red shift and broadening of MEM spectra compared to COM ones when calculated for both bare (solid lines) disks and coated (dashed lines) nanodisks with 30-nm thickness. A possible physical origin behind the low accuracy of MEM could be attributed to the limited size-dependent ($\sim x^4$) correction for such large disks. Such limitations become quite evident if we consider the case when both the wave vector and electric field are perpendicular to the disk symmetry axis (Figure 7B). Specifically, panel B clearly demonstrates strong excitation of the quadrupole absorption mode near 650 nm, making a dominant contribution to the extinction spectrum (black line). Note that for randomly oriented disks, this quadrupole peak is not seen (green spectrum).

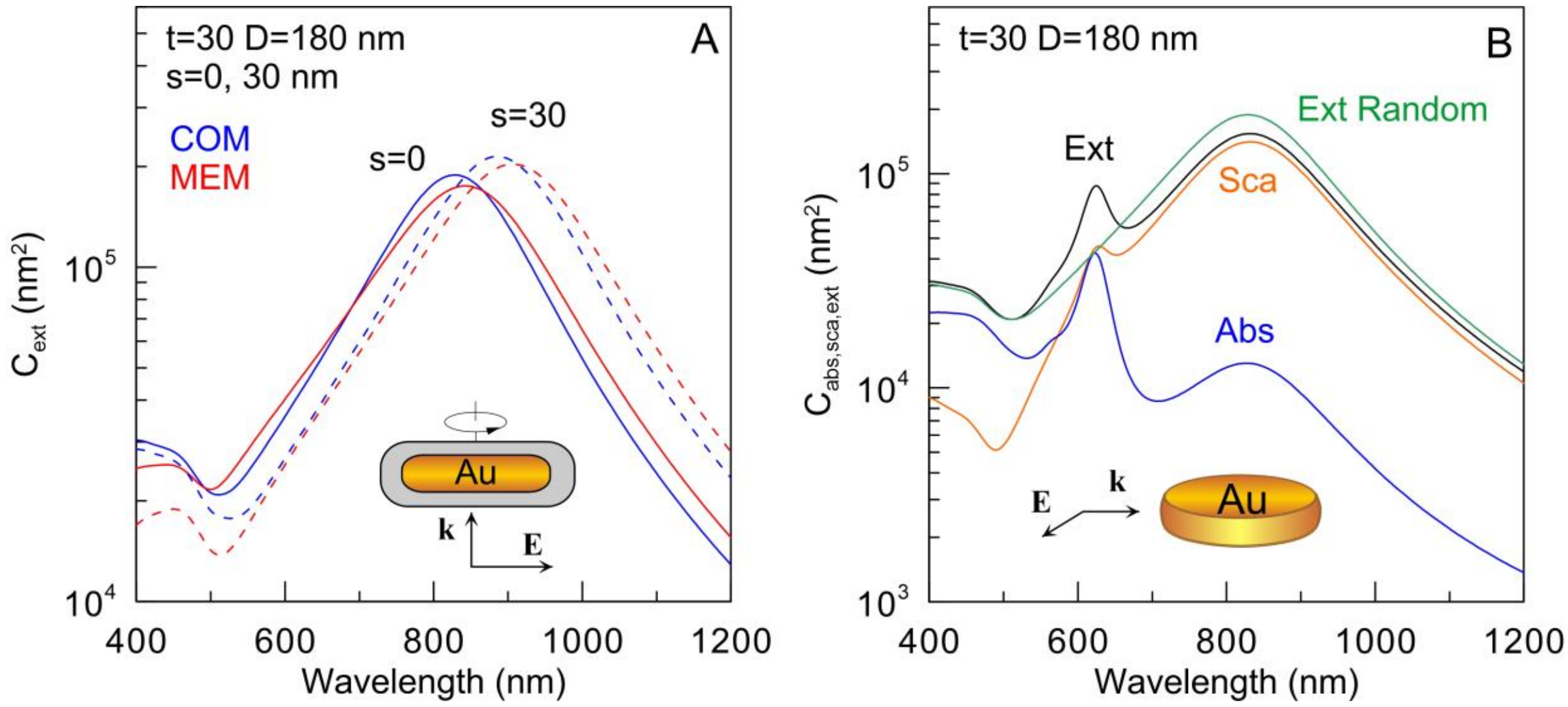

**Figure 7**. (A) Extinction spectra calculated by COMSOL (blue) and MEM (red) for bare (solid lines) and coated (dashed lines, $s = 30$ nm) nanodisks; the incident wave vector is parallel to the symmetry axis. (B) Absorption (blue), scattering (orange), and extinction (black) spectra for perpendicular incidence. The green spectrum corresponds to random particle orientations.

Thus, the MEM approximation can be used to simulate the extinction and scattering spectra of oblate gold particles if their diameter does not exceed 150 nm and the coating thickness is less than 20-30 nm, depending on the disk diameter.

To verify the accuracy of COMSOL data for silver disks, we compared benchmark TMM spectra with those calculated by COMSOL. Figures S12 and S13 illustrate almost perfect agreement between TMM and COMSOL spectra for 20-nm and 30-nm silver disks with aspect ratios from 3 to 6. Some minor differences are observed only for the short-wavelength multipole peaks for the largest 30x180 nm disks (Figures S13 C, D).

Figure 8 shows extinction and scattering spectra for 20-nm silver disks at in-plane excitation. Similar to the data in Figure 6, we observe excellent agreement between MEM and COMSOL spectra when the disk diameter is less than 60 nm. For larger disks with diameters of 100 nm (the aspect ratio is 5), we note more evident deviations of MEM spectra from COMSOL ones. Similar data are presented in the SI file for thin 10-nm and thick 30-nm disks (Figure S14).

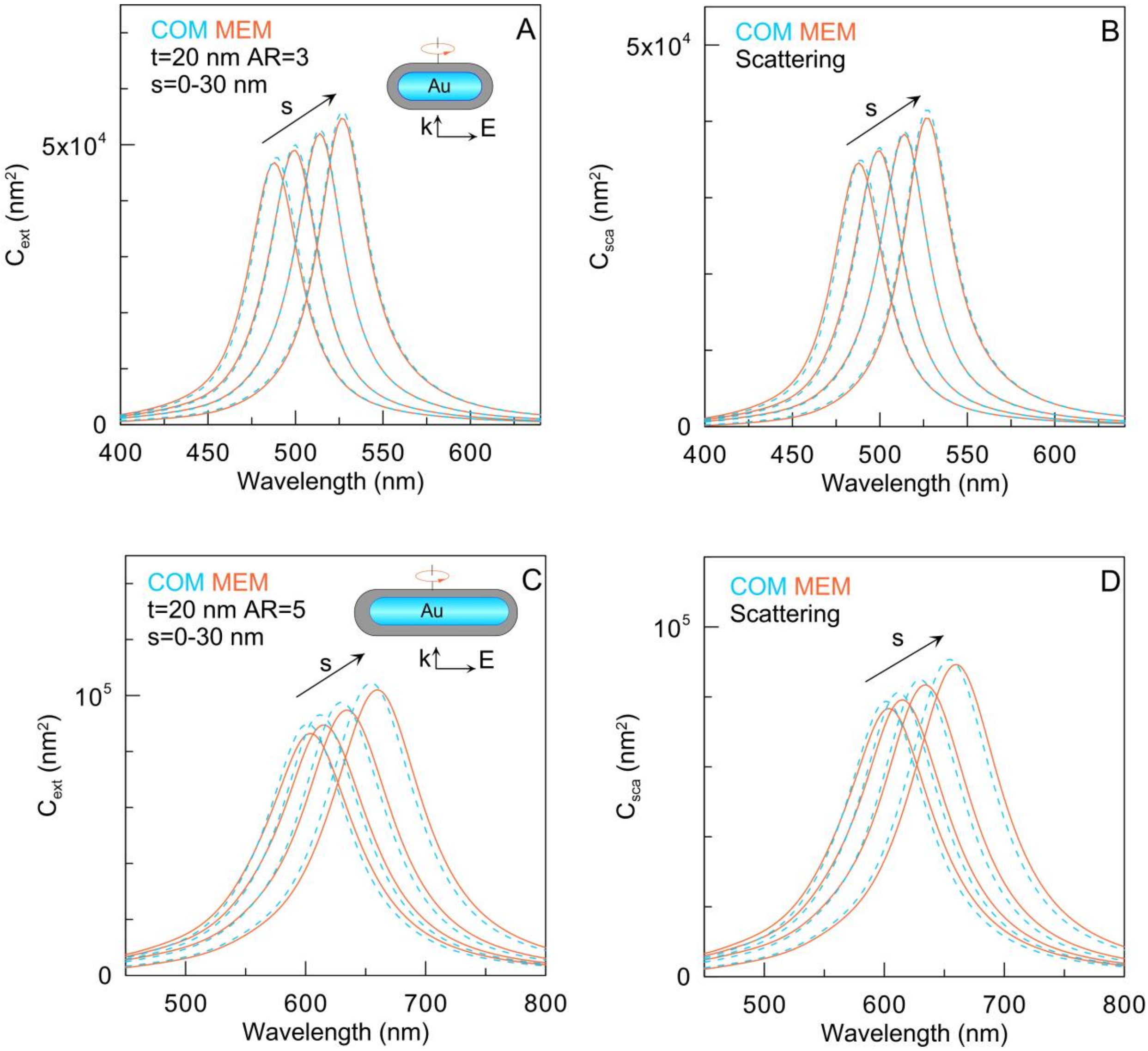

**Figure 8**. Extinction (A, C) and scattering (B, D) spectra of silver bare and coated nanodisks were calculated using MEM (red lines) and COMSOL (blue dashed lines). The particle thickness is $L = 20$ nm, and their aspect ratios are 3 (A, B) and 5 (C, D). The coating thickness varies from 0 to 30 nm; the coating refractive index is 1.5. The wave vector is directed along the disk axis.

In summary, the MEM approximation can be used to simulate the extinction and scattering spectra of silver disk-shaped particles if their thickness and diameter do not exceed 20 and 100 nm and the coating thickness is less than 30 nm, depending on the disk diameter.

### 4.3. Gold and silver nanoprisms

Figures S15-S16 compare COM and MEM extinction and scattering spectra for thin AuNTs with thicknesses ranging from 10 to 20 nm and aspect ratios ranging from 3 to 15. Note that MEM does not reproduce multipole peaks but approximates major plasmonic peaks well. We derived new MEM parameters that ensure good accuracy within much broader size and shape ranges compared to those in Ref.[31]. Similar conclusions can also be drawn for AgNTs (Figure S17). Our nanoprism model has the same smooth edges with a small rounding radius as proposed in Ref.[31]. Effects of larger rounding radii have been reported in Ref.[59]

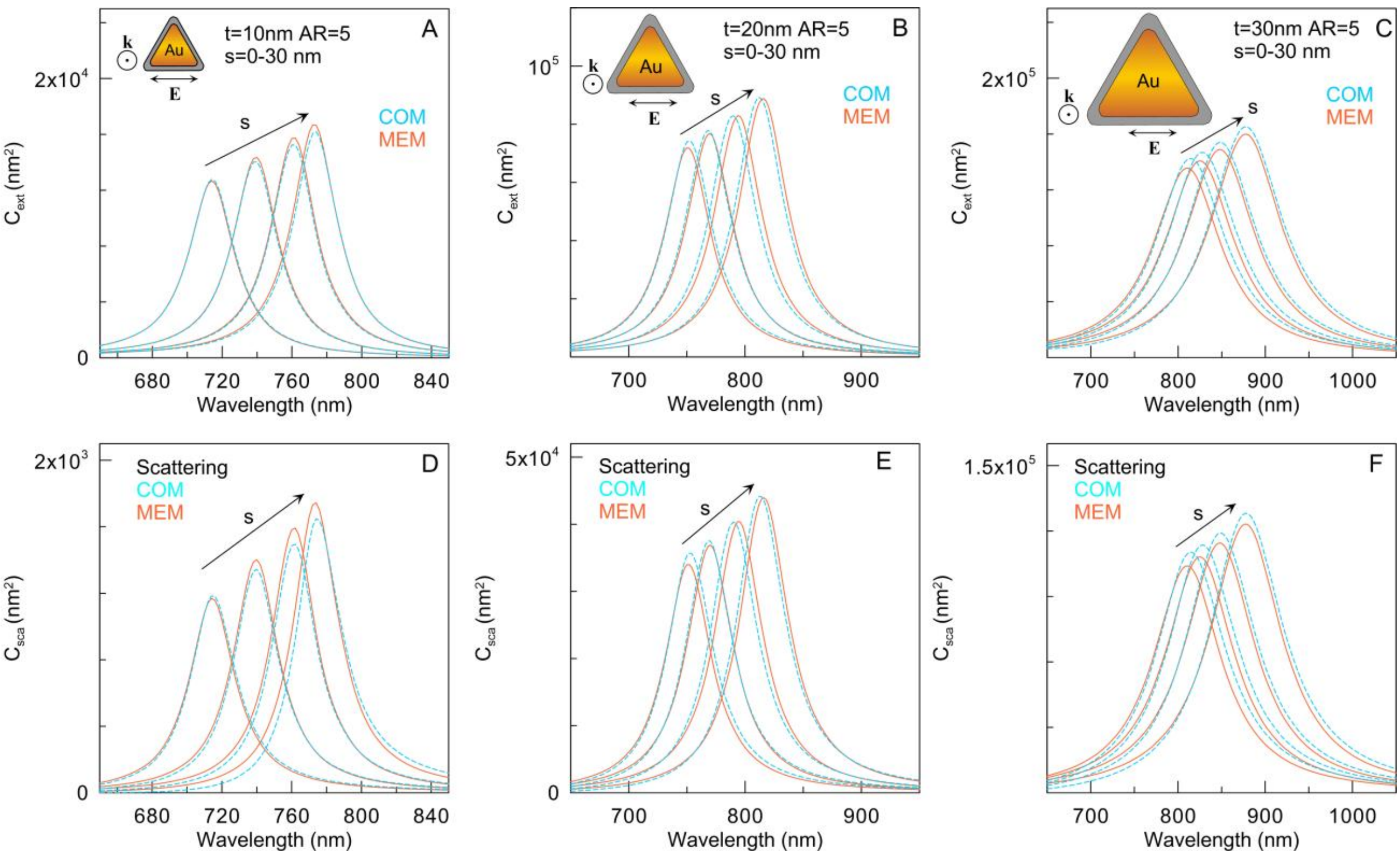

**Figure 9**. Extinction (A–C) and scattering (D–F) spectra of gold bare and coated nanoprisms were calculated using MEM (orange) and COMSOL (blue). The prism thickness $t = 10$ (A, D), 20 (B, E), and 30 nm (C, F); the core aspect ratio $AR = L/t$ is constant and equals 5, and the coating thicknesses $s$ are 0, 3, 10, and 30 nm. The wave vector is directed along the nanoprism side (in-plane excitation.

Figure 9 illustrates the effects of coating on the extinction and scattering plasmonic peak position and amplitude as predicted by COMSOL and MEM. These simulations show that MEM approximation gives good accuracy provided that the prism sizes are within $30\times150$ nm and the coating thickness is less than 30 nm. These limits satisfy typical parameters of gold nanotriangles fabricated by modern chemical protocols.[60] The same conclusion applies to coated silver nanoprisms (Figure 10).

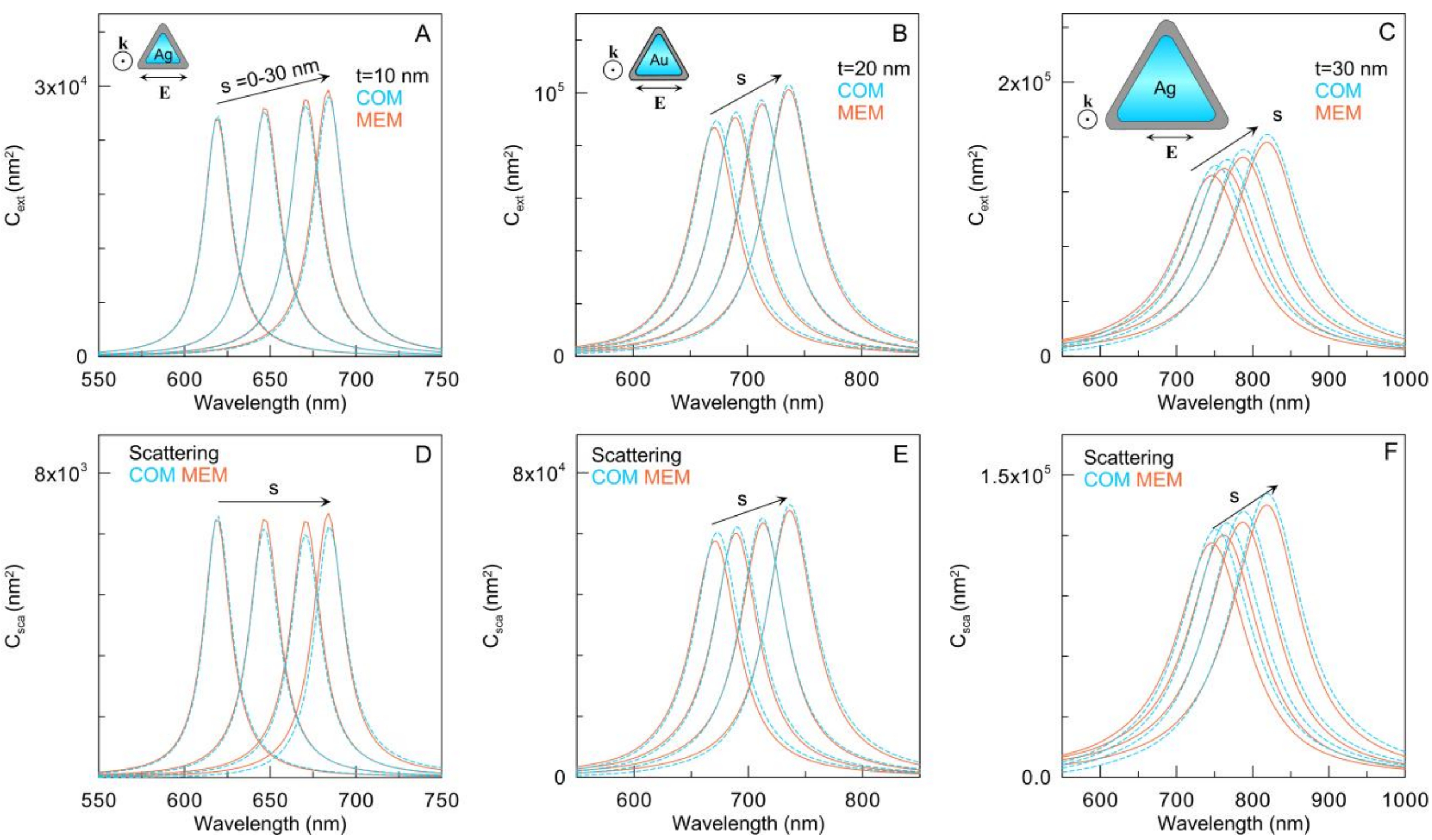

**Figure 10**. Extinction (A–C) and scattering (D–F) spectra of silver bare and coated nanoprisms were calculated using MEM (orange) and COMSOL (blue). The prism thickness $t = 10$ (A, D), 20 (B, E), and 30 nm (C, F); the core aspect ratio $AR = L/t$ is constant and equals to 5, and the coating thicknesses $s$ are 0, 3, 10, and 30 nm. The wave vector is directed along the nanoprism side (in-plane excitation.

### 4.3. Gold bicones and bipyramids

Pentagonal gold BPs are a new class of popular anisotropic nanoparticles whose extinction and scattering spectra resemble those of gold nanorods.[61-64] During the chemical etching of as-prepared initial AuBPs, their shape subsequently transforms to rounded BPs, then to rounded bicones (BCs), and finally to nanospheres.[65] Keeping in mind these intermediate shapes, we include AuBCs in our consideration.

In line with the previous report,[31] our analytical MEM model accurately describes the numerical LPR peaks for gold bicones (Figure S18). We showed above that for rods, disks, and nanoprisms, the analytical MEM+DEM method well describes the extinction and scattering spectra of coated particles. However, we found noticeable discrepancies in the plasmon resonance shifts predicted by the MEM+DEM and COMSOL numerical calculations for coated bicones (Figure S19). Analytical theory for bilayer gold and silver nanospheres[57] predicts a linear relationship between the plasmon resonance shift and the volume fraction of the dielectric shell $g = V_{shell} / (V_{core} + V_{shell})$. Following this prediction, in Figure 11, we show the dependence of the plasmon extinction shift on the shell volume fraction, calculated for bicones with aspect ratios 2 and 4. It can be seen that the analytical orange curves (MEM) are very different from the numerical calculation data (black crosses). The physical reasons for this discrepancy can be associated with a strong field inhomogeneity inside and around the particle (Figure 12). This contradicts the principle of dipole equivalence based on the equality of dipole moments and polarizabilities of multilayer and a homogeneous equivalent particle. Physically, such equivalence implicitly assumes a uniform field distribution in the equivalent particle, which is violated in the case of bicones, for which the vertices are singular points of strong field concentration.

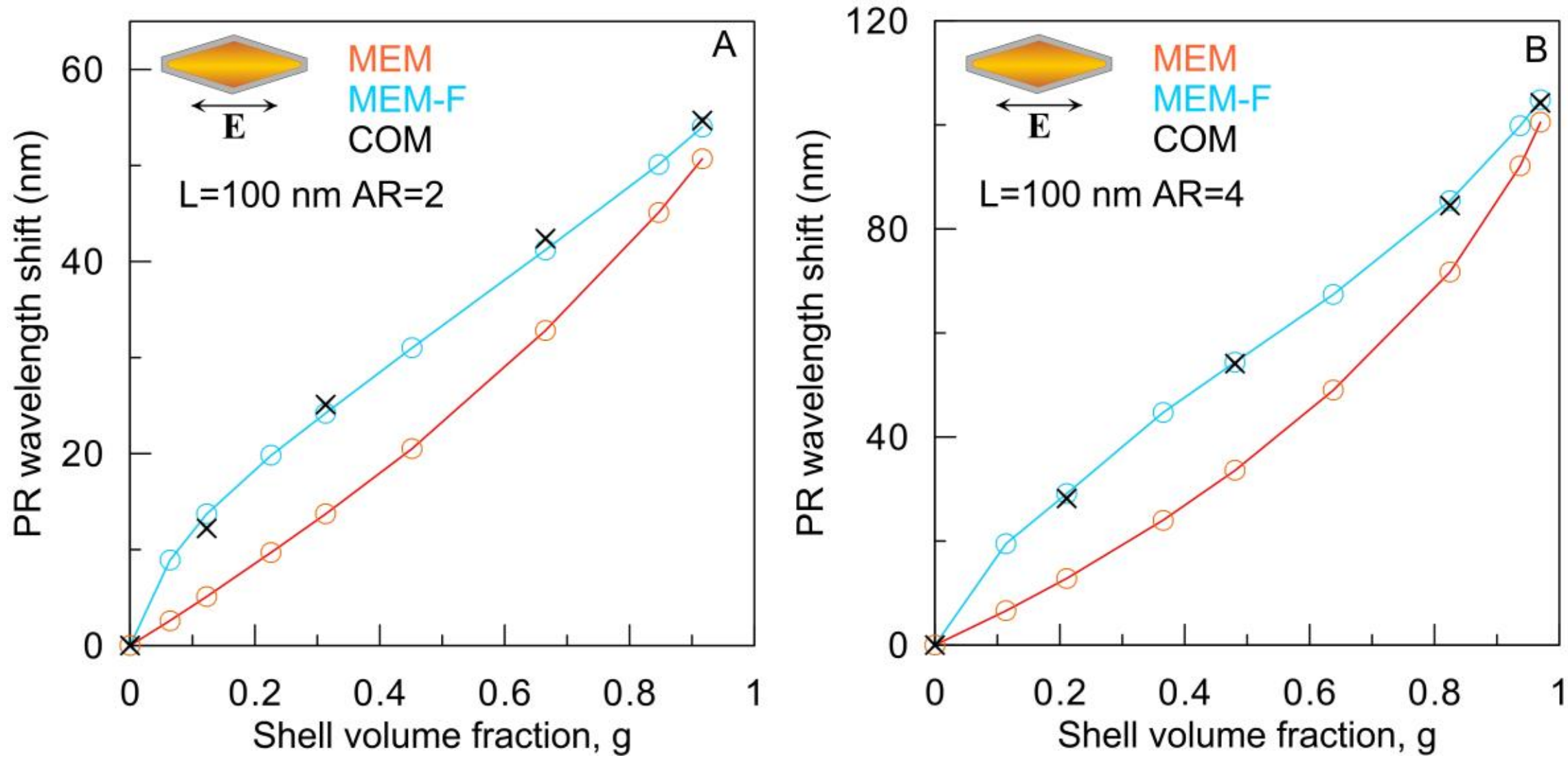

**Figure 11**. Dependences the plasmon resonance shift on the shell volume fraction $g$ calculated by MEM (orange curve) and COMSOL (black crosses) for bicones with the total length $L = 100$ nm and aspect ratio $AR = 2$ (A) and 4 (B). Blue lines show calculations by the improved version of MEM-F.

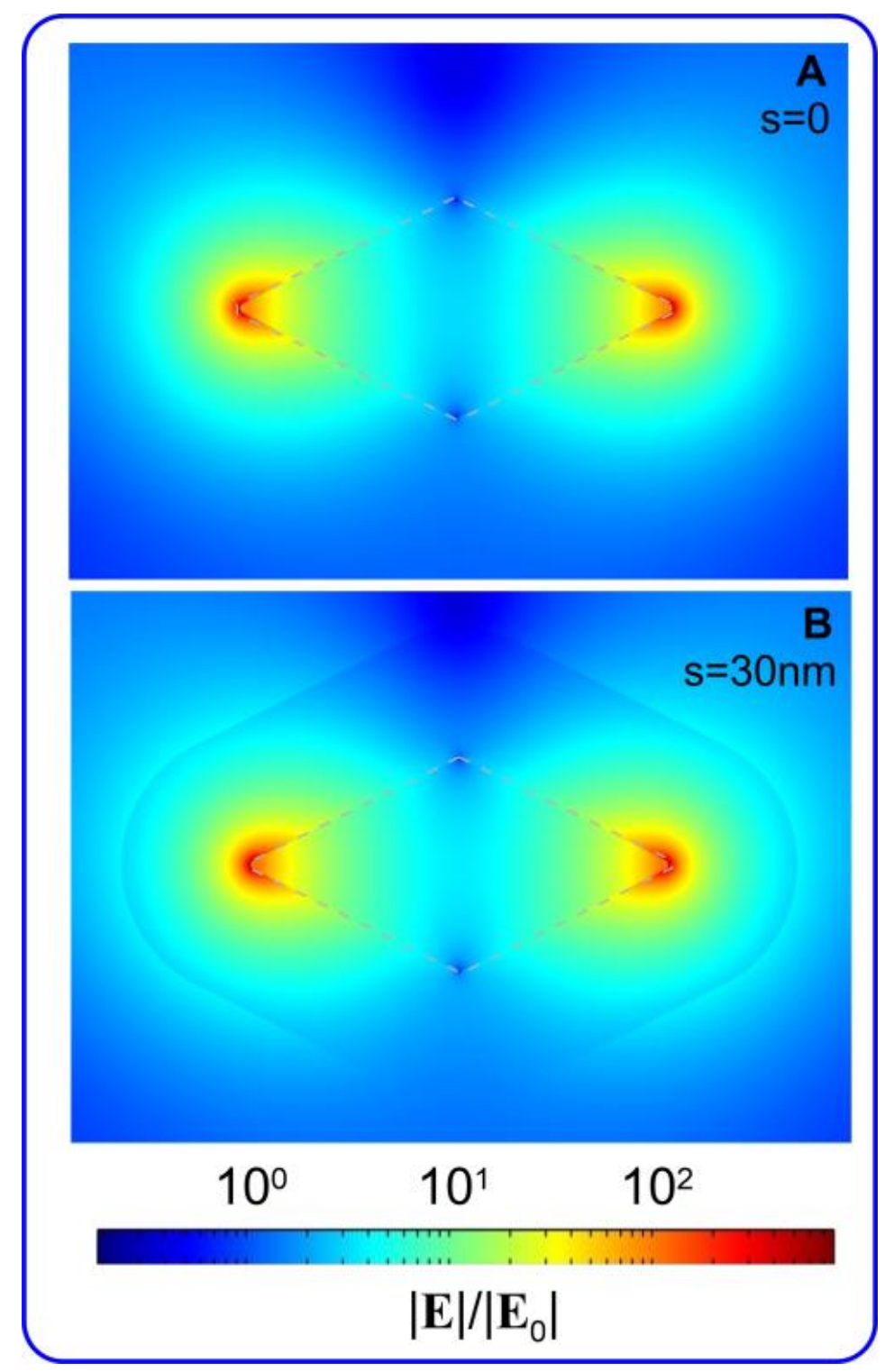

**Figure 12**. Field distribution for bare (A; s=0) and coated (B, s=30 nm) gold bicones at the longitudinal excitation along the symmetry axis. The wavelengths correspond to the major plasmonic peaks.

To eliminate this drawback of MEM, we assumed that the MEM parameters $\eta_1$ and $V_1$ in equation (1) can be represented in a factorized form (for simplicity, we consider below only the first mode)

$$\eta_1 = \eta_1(AR_i) f_\eta(s)\,,\; V_1 = V_1(AR_i) f_V(s)\,, \qquad (21)$$

where the first factors correspond to the MEM parameters calculated for the aspect ratio of the core ($AR_1$) or shell ($AR_2$) according to the formulas specified in the SI. The second factor in equation (21) considers the specific shell effects and satisfies the apparent conditions $f_\eta(0)=1$ and $f_V(0)=1$. Explicit formulas for functions $f_\eta(s)$ and $f_V(s)$ are given in the supporting information file. We emphasize that the factorized correction (21) for the shell effect is universal and does not depend on the particle size, shape, and metal composition. Otherwise, the analytical description would be reduced to a non-universal fitting of the spectra for a specific particle shape and a specific shell thickness.

The blue curves in Figure 11 were calculated using the described scheme (called factorized MEM or MEM-F here). We see excellent agreement between the LPR shift values from analytical theory and numerical calculations. Similarly, the analytical extinction and scattering spectra (Figure 13) reproduce COMSOL in peak positions and amplitudes. One important note is in order here. Due to the strong localization of the local field near bicone tips, we observe extraordinary sensitivity of the LPR peak positions to the local dielectric environment. In particular, even for the smallest shell thickness of 1 nm, we observed the LPR shift to about 30 nm for AR=4. By contrast, the LPR shift is one order lower for spherical particles with similar shell volume fraction, less than 3 nm.[57] We also note different coating effects on the scattering cross section for small (panel E) and large (panel H) bipyramids at a constant bicone length.

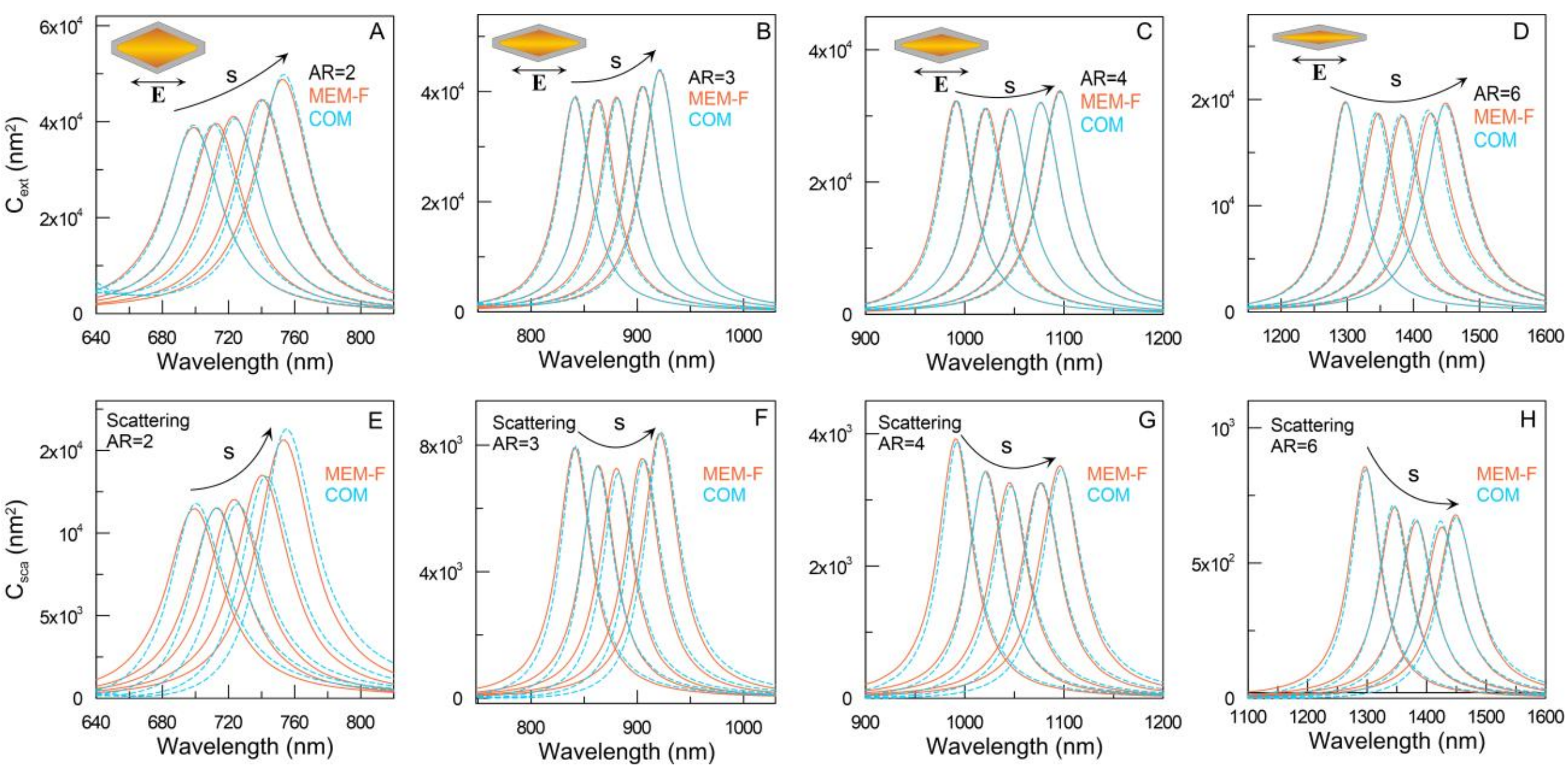

**Figure 13**. Extinction (A-D) and scattering (E-H) spectra were calculated by MEM-F (orange) and COMSOL (blue) for coated Au bicones with a total length of 100 nm and aspect ratios 2, 3, 4, and 6. The shell thicknesses are 0, 1, 3, 10, and 30 nm.

Now we proceed to the coated bypiramids. In the original MEM papers by Yu et al.[31], the rounding radii of the BP tip and base were too small and did not match experimental particles. From experimental and geometrical inversion data published by Montaño-Priede *et al.*[65], the tip rounding radius increases with a decreased rounded BP length. Similarly, the base rounding radius also increases during the chemical etching. In this work, we included both rounding effects in our geometrical BP model (for details, see the SI file, Section S1.5). Note that in the cited work[65], the authors disregarded the base rounding of their BPs. It follows from the above-cited papers that the average BP width $w \equiv d$ ranges from 20 to 35 nm, and the maximal aspect ratio $AR = L / d$ is typically less than 6. The readers are referred to the SI file for all analytical derivations concerning the rounded BC and BP shapes and volumes. Here, we note only one key point. The initial (non-rounded) and rounded volumes of coated particles can be calculated analytically for bicone shape. The BP shape is characterized by a circle's diameter embedded in the pentagonal base and the total length of the embedded bicone. An analytical calculation of the rounded BP volume is not trivial, especially for the coated particles. To avoid this difficulty, we assumed that the volume of a coated rounded BP can be approximated as follows:

$$V_{\mathrm{BPr}}(s) = V_{\mathrm{BPr}}(s=0)\frac{V_{\mathrm{BCr}}(s)}{V_{\mathrm{BCr}}(s=0)}, \qquad (22)$$

where the symbol 'r' stands for rounded particle shape. Direct numerical calculations showed excellent accuracy of the above approximation (22) with relative errors less than 0.002%.

Figure 14 shows the extinction and scattering spectra of coated BPs for three aspect ratios that ensure the LPR peak positions within the most important biological window, 650-900 nm. After dielectric coating with a thickness of 1 to 30 nm, the LPR positions shift to the red by 70 nm at AR=3 and 100 nm at AR =5. One can see excellent accuracy of the MEM+DEM approximation without any factorization, which we had to apply for more sharp bicones. Additional spectra can be found in the SI file.

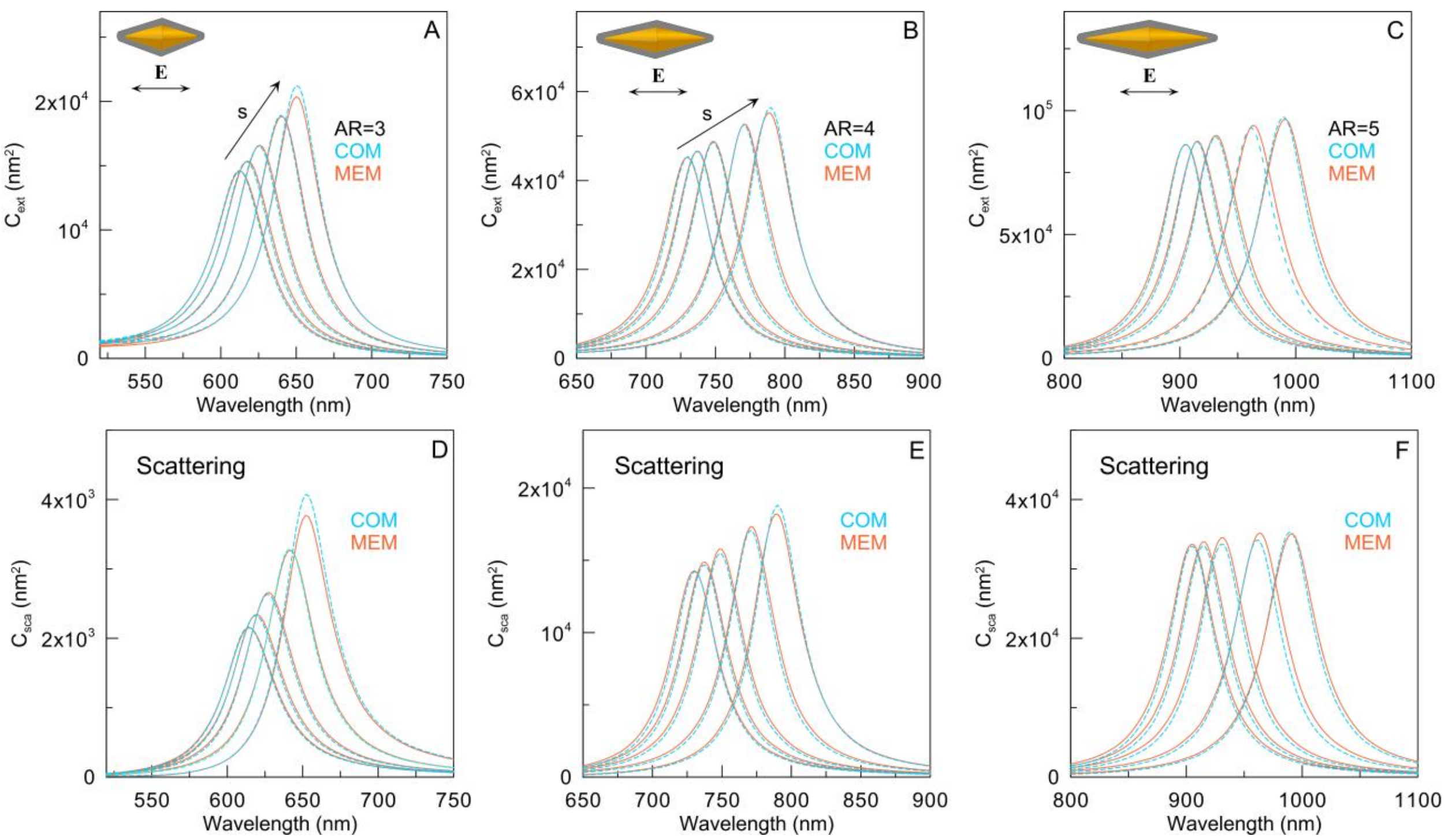

**Figure 14**. Extinction (A-C) and scattering (D-F) spectra were calculated by MEM (orange) and COMSOL (blue) for coated rounded gold bipyramids with a constant width of 30 nm and aspect ratios 3, 4, and 5. The shell thicknesses are 0, 1, 3, 10, and 30 nm.

In contrast to gold BPs, silver pentagonal BPs are rare examples in the plasmonic nanoparticle family.[66] Nevertheless, several works reported synthesizing and characterizing anisotropic silver nanoparticles similar to gold BPs, mainly silver right bipyramids.[67-69] Figure 15 illustrates an application of the MEM+DEM method to the coated silver pentagonal BPs. Because of sufficiently large rounding radii, there was no need for additional factorization to reproduce COMSOL-simulated extinction and scattering spectra. The only difference from the AuBP case is that we modified MEM parameters slightly (see SI file, Section 1.5). Let us note a change in the tendency in the scattering peaks from increasing to decreasing with increasing thickness of the coating of short (E) and long (H) bipyramids. Clearly, the amplitudes of absorption peaks increase in all cases with an increase in the coating thickness.

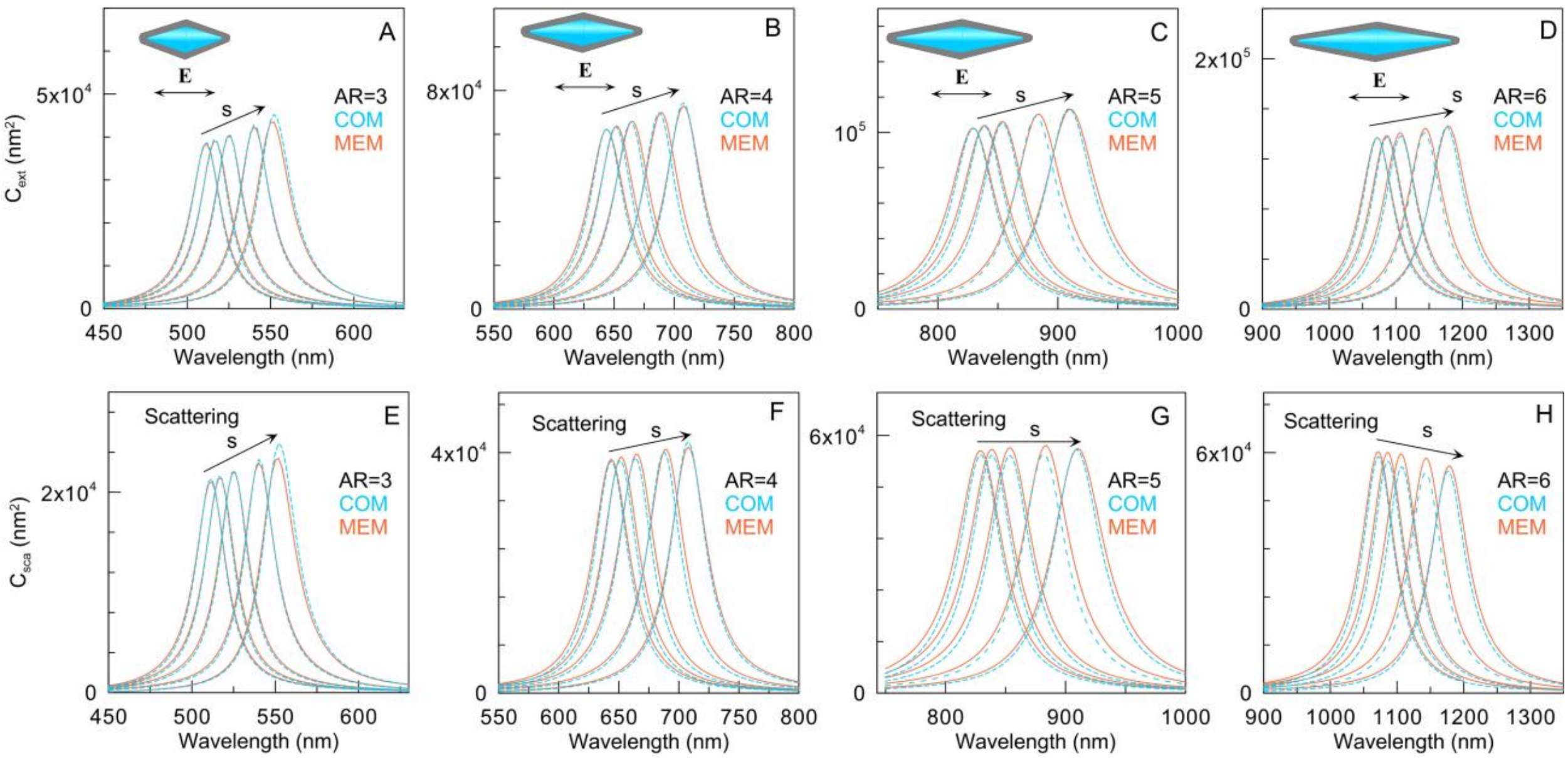

**Figure 15**. Extinction (A-D) and scattering (E-H) spectra were calculated by MEM (orange) and COMSOL (blue) for coated rounded silver bipyramids with a constant width of 30 nm and aspect ratios from 3 to 6. The shell thicknesses are 0, 1, 3, 10, and 30 nm.

## 5. Experimental example

As an experimental example, we chose the stabilization of gold nanorods with a thin layer of adsorbed CTAB molecules. The presence of such a stabilizing layer should lead to a slight systematic redshift of the plasmon resonance. Although this type of stabilization of AuNR colloids is well known, the influence of CTAB on the dependence of the plasmon resonance on the aspect ratio, as far as we know, has not been previously described.

Figure 16 shows TEM images of 12 samples S1-S12 obtained by chemical etching of the initial first sample and their normalized extinction spectra. During the etching process, the average particle diameter remains approximately constant and equal to $d_{av} = 13.8 \pm 0.3$ nm, and the average length or average aspect ratio $AR = L_{av} / d_{av}$ progressively decreases (Figure 17A). Accordingly, the etching process is accompanied by the well-known shift of the longitudinal plasmon resonance towards shorter wavelengths (Figure 16B).

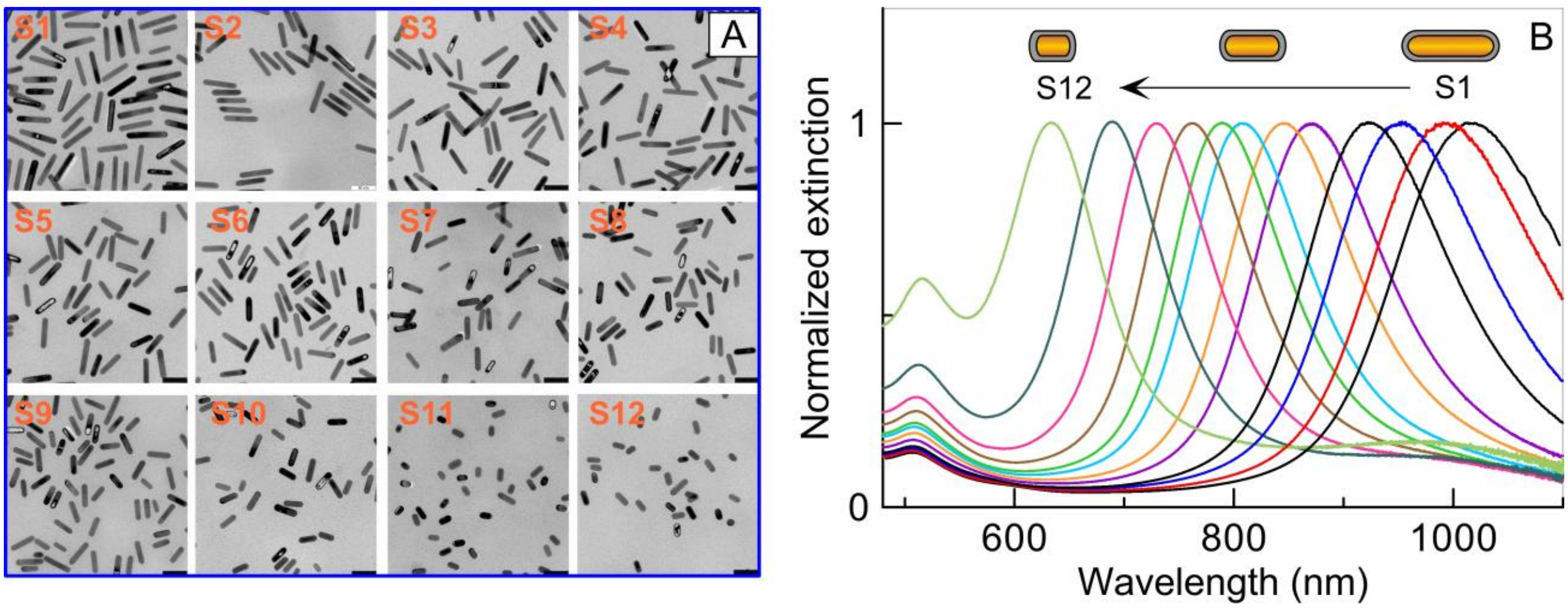

**Figure 16**. TEM images (A) and normalized extinction spectra (B) of etched AuNR samples S1-S12. The bars are 50 nm.

Figure 17A shows the decrease in the average aspect ratio for 12 etched gold nanorod samples. The vertical bars show the standard deviation for an ensemble of 300 particles. Figure 17B shows the experimental dependence of the longitudinal plasmon resonance wavelength. As is well known, this dependence is almost linear.[70]

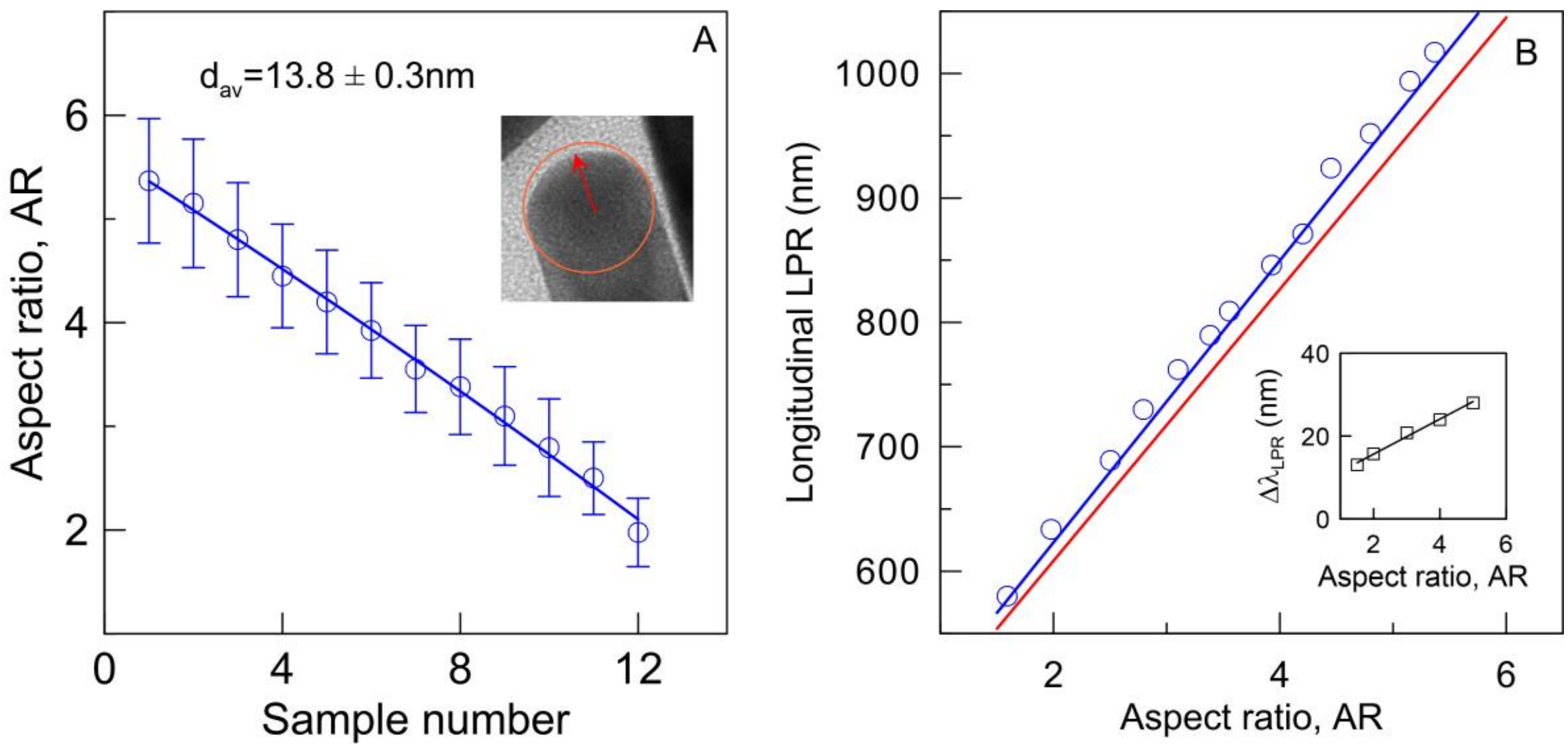

**Figure 17**. (A) The average diameters and standard deviations were plotted vs sample numbers from 1 to 12. The data are obtained from statistical analysis of TEM images shown in Figure 16A. The inset shows an enlarged image of the tip morphology. Also, the average constant diameter is indicated. (B) Experimental dependence of the LPR peak position as a function of the average aspect ratio (blue circles). The red line shows the TMM simulation obtained for polydisperse ensembles of bare particles. The blue line was obtained by MEM+DEM correction for the 3-nm CTAB coating (see the inset).

The theoretical model was a polydisperse ensemble of nanorods with a constant average diameter of 13.8 nm and a Gaussian distribution of aspect ratios. The width of the distribution was determined from the standard deviation obtained from the statistics of TEM measurements. Since the position of the plasmon resonance depends on the details of the particle end morphology,[71-73] we took into account a slight flattening of the ends compared to the hemispherical model (Fig. 17A, inset). The end-cap shape parameter[74] $\chi_c$ was set to 0.8. The spectral dependence of the optical constants of bulk gold was obtained from.[24] The dimensional correction was applied according to the Lorentz-Drude model with a surface scattering constant, $A_s = 0.33$[74] The spectral dependence of the refractive index of water was calculated as described in Ref.[74] The absorption, scattering, and extinction spectra for a polydisperse ensemble of particles with random orientations were calculated using the TMM.[35]

Even though the theoretical model was developed with the utmost care, the calculation of the position of the longitudinal resonance peaks as a function of the average aspect ratio yielded a linear dependence (red line in Fig. 17B)

$$\lambda_{LPR}(\text{TMM}) = 109.1 \times AR + 389.5\,, \tag{23}$$

which is clearly inconsistent with experimental data. A similar discrepancy between the experimental and calculated data was also observed in the simplest electrostatic simulation based on the spheroid model.[70] Based on the calculations of the extinction spectra of coated nanoparticles (Section 2), we suggested that the reason for the obtained discrepancy could be related to the influence of the shell of the CTAB stabilizer molecules on the position of the PR peaks.

To analyze the effect of the CTAB coating, a refined theoretical model is needed that considers the geometric and optical parameters of the CTAB layer on the surface of the nanorods. Two CTAB shell parameters determine the LPR shift: the shell thickness, $t_s$, and its refractive index. According to the Powerful Online Chemical Database,[75] the refractive index of CTAB is $n_s$ =1.526. For CTAB solutions, Ekwall et al.[76] reported the refractive index increment of 0.150 mL/g, which, together with the specific volume of 0.984 g/mL and the refractive index of water 1.334, gives $n_s$ =1.486. Two other papers[77,78] indicated a somewhat smaller value of 1.435. In our simulations, we used the average value $n_s$ =1.5.

The CTAB shell thickness has been measured in AuNR colloids with SAXS[79,80] and SANS[81] techniques to give the average value of 3.2 nm.[82] However, the single-particle measurements with the electron energy loss spectroscopy in an aberration-corrected scanning transmission electron microscope (STEM-EELS) revealed a somewhat smaller thickness of 2.4 nm. Here, we used an average value of 3 nm to characterize the shell thickness of CTAB molecules adsorbed on the side surface and tips of AuNRs. We assumed a homogeneous distribution of CTAB molecules on the nanorod surface instead of an asymmetric dipole-like distribution discussed in Ref.[83]

From MEM+DEM simulations (confirmed by COMSOL), we derived the following correlation between the CTAB-coating plasmonic shift and the aspect ratio of nanorods (see the inset in Figure 17B)

$$\Delta\lambda_{LPR} \equiv \lambda_{LPR}(s=3\text{nm}) - \lambda_{LPR}(s=0) = 4.2\times AR + 7.2\,, \tag{24}$$

After inserting this correction into Eq. (23), we get the corrected equation (blue line in Figure 17B)

$$\lambda_{LPR}(\text{MEM+DEM}) = 113.3\times AR + 396.7\,, \qquad (25$$

which is in good agreement with the experimental points. Thus, including the CTAB shell in the simulation model allows for quite satisfactory prediction of the LPR peak position as a function of the aspect ratio.

**Conclusions**

We have extended the MEM approximation previously developed for bare plasmonic nanoparticles to the coated ones. Remarkably, our MEM+DEM extension requires no additional numerical simulations to obtain analytical models for multilayered nanostructures. Actually, it is sufficient to find an analytical model for specific particle morphology and then apply DEM to extend the solution for a multilayered structure. We have found some disagreement between MEM+DEM spectra and numerical ones for sharp tip-shaped particles like bicones. However, our method works almost perfectly for bipyramide models with realistic rounding of both tips and bases. In line with the original MEM theory,[31] the MEM+DEM approach can be easily applied to other particle shapes like ellipsoids, cubes, nanocages, etc.

We primarily tested 30-nm maximum coatings for two reasons. First, thicker coatings are generally impractical because the overall particle size becomes too large, exceeding 100 nm. Second, for a thick dielectric shell, the coated plasmonic particle is physically embedded in the shell material's dielectric environment, causing all shell effects to become saturated. In particular, Figure S20 demonstrates excellent agreement between COMSOL and MEM+DEM spectra for AuNRs coated with a 60-nm shell. Even for a maximum 100-nm shell, MEM+DEM method still performs effectively and predicts not only the main plasmonic peak but also the increased extinction at shorter wavelengths caused by enhanced scattering.

Due to its simplicity and reasonable accuracy, the developed analytical models can find valuable applications in machine learning to predict various plasmonic responses from coated particles, such as UV-vis extinction and scattering spectra, SERS, and MEF enhancement. In many laboratories worldwide, researchers need to monitor the functionalization process *in situ*, comparing experimental changes in plasmonic peaks with theoretical expectations. The simple analytical models proposed here may prove extremely useful for such purposes.

**Acknowledgments**

This research was supported by the Russian Science Foundation (project no. 24-22-00017). The authors thank A. M. Burov and B. N. Khlebtsov (IBPPM RAS) for their help with experiments.

**Supporting Information**. Section S1: MEM parameters for selected particle shapes; Figures S1-S4. Section S2. Scattering cross section of small coated spheres. Figure S5. Section S3: Additional extinction and scattering spectra of bare and coated gold and silver nanorods, nanodisks, nanotriangles, bicones, and bipyramids; Figures S6-S20.

# Supporting Information

# Universal Analytical Modeling of Coated Plasmonic Particles

Nikolai G. Khlebtsov[1,2,*], Sergey V. Zarkov[1,3]

[1]Institute of Biochemistry and Physiology of Plants and Microorganisms, "Saratov Scientific Centre of the Russian Academy of Sciences," 13 Prospekt Entuziastov, Saratov 410049, Russia

[2]Saratov State University, 83 Ulitsa Astrakhanskaya, Saratov 410012, Russia

[3]Institute of Precision Mechanics and Control, "Saratov Scientific Centre of the Russian Academy of Sciences," 24 Ulitsa Rabochaya, Saratov 410028, Russia

[*]To whom correspondence should be addressed. E-mail: (NGK) khlebtsov@ibppm.ru

**Section S1. MEM parameters for selected particle shapes**

For rods and discs, MEM parameters coincide with those given in Ref.[1] Coated rods and disks' geometrical sizes should be increased by twice the shell thickness $2s$. Accordingly, the aspect ratios of coated rods and disks are calculated as $AR_2 = \dfrac{L_1 + 2s}{d_1 + 2s}$ and $AR_2 = \dfrac{D_1 + 2s}{t_1 + 2s}$, respectively.

Below, the symbol $h$ designates the generalized aspect ratio $AR_1$ for the metal core and $AR_2$ for shell. It should be emphasized here that we use the same MEM functions for metal cores and shells due to DEM philosophy.

**S1.1 Rods**

$$\eta_1(h) = -1.73h^{1.45} - 0.296\,,\ a_{12}(h) = 6.92/\left[1-\eta_1(h)\right],\ a_{14}(h) = -11h^{-2.49} - 0.0868\,, \qquad \text{(S1)}$$

$$V_1^m / V_{NR} = 0.896\,,\; V_{NR} / L_1^3 = \frac{\pi(3h-1)}{12h^3}\,. \tag{S2}$$

**S1.2. Disks**

$$\eta_1(h) = -1.36h^{0.872} - 0.479\,,\; a_{12}(h) = 7.05 / \left[1-\eta_1(h)\right],\; a_{14}(h) = -10.9h^{-0.98}\,, \tag{S3}$$

$$V_1^m / V_{ND} = 0.944\,,\; V_{ND} / L_1^3 = \frac{\pi}{24h^3}\left[4+3(h-1)(2h+\pi-2)\right]. \tag{S4}$$

**S1.3. Triangle prism**

The generalized aspect ratios of bare and coated nanoprisms are characterized by similar expressions in $AR_i \equiv h_i = L_i / t_i$ , $i = 1,2$, where the indices 1 and 2 stand for the metal core and shell, respectively; $L_2 = L_1 + 2\sqrt{3}s$ , and $t_2 = t_1 + 2s$ . Here, $t_1$ and $t_2$ designate the bare and coated prism thicknesses, respectively. Note that both lengths here are given in terms of nonrounded particles. By fitting MEM spectra to COMSOL ones, we derived the following MEM parameters for rounded triangle nanoprisms:

$$\eta_1(h) = -3.2483 - 1.2047h + 0.00632112h^2\,, \tag{S5}$$

$$a_{12}(h) = 5.57 / \left[1-\eta_1(h)\right],\; a_{14}(h) = -6.83 / \left[1-\eta_1(h)\right], \tag{S6}$$

$$V_1^m / V_{NT} = 0.55 + 0.155\left[1 - e^{-(h-2)/5}\right], \tag{S7}$$

$$V_{NTr} / L_1^3 = \frac{\sqrt{3}}{4h} - \frac{(h+1)^2}{1600h^2}\left(\frac{3\sqrt{3}-\pi}{h} + 6\left(1-\tfrac{1}{4}\pi\right)\right). \tag{S8}$$

In all the above equations, the symbol $h$ stands for the aspect ratio of rounded core ( $AR_1$ ) or shell ( $AR_2$ ), respectively.

In Eq. (S8), the nanoprism volume $V_{NTr}$ was calculated by considering rounding for three base corners and six lateral edges. The rounding radius equals to $r = (1+1/AR) \times L / 40$ (Ref.[1]). Below we provide a short derivation.

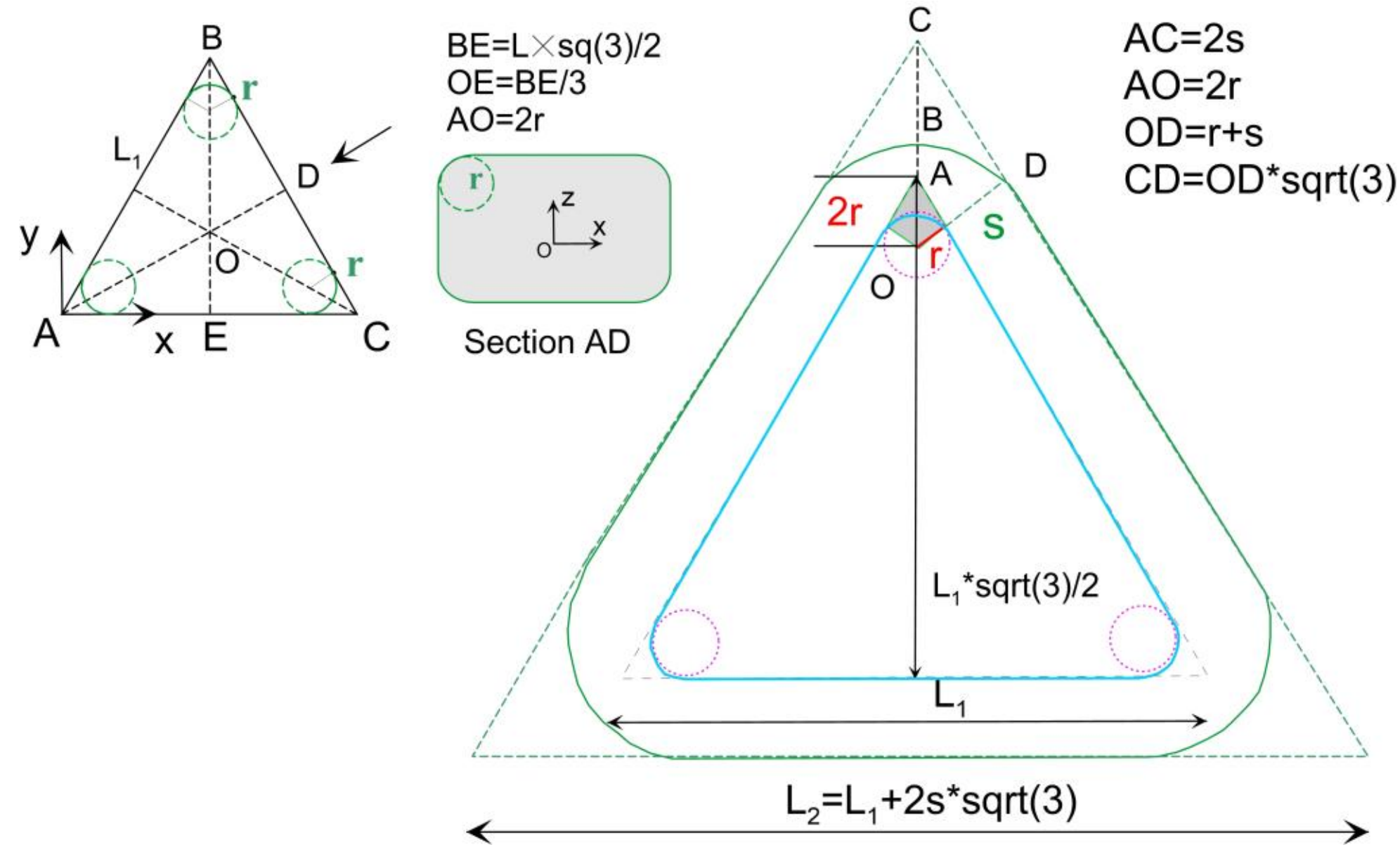

**Figure S1**. Scheme of the rounded bare and coated triangle prisms

Let's start with three corners in the triangle plane (Figure S1).

$S_0 = (L/2)\times L\sqrt{3}/2 = L^2\sqrt{3}/4$ is the area of the initial triangle without rounding

$S_1 = 2\times(r/2)r\sqrt{3} = r^2\sqrt{3}$ is the area of two shadowed triangles at the upper corner

$S_2 = \pi r^2/3$ is the area of the shadowed portion of the corner circle

The total area of three corner potions that should be excluded from the initial triangle is

$$S_{exc} = S_1 - S_2 = 3\times r^2(\sqrt{3}-\pi/3) = r^2(3\sqrt{3}-\pi)$$

The total area of the rounded triangle

$$S_r = S_0 - S_{exc} = \left[L^2\sqrt{3}/4 - r^2(3\sqrt{3}-\pi)\right]$$

Considering the above expression for the rounding radius $r$ and multiplying by the particle thickness, we obtain the total volume of the triangle prism with rounded corners:

$$V_{rc} = \left[L^2\sqrt{3}/4 - (1+1/AR)^2\times L^2/1600\times(3\sqrt{3}-\pi)\right]L/AR$$

Now, we consider six lateral edges. Because of rounding, we have to exclude six volumes between the square area $r^2$ and ¼ of the circle area $\pi r^2$ multiplied by six edge length $L$:

$$V_{1exc} = 6L\times(r^2 - \tfrac{1}{4}\pi r^2) = 6L^3(1+1/AR)^2(1-\pi/4)/1600.$$

A similar expression can be written for three angular edges of a thickness $t$:

$$V_{2exc} = 3t\times(r^2-\tfrac{1}{4}\pi r^2) = (3/AR)L\times(r^2-\tfrac{1}{4}\pi r^2) = 3L^3(1+1/AR)^2(1-\pi/4)/1600AR$$

As we consider thin prisms with AR>3, the last contribution $V_{2exc}$ is one order less than $V_{1exc}$. Thus, one can neglect $V_{2exc}$ compared to $V_{1exc}$.

Combining all previous results, we arrive at the final expression (S8) for the rounded triangle prism volume (remember that $h = AR$). Note that the first term coincides with that given in Fig. 3 caption by Yu *et al.*[1]:

$$\frac{V^{Yu}}{L^3} = \frac{0.433}{AR} - \frac{0.00544}{AR^2}, \tag{S9}$$

but the other terms differ. However, as the rounding radius is small, the above difference should not affect the main conclusions made from calculations by Yu *et al.*[1]

It follows from Figure1 that the volume of a rounded coated nanoprism can be calculated by the above Eq. (S8) with $L \equiv L_2 = L_1 + 2\sqrt{3}s$ and $h = AR_2 = L_2/(t_1 + 2s)$

## S1.4. Rounded bicone

### S1.4.1. Geometrical parameters

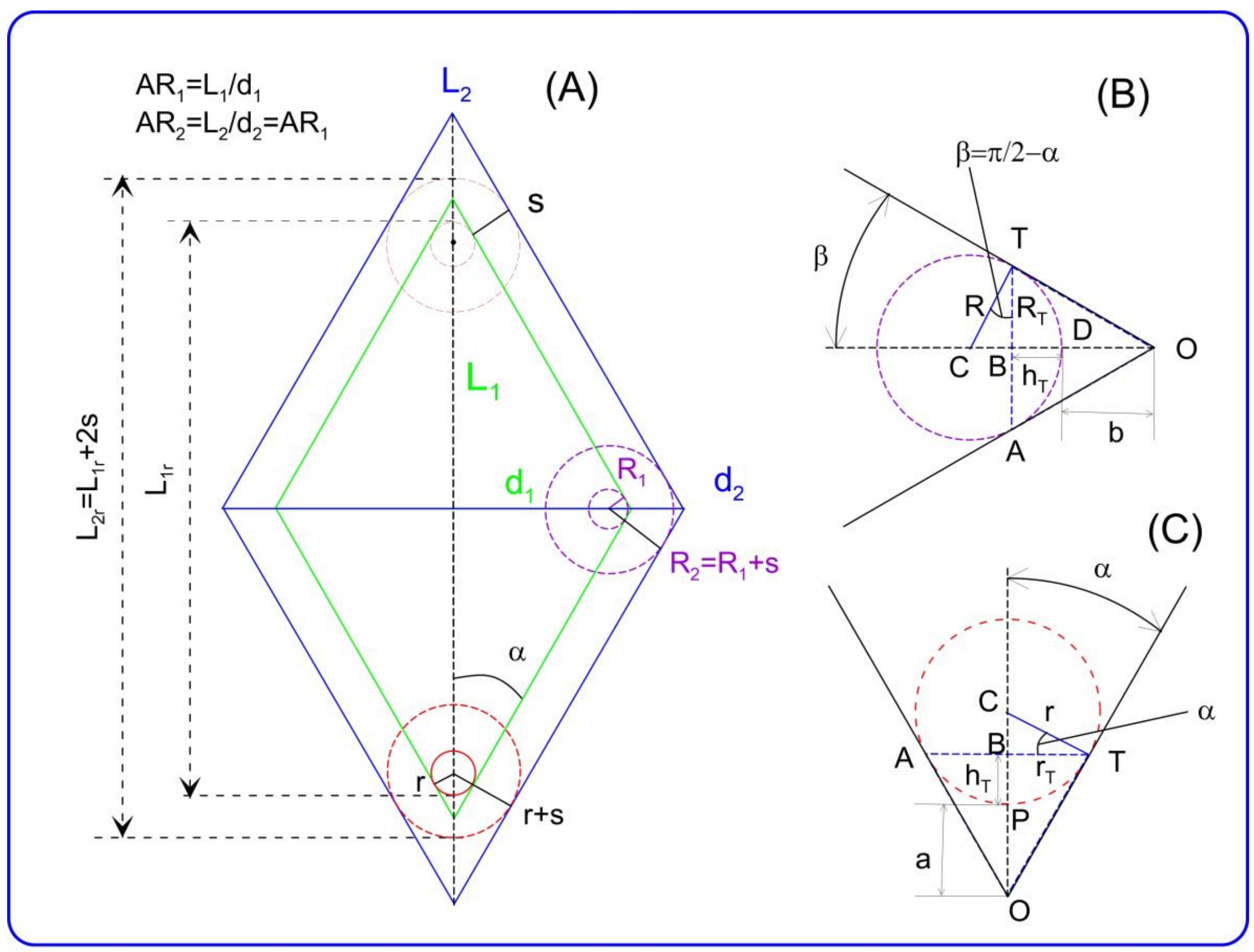

**Figure S2**. Scheme of the rounded bare and coated bicones

The size and shape of the initial bicone (unrounded metal core) is characterized by the total $L_1$ length, diameter $d_1$, and aspect ratio $AR_1 = L_1/d_1$. The rounded core is characterized by the rounded length $L_{1r} = p_1$ and the rounded base diameter $d_{1r}$. The unrounded shell bicone is characterized by similar parameters $L_2$, $d_2$, and $AR_2 = L_2 / d_2 = AR_1$. From Figure S2, we have

$$L_2 = L_1 + 2s / \sin\alpha = L_1 + 2s\sqrt{1+AR_1^2} \, , \tag{S10}$$

$$d_2 = d_1 + 2s / \cos\alpha = d_1 + 2s\sqrt{1+AR_1^2} / AR_1 \, . \tag{S11}$$

The constant shell thickness is $s$; therefore, the rounded shell length and diameter are $L_{2r} \equiv p_2 = p_1 + 2s \equiv L_{1r} + 2s$ and $d_{2r} = d_{1r} + 2s$, respectively. The rounding radii for metal and coated bicones are $r_1$ and $r_2 = r_1 + s$, respectively. Similarly, the rounding radii of the bases are $R_1$ and $R_2 = R_1 + s$, respectively. The rounded length and diameter of the core are

$$L_{1r} = p_1 = L_1 - 2a_1 \, , \tag{S12}$$

$$d_{1r} = d_1 - 2b_1 \, , \tag{S13}$$

where

$$a_1 = qr_1 \, , \; b_1 = gR_1 \tag{S14}$$

$$q = (1-\sin\alpha)/\sin\alpha = \sqrt{1+AR^2} - 1 \, , \tag{S15}$$

$$g = (1-\cos\alpha)/\cos\alpha = \sqrt{AR^{-2}+1} - 1 \tag{S16}$$

$$\alpha = \mathrm{arctg}(1/AR_1) \, , \; \sin\alpha = 1/\sqrt{1+AR^2} \, , \cos\alpha = AR/\sqrt{1+AR^2} \, . \tag{S17}$$

Note that the aspect ratios of rounded core and shell differ from their initial (unrounded) counterparts

$$AR_{1r} = \frac{L_{1r}}{d_{1r}} = \frac{L_1 - 2a_1}{d_1 - 2b_1} \le \frac{L_1}{d_1} \, , \tag{S18}$$

$$AR_{2r} = \frac{L_{2r}}{d_{2r}} = \frac{L_{1r} + 2s}{d_{1r} + 2s} \le \frac{L_{1r}}{d_{1r}} \tag{S19}$$

### S1.4.2. Calculation of the rounded bicone volume

The volume of initial bicone with the total length $L$ and diameter $d$ equals

$$V_i = \frac{2}{3}\pi\left(d_i/2\right)^2 \frac{L_i}{2} = \frac{\pi L_i^3}{12 AR_i^2} \, , \; AR_i = L_i / d_i \, , i = 1,2 \, , \tag{S20}$$

where the indices 1 and 2 stand for the core and shell, respectively. To exclude the rounded tip volumes, we have to substrate two volumes located between the spherical segment ATP and ATO cone (Figure S2C):

$$V_{exc}^{tips} = \frac{\pi}{3}\left(2r_T^2 a - r_T^2 h_T - h_T^3\right), \tag{S21}$$

$$h_T = r(1-\sin\alpha), \tag{S22}$$

$$r_T = r\cos\alpha, \tag{S23}$$

where the rounding radius $r$ can be $r_1$ or $r_2$. After some algebra, we get

$$V_{exc}^{tip} = \frac{2\pi}{3} r^3 \frac{(1-\sin\alpha)^2}{\sin\alpha}. \tag{S24}$$

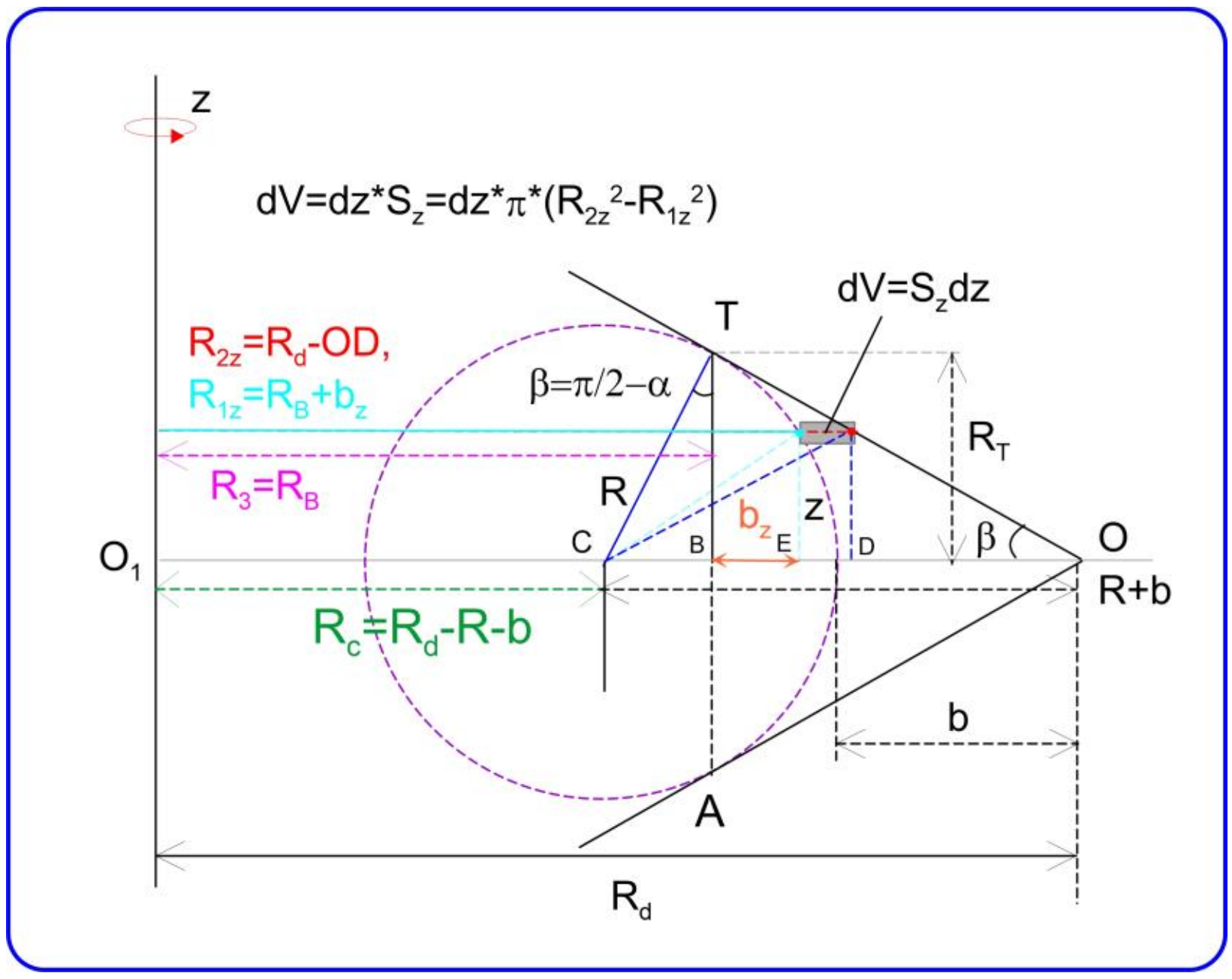

**Figure S3.** Calculating the base excluded volume.

For a rounded base, the excluded volume can be calculated as (see Figure S3)

$$V_{exc}^{base} = V_{21} - V_{11} = 2\pi \int_0^{R_T} \left(R_{2z}^2 - R_{1z}^2\right) dz \tag{S25}$$

where the following relations define all related quantities:

$$R_{2z}^2 = \left\lfloor R_d - z\,\text{tg}(\alpha)\right\rfloor^2 = R_d^2 - 2R_d\,\text{tg}(\alpha)z + z^2\text{tg}^2(\alpha),$$

$$R_{1z}^2 = \left(R_3 + b_z\right)^2 = \left(R_3 - \sqrt{R^2 - R_T^2} + \sqrt{R^2 - z^2}\right)^2 = \left(R_3 - R_4 + \sqrt{R^2 - z^2}\right)^2 = \left(R_{34} + \sqrt{R^2 - z^2}\right)^2$$

$$R_{2z}^2 - R_{1z}^2 = R_d^2 - 2R_d \text{tg}(\alpha) z + z^2 \text{tg}^2(\alpha) - R_{34}^2 - 2R_{34}\sqrt{R^2 - z^2} - R^2 + z^2$$

$$R_{2z}^2 - R_{1z}^2 = A - 2R_d \text{tg}(\alpha) z + z^2 / \cos^2(\alpha) - 2R_{34}\sqrt{R^2 - z^2}$$

$$A = R_d^2 - R_{34}^2 - R^2,\ R_{34} = R_3 - R_4 = R_3 - \sqrt{R^2 - R_T^2} = R_3 - R\cos\alpha$$

$$R_3 = R_B = R_C + R\sin\beta = R_d - R/\cos\alpha + R\cos\alpha = R_d - R\text{tg}\alpha\sin\alpha$$

$$R_{34} = R_d - R\tan\alpha\sin\alpha + R\cos\alpha = R_d - R/\cos\alpha\ .$$

With using integral

$$\int_0^{R_T} \sqrt{R^2 - z^2}\, dz = \frac{R_T\sqrt{R^2 - R_T^2}}{2} + \frac{R^2}{2}\arcsin\frac{R_T}{R},$$

we get:

$$V_{21} = R_T\left[R_d^2 - R_d R_T \tan(\alpha) + \tfrac{1}{3}R_T^2\tan^2(\alpha)\right] \tag{S26}$$

$$V_{11} = R_T\left[{R_{34}}^2 + R^2 - \tfrac{1}{3}R_T^2\right] + R_{34}\left[R_T R\cos(\alpha) + R^2\arcsin\frac{R_T}{R}\right] \tag{S27}$$

$$V_{exc}^{base} = 2\pi\left(V_{11} - V_{21}\right) \tag{S28}$$

After some algebra, we arrive at the final expression:

$$V_{exc}^{base} = 2\pi R^2\left[R_d\left(\tan\alpha - \alpha\right) + R\left(-\frac{\sin\alpha}{3}\left(1 + \frac{2}{\cos^2\alpha}\right) + \frac{\alpha}{\cos\alpha}\right)\right], \tag{S29}$$

$$\alpha = \arctan\frac{1}{AR_1}. \tag{S30}$$

Note, the angle $\alpha$ is the same for the core and shell. In Eq. (S29), the rounding radius $R$ can be $R_1$ for the core and $R_2 = R_1 + s$ for the shell.

The above analytical expressions for the rounded bicone volume were checked with a direct numerical integration by COMSOL. For numerical simulations, we used the rounding radii $r_i = L_i / 40AR_i = d_i / 40, i = 1,2$ for the core and shell tips and $R_i = d_i / 100, i = 1,2$ for the core and shell bases. For metal bicones, this parametrization was used by Yu *et al.*[1]

### S1.4.3. MEM parameters for bare and coated rounded bicones.

The rounded bare and coated bicones volume is calculated as a difference between initial and excluded base and tip volumes. $V_{BCr} = V_{BC} - \left(V_{exc}^{tip} + V_{exc}^{base}\right)$. All related expressions have been provided in the previous subsections 1.4.1-1.4.2. To calculate the modal volumes (mode 1), we used the following factorization model

$$V_1^m = V_{BCr} \times f_{Vh}(h_r) f_{Vs}\left(s\right), \qquad \text{(S31)}$$

$$f_{Vh}(x) = 0.32259 + 0.086858x - 0.015968x^2 + 0.0013815x^3 - 4.55693\text{E-}005x^4, \qquad \text{(S32)}$$

$$f_{Vs}(x) = \begin{cases} 1, \quad s = 0 \\ \exp\left(0.083478 + 0.033629x - 0.0057969x^2 - 0.0020217x^3\right), \; x = \ln(s), 1 \le s \le 30 \end{cases}. \qquad \text{(S33)}$$

The MEM parameters $\eta_j^p(h_r)$ for the first mode $j=1$ and longitudinal excitation along the bicone symmetry axis ( $p = l \equiv ||$ ) are calculated by the following factorized approximation

$$\eta_1 = -f_{\eta h}(h_r) \times f_{\eta s}(s), \;, \qquad \text{(S34)}$$

$$f_{\eta h}(x = h_r) = 0.29105 + 2.6267x + 0.84365x^2 - 0.0057085x^3 - 0.00043102x^4, \qquad \text{(S35)}$$

$$f_{\eta s}(s) = \begin{cases} 1, \quad s = 0 \\ 1.29478 s^{0.114969}, \quad 1 \le s \le 30 \end{cases}, \qquad \text{(S36)}$$

Finally, the last two MEM parameters for the first longitudinal mode are[1]

$$\alpha_{12}(h_r) = \frac{1.34}{1-\eta_1}, \; \alpha_{11}(h_r) = -\frac{1.04}{1-\eta_1}. \qquad \text{(S37)}$$

In all the above equations, $h_r$ stands for the aspect ratio of rounded core ( $AR_{1r}$ ) or shell ( $AR_{2r}$ ), respectively.

If the rounding radii of core and base are large enough, the functions $f_{Vs}(s)$ and $f_{\eta s}(s)$ can be replaced with 1, and the functions $f_{Vh}(x)$ and $f_{\eta h}(x)$ should be properly defined from the fitting to numerical data. In other words, the factorized MEM-F representations (S31) and (S34) are reduced to standard MEM ones.

### S1.5. Rounded bipyramid

The geometrical parameters of our rounded bipyramid (BP) model are based on the embedded bicone model. This model is characterized by the diameter of a circle inscribed in the pentagon of the BP base and its height coinciding with the height of the BP. It follows from experimental data (e.g., Ref.[2] and references therein) that the tip rounding radius $R_{tip}$ increases with a decrease in the actual BP length $L_{1r}$. Similarly, the base rounding radius $R_{base}$ also increases during the chemical etching. Unfortunately, in the theoretical model, the length of a rounded BP and its aspect ratio depend on the rounding radii, so we have a nonlinear relationship. To avoid this difficulty, we assume the linear relations between the rounding radii and the aspect ratio of the initial (unrounded) BP

$$R_{tip}(\text{nm}) = 2 + 2(6 - AR_1), 1 \le AR_1 \le 6\,, \tag{S38}$$

$$R_{base}(\text{nm}) = 16 - 2AR_1\,. \tag{S39}$$

In our simulations, we fixed the inscribed bicone diameter as the BP width and varied the aspect ratio from 2 to 6. This resulted in a progressive decrease in the rounding radii, consistent with the experimental measurements presented in Ref.[2]

Given the bipyramid diameter, aspect ratio, and rounding radii (S38) and (S39), we numerically calculated the rounded BP volume and all geometrical parameters of the embedded bicone analytically. Then, we approximate the BP volume by a polynomial (Figure S4) of the rounded aspect ratio

$$V_{B\mathrm{Pr}}(s=0) = \text{-}7486.03 + \ 18338.49AR_{1r} \ \text{-} \ 1433.54AR_{1r}^2\,, \tag{S40}$$

where the shell thickness s is assumed to be zero. Then, for a given shell thickness $s$, we calculated analytically the ratio $F_V(s) = V_{BCr}(s)/V_{BCr}(s=0)$. Finally, the volume of a rounded coated BP was calculated as $V_{BCr}(s) = F_V(s)V_{BCr}(s)/V_{BCr}(s=0)$. Panel B in Figure 4 illustrates the excellent accuracy of this approximation, better than 0.002%.

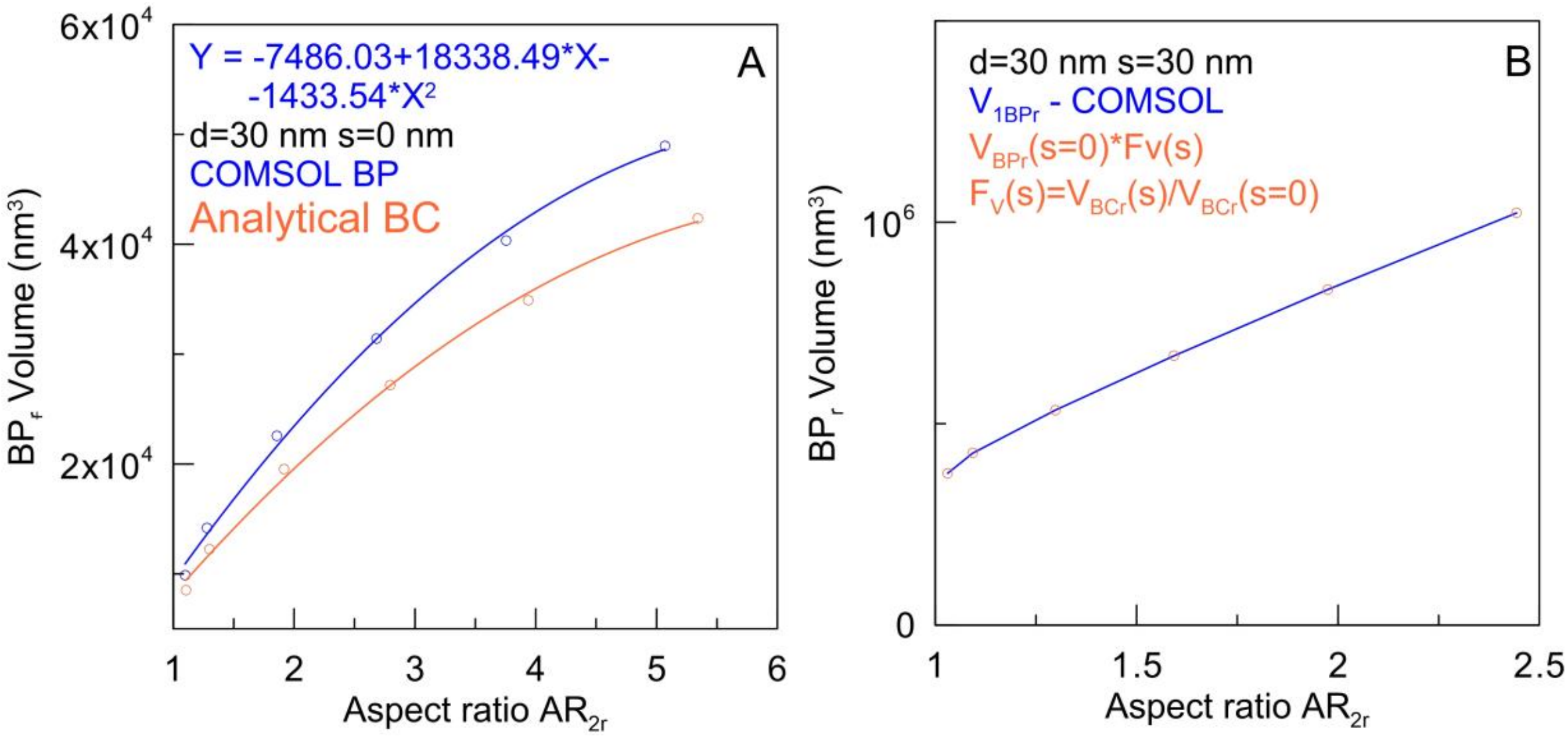

**Figure S4**. (A) Polynomial approximation of the BP (blue, numerical) and BC (orange, analytical) as a function of the rounded bicone aspect ratio. (B) Comparison of numerical (blue line) and analytical (orange circles) $BP_r$ volumes.

With the rounded BP volume in hand, we calculated other MEM parameters as follows:

$$\eta_1(h_r) = -f_\eta(h_r), \\ f_\eta(x) = -1.36762 + 3.0204378x + 0.09294389x^2 + 0.07895368x^3, \quad \text{(S41)}$$

$$a_{12}(h_r) = 5.57/\left[1-\eta_1(h_r)\right], \; a_{14}(h_{ir}) = -6.83/\left[1-\eta_1(h_{ir})\right], \quad \text{(S42)}$$

$$V_1^m = f_V(h_r)V_{B\text{Pr}}, \\ f_V(x) = 0.976704 - 0.0237377x - 0.00453238x^2. \quad \text{(S43)}$$

For silver bipyramids, Eqs. (S41) and (S43) shoud be modified as follows:

$$\eta_1(h_r) = -f_\eta(h_r), \\ f_\eta(x) = -0.323882 + 1.42211x + 1.01670x^2 - 0.129240x^3 + 0.0160452x^4, \quad \text{(S44)}$$

$$V_1^m = f_V(h_r)V_{B\text{Pr}}, \\ f_V(x) = 0.519999 + 0.562648x - 0.279774x^2 + 0.0561240x^3 - 0.00413525x^4. \quad \text{(S45)}$$

In all the above equations, the symbol $h_r$ stands for the aspect ratio of rounded core ($AR_{1r}$) or shell ($AR_{2r}$), respectively.

### Section S2. Scattering cross section of small coated spheres

The limiting case of nanorods with a small aspect ratio is a sphere. It is natural to assume that for coated spheres and nanorods with a small AR, the dependence of the scattering cross-section peaks on the coating will be qualitatively similar. Therefore, for small AR, we will consider the limiting case of a coated sphere (Figure S5).

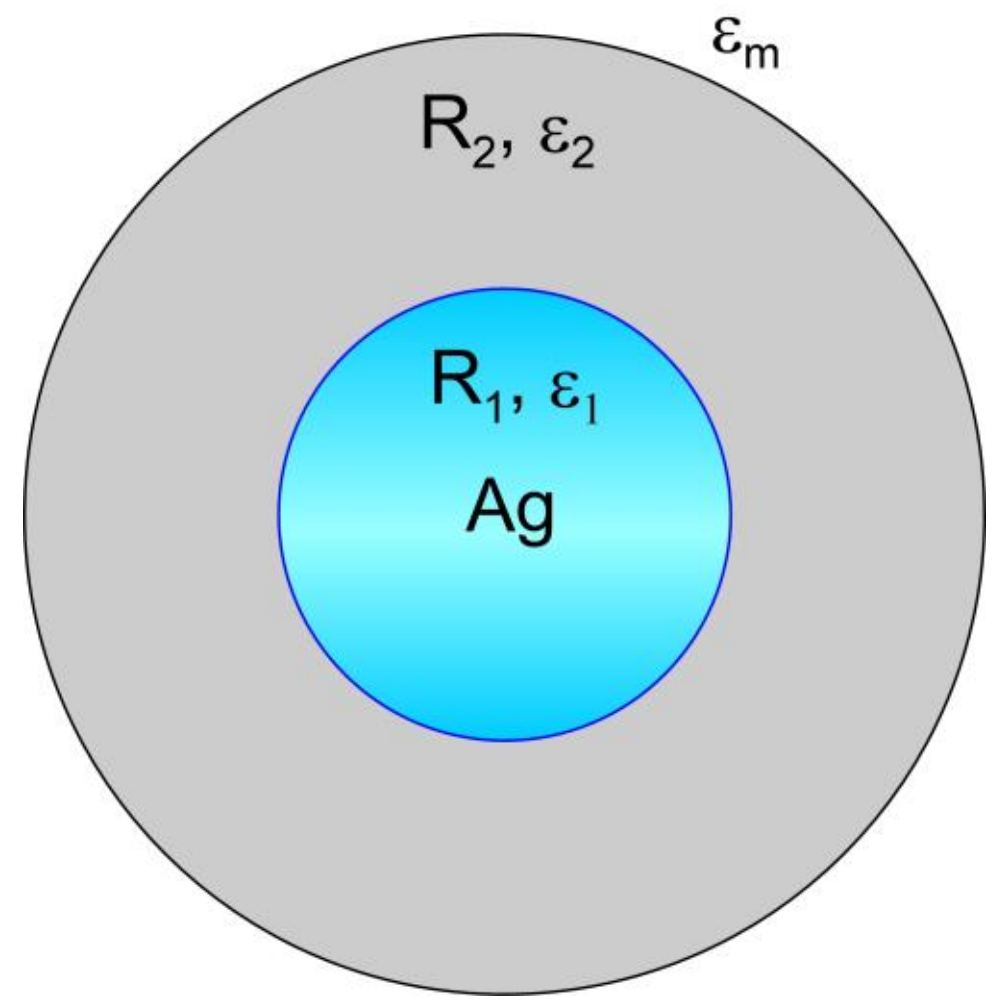

**Figure S5**. Schematics of a coated Ag sphere.

Let's use the article[3] as the basis, since it presents the polarizability of a coated spherical particle in a form that is more convenient and simpler than that in the book[4]

$$C_{sca,2} = \frac{8\pi}{3} k^4 \left|\alpha_{av}\right|^2 = \frac{8\pi}{3} k^4 R_2^6 \left| \frac{\varepsilon_{av} - \varepsilon_m}{\varepsilon_{av} + 2\varepsilon_m} \right|^2 \equiv \pi R_2^2 Q_{sca,2}, \quad \text{(S46)}$$

$$Q_{sca,2} = \frac{8}{3} \left(kR_2\right)^4 \left| \frac{\varepsilon_{av} - \varepsilon_m}{\varepsilon_{av} + 2\varepsilon_m} \right|^2, \quad \text{(S47}$$

$$\varepsilon_{av} = \varepsilon_2 \frac{\varepsilon_1 + 2\varepsilon_2 + 2f(\varepsilon_1 - \varepsilon_2)}{\varepsilon_1 + 2\varepsilon_2 - f(\varepsilon_1 - \varepsilon_2)}, \quad \text{(S48)}$$

where $k \equiv k_m = 2\pi\sqrt{\varepsilon_m} / \lambda$ is the wave vector modulus in the medium, $\varepsilon_m$ is the dielectric permittivity of the surrounding medium, $\varepsilon_{av}$ is an equivalent (“average”) dielectric permittivity of a two-layer sphere with the dielectric functions $\varepsilon_1$ (core) and $\varepsilon_2$ (shell); $f = \left(R_1 / R_2\right)^3$ and $g = 1 - f$ are the volume fractions of core and shell with the radii $R_1$ and $R_2$, respectively. For a metallic particle, $f = 1$, $\varepsilon_{av} = \varepsilon_1$, $R_2 = R_1$ and Eq. (S46) reduces to usual formulae for small sphere

$$C_1 \equiv C_{sca,1} = \frac{8\pi}{3} k^4 |\alpha_1|^2 = \frac{8\pi}{3} k^4 R_1^6 \left| \frac{\varepsilon_1 - \varepsilon_m}{\varepsilon_1 + 2\varepsilon_m} \right|^2 . \tag{S49}$$

Using the LPR condition in the following form

$$\varepsilon_1 \simeq -2\varepsilon_m + i\varepsilon_1'' , \tag{S50}$$

we obtain the LPR scattering cross section

$$C_{1LPR} \frac{8\pi}{3} k^4 R_1^6 \left| \frac{i\varepsilon_1'' - 3\varepsilon_m}{i\varepsilon_1''} \right|^2 = \frac{8\pi}{3} k^4 R_1^6 \frac{9\varepsilon_m^2 + (\varepsilon_1'')^2}{(\varepsilon_1'')^2} , \tag{S51}$$

where the dielectric functions of silver and water are taken near the wavelength LPR of silver spheres in water at 385 nm. Note that at this wavelength, for bulk silver, we have $\varepsilon_1 \approx -3.65 + i0.3$, which is in excellent agreement with (S50), since for water at this wavelength , $2\varepsilon_m = 2 \times 1.344^2 = 3.61$. Since in formula (S51) $9\varepsilon_m^2 \approx 16 \gg (\varepsilon_1'')^2 \approx 0.09$, it can be rewritten as:

$$C_{1sca}^{LPR} = 24\pi k^4 R_1^6 \left( \frac{\varepsilon_m}{\varepsilon_1''} \right)^2 = \pi R_1^2 Q_{1sca}^{LPR} , \tag{S52}$$

$$Q_{1sca}^{LPR} = 24 x_1^4 \left( \varepsilon_m / \varepsilon_1'' \right)^2 , \tag{S53}$$

where $x_1 = kR_1$. As we can see, the resonant scattering is inversely proportional to the square of the imaginary part of the dielectric function of silver, and therefore it is extraordinarily large—approximately $3 \times 10^4$ in magnitude—compared to that of a regular dielectric sphere ($n = \sqrt{\varepsilon_1} = 1.5$) of the same size. Resonant scattering also increases in optically denser matrices than water.

Let's consider a numerical example for a silver 15-nm sphere in water. For LPR wavelength 387 nm we have $\varepsilon_m \approx 1.8$ and $\varepsilon_1'' \approx 0.298$. Calculations by (S53) give the scattering efficiency factor $Q_{1sca}^{LPR} \approx 0.628$, in good agreement with Mie calculations $Q_{1sca}^{LPR} \approx 0.616$. Note that Eq. (S53) gives excellent approximation for the electrostatic approximation (EA) of the scattering cross section which gives 0.648 for EA LPR 384 nm almost identical to 0.647 by (S53).

Now we proceed to the resonance scattering by a small silver sphere covered by a thick dielectric shell with $f = \left(R_1 / R_2\right)^3 \ll 1$. For a 15-nm silver sphere and 30-nm shell the condition $f = 0.008 \ll 1$ is fully satisfied. In fact, even for a thinner 10-nm shell we have $f = 0.078 \ll 1$. Evident advantage of our solution (S46) is that the LPR condition has well known form[4]

$$\varepsilon_{av} \simeq -2\varepsilon_m + i\varepsilon''_{av} . \tag{S54}$$

In Ref.[3] we showed that the condition (S54) is equivalent to the following condition for the dielectric function of the core

$$\varepsilon_1 = -2\varepsilon_2 \frac{1 - f\alpha_{2m}}{1 + 2f\alpha_{2m}} = -2\varepsilon_2 \frac{\varepsilon_2(1-f) + \varepsilon_m(2+f)}{\varepsilon_2(1+2f) + 2\varepsilon_m(1-f)} , \tag{S55}$$

where parameter

$$\alpha_{2m} = \frac{\varepsilon_2 - \varepsilon_m}{\varepsilon_2 + 2\varepsilon_m} \tag{S56}$$

corresponds to the polarizability of a sphere with the shell permittivity $\varepsilon_2$ embedded in the external medium with the permittivity $\varepsilon_m$. In the case of small $f << 1$, Eq. (S55) reduces to

$$\varepsilon_1 = -2\varepsilon_2(1 - 3f\alpha_{2m}) , \tag{S57}$$

which differs from previous result.[5] It should be emphasized that the condition (S55) defines only the real part of $\varepsilon_1$ at the resonance wavelength, as the real quantity is on the right side of Eq. (S55). The physical meaning of condition (S57) is that, for a thick dielectric shell, the LPR wavelength "freezes" around its limiting value, which corresponds to an infinite homogeneous medium with dielectric function $\varepsilon_2$.[4]

To obtain an analytical estimate of the scattering cross-section, a careful analysis is required. The fact is that in this problem, there is no small parameter. The denominator of expression (S48) contains the difference of two small parameters: at resonance, the first term is close to zero according to condition (S57), and the second term is proportional to the small volumetric fraction of the core. Near resonance, we can write (S57) as $\varepsilon_1 \cong -2\varepsilon_2 + i\varepsilon''_1$, which, after substitution into (S48), yields

$$
\frac{\varepsilon_{av}}{\varepsilon_2} = \frac{i\varepsilon_1''(1+2f) - 6\varepsilon_2 f}{i\varepsilon_1''(1-f) + 3\varepsilon_2 f} = \frac{\left[i\varepsilon_1''(1+2f) - 6\varepsilon_2 f\right]\left[-i\varepsilon_1''(1-f) + 3\varepsilon_2 f\right]}{\left(\varepsilon_1''\right)^2 (1-f)^2 + 9\varepsilon_2^2 f^2} \cong
$$

$$
\cong \frac{\left(\varepsilon_1''\right)^2 (1+f) + i\varepsilon''(1+2f)_1 3\varepsilon_2 f + i\varepsilon'' 6\varepsilon_2 f (1-f) - 18\varepsilon_2^2 f^2}{\left(\varepsilon_1''\right)^2 (1-2f)} \cong
$$

$$
\cong \frac{\left(\varepsilon_1''\right)^2 (1+f) + i9\varepsilon_2 f / \varepsilon''}{\left(\varepsilon_1''\right)^2 (1-2f)} \cong 1 + 3f + i9\varepsilon_2 f / \varepsilon''.
$$

Thus, we finally get

$$
\varepsilon_{av} \cong \varepsilon_2 \left(1 + 3f + i\frac{9\varepsilon_2 f}{\varepsilon_1''}\right), \tag{S58}
$$

where we retained all the terms linear in the volumetric fraction of the core and neglected quadratic terms. Let's consider a numerical example, similar to the one discussed above, but for a shell 30 nm thick with a refractive index of 1.5. The calculation of the parameter $\beta_{av} = \frac{\varepsilon_{av} - \varepsilon_m}{\varepsilon_{av} + 2\varepsilon_m} = 0.12 + i0.183$ after substitution into (S47) yields a scattering efficiency factor of 0.048, which is in excellent agreement with the exact calculation for a two-layer spherical particle, which gives 0.052. Although the scattering efficiency factor of the coated silver sphere is almost an order of magnitude smaller than that of the pure silver sphere, the scattering cross-section of the coated sphere turns out to be nearly twice as large.

$$
\frac{C_{2sca}}{C_{1sca}} \simeq \left(\frac{R_2}{R_1}\right)^2 \frac{Q_{2sca}}{Q_{1sca}} = \left(\frac{37.5}{7.7}\right)^2 \frac{0.05}{0.62} \simeq 2. \tag{S59}
$$

This explains why the scattering always increases after coating plasmonic spheres and short rods.

**Section S3. Additional extinction and scattering spectra of bare and coated gold and silver nanorods, nanodisks, nanotriangles, bicones, and bipyramids.**

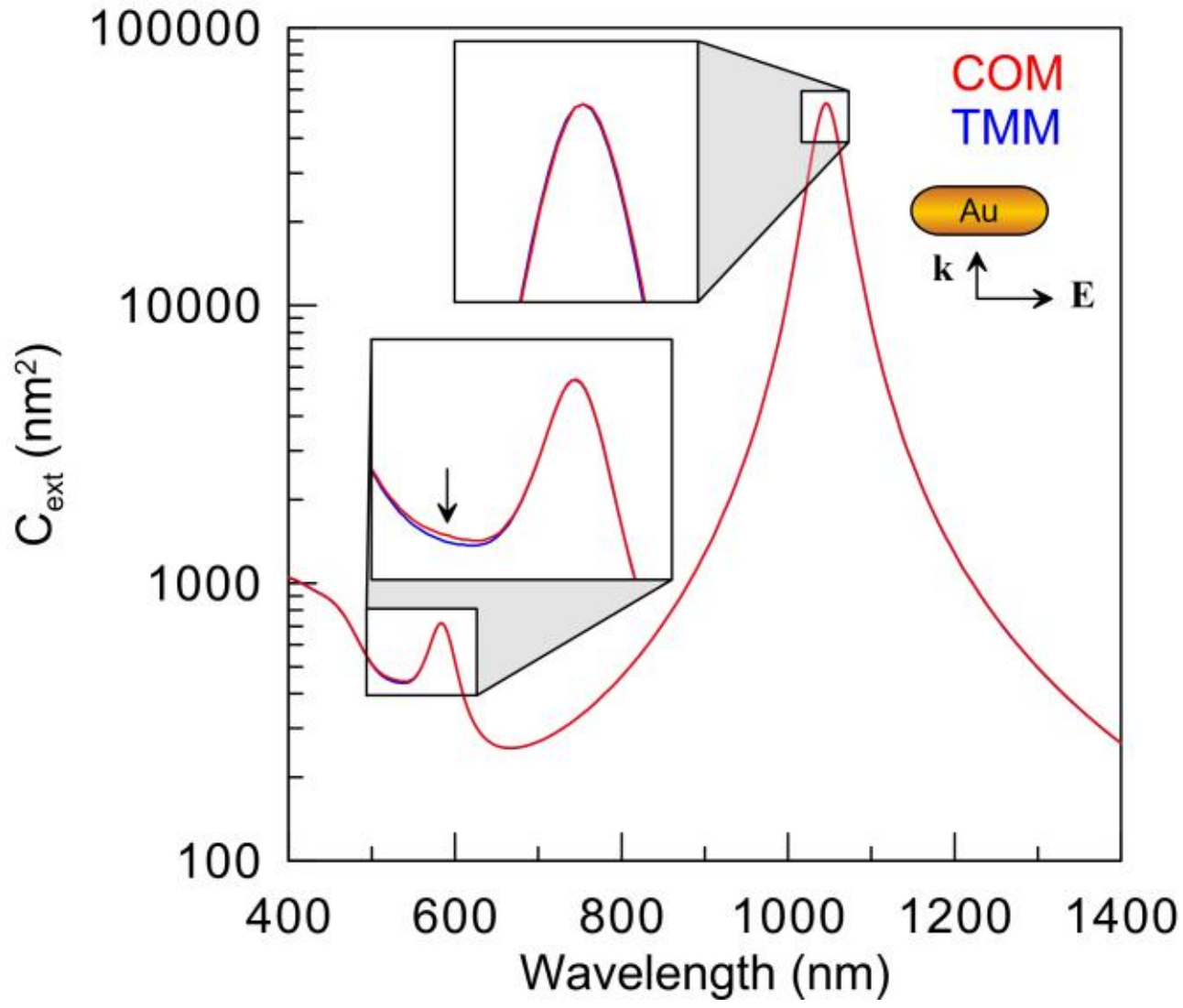

**Figure S6**. Comparison of extinction spectra calculated by COMSOL (2.5D, red curve COM) and T-matrix method (TMM, blue) for a longitudinal excitation of an AuNR with length 90 nm and diameter 15 nm. Both spectra have excellent agreement except for a slight difference near 550 nm. A small peak near 580 nm corresponds to the quadrupole excitation.

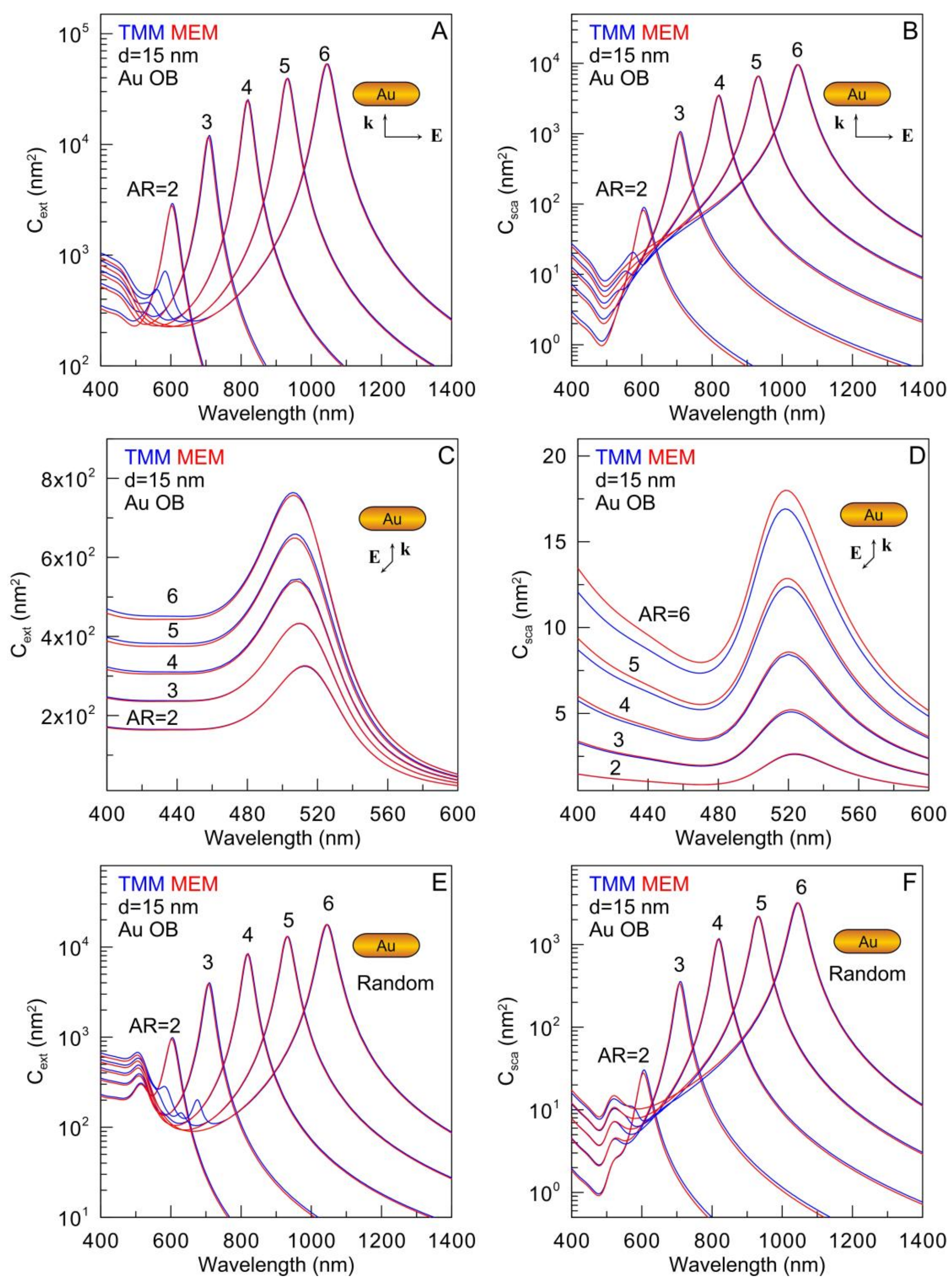

**Figure S7**. Comparison of extinction and scattering spectra calculated by the T-matrix method (TMM, blue) and MEM (red) for longitudinal (A, B) and perpendicular (C, D) excitations and for randomly oriented AuNRs (E, F) with length 90 nm and diameter 15 nm. There is excellent agreement between TMM and MEM spectra around main plasmonic peaks and satisfactory agreement for short wavelength perpendicular peaks except for quadrupole resonances that MEM cannot reproduce.

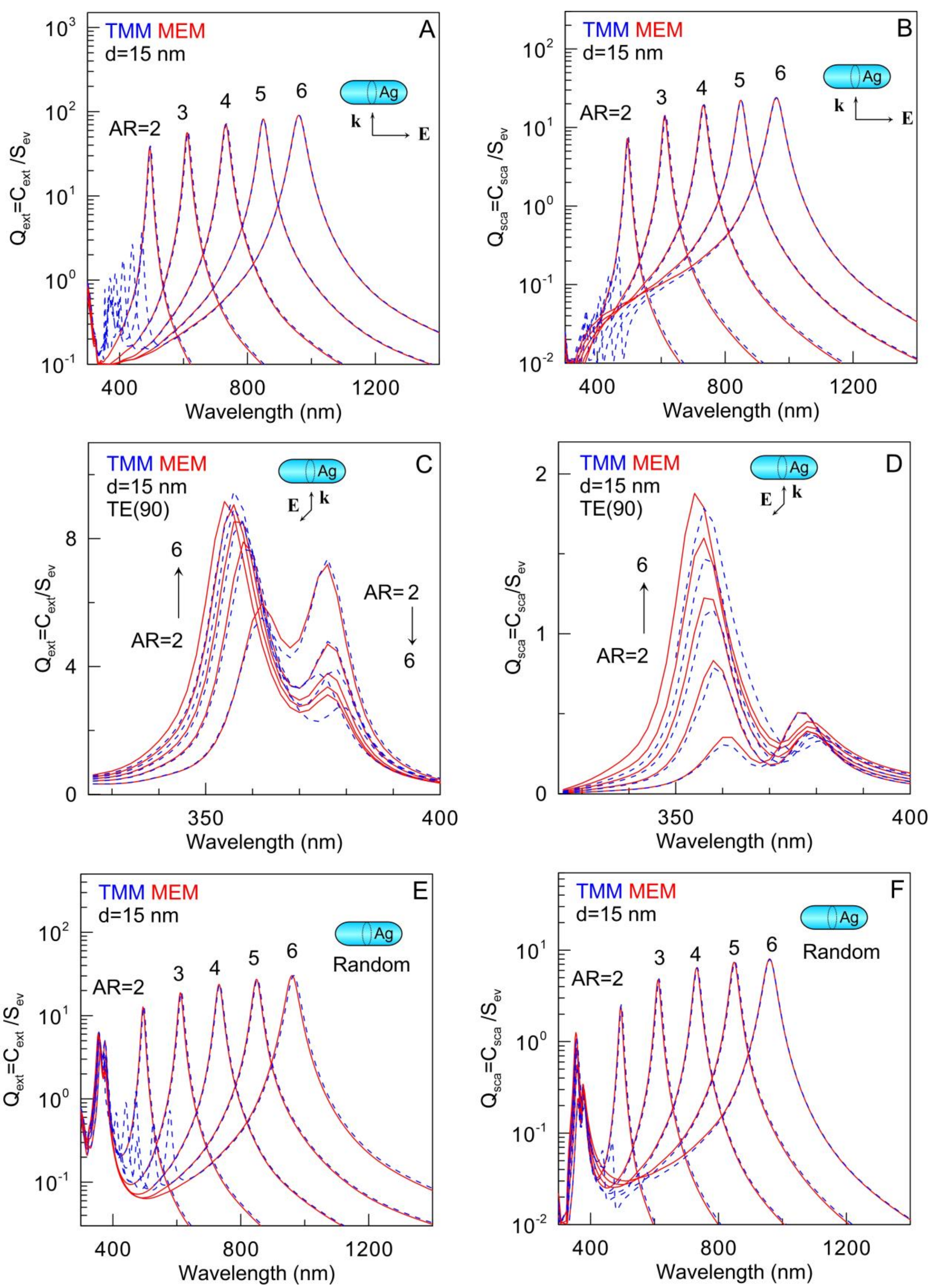

**Figure S8**. Comparison of extinction and scattering spectra calculated by the T-matrix method (TMM, blue) and MEM (red) for longitudinal (A, B) and perpendicular (C, D) excitations and for randomly oriented AgNRs (E, F) with length 90 nm and diameter 15 nm. There is excellent agreement between TMM and MEM spectra around main plasmonic peaks in panels A, B, E, and F. Two minor peaks near 355 and 375 nm are reproduced by MEM with acceptable accuracy. In addition, spectra in panels A, B, E, and F exhibit multipole resonances that MEM cannot reproduce.

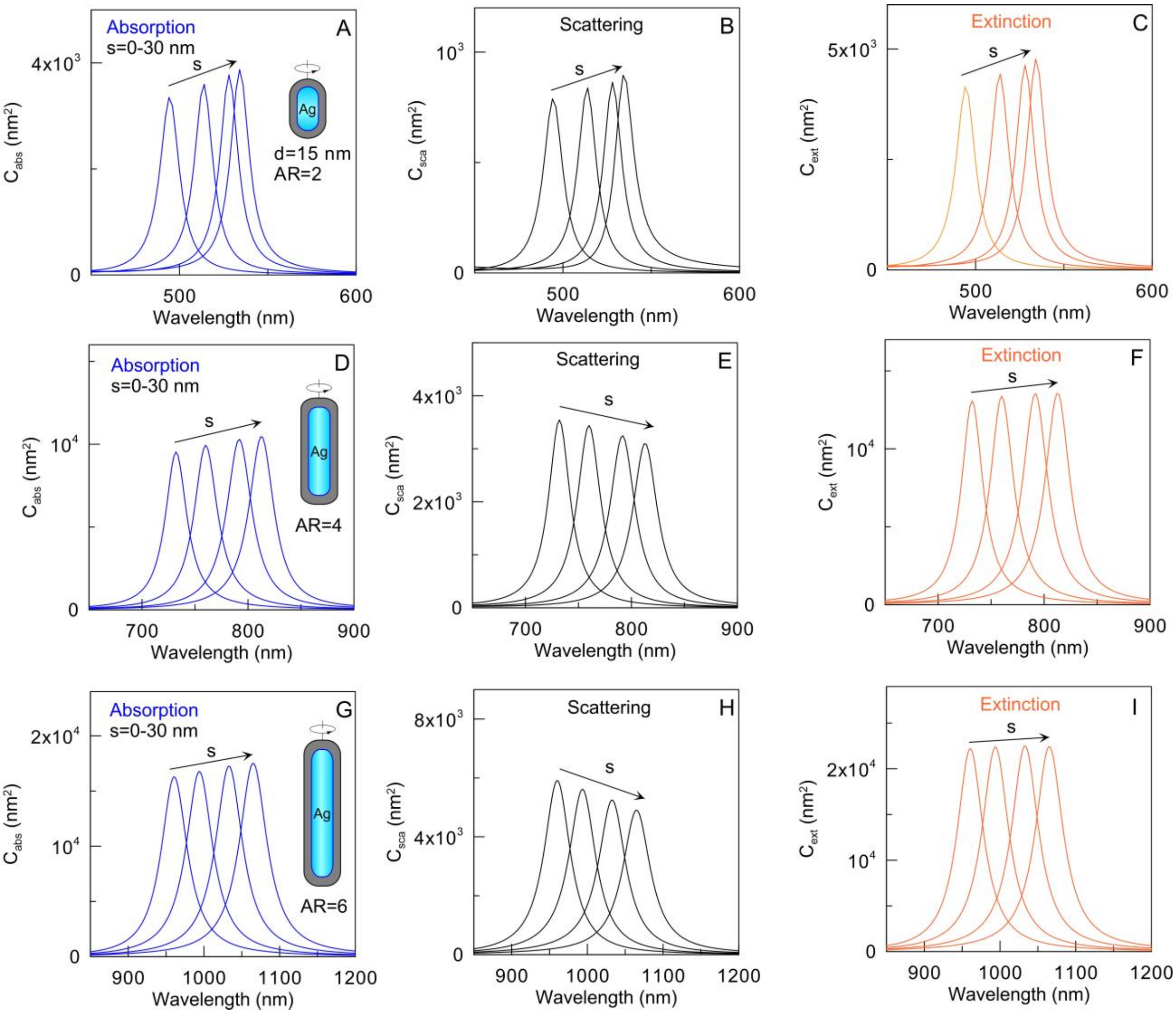

**Figure S9**. Absorption (A-C), scattering (D-F), and extinction (G-I) spectra of randomly oriented Ag bare and coated nanorods calculated by MEM. The particle diameter is 15 nm, and their aspect ratio is 20 (A-C), 4 (D-F), and 6 (G-I). The coating thickness is 0, 3, 10 , and 30 nm; the coating refractive index is 1.5.

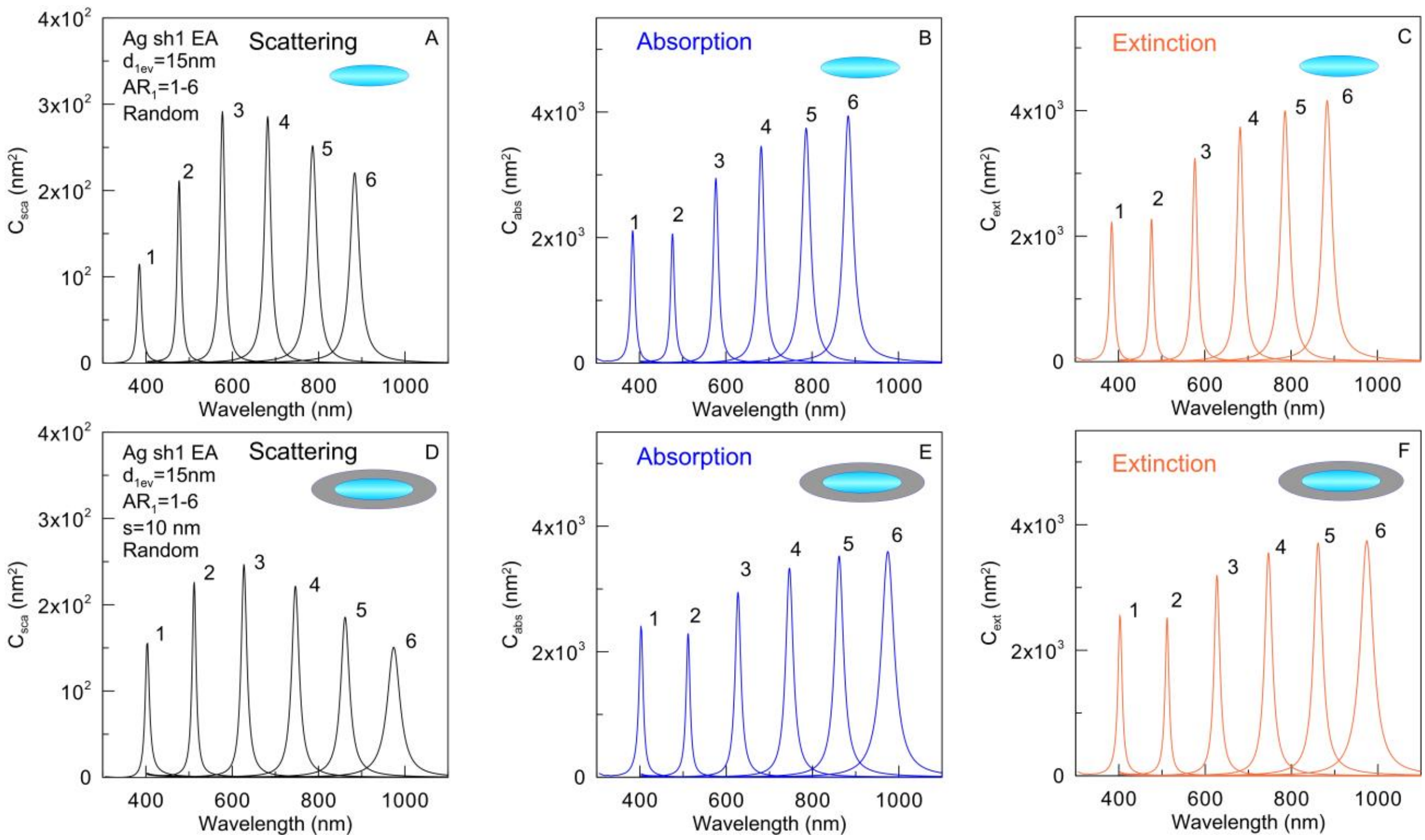

**Figure S10**. Absorption (A-D), scattering (D-E), and extinction (C-F) spectra of randomly oriented Ag bare and coated nanospheroids calculated by MEM. The equivolume diameter of particles is constant and equal to15 nm, whereas the aspect ratio varies from 1 to 6. In panels D-F, the coating thickness and refractive index are 10 nm and 1.5, respectively.

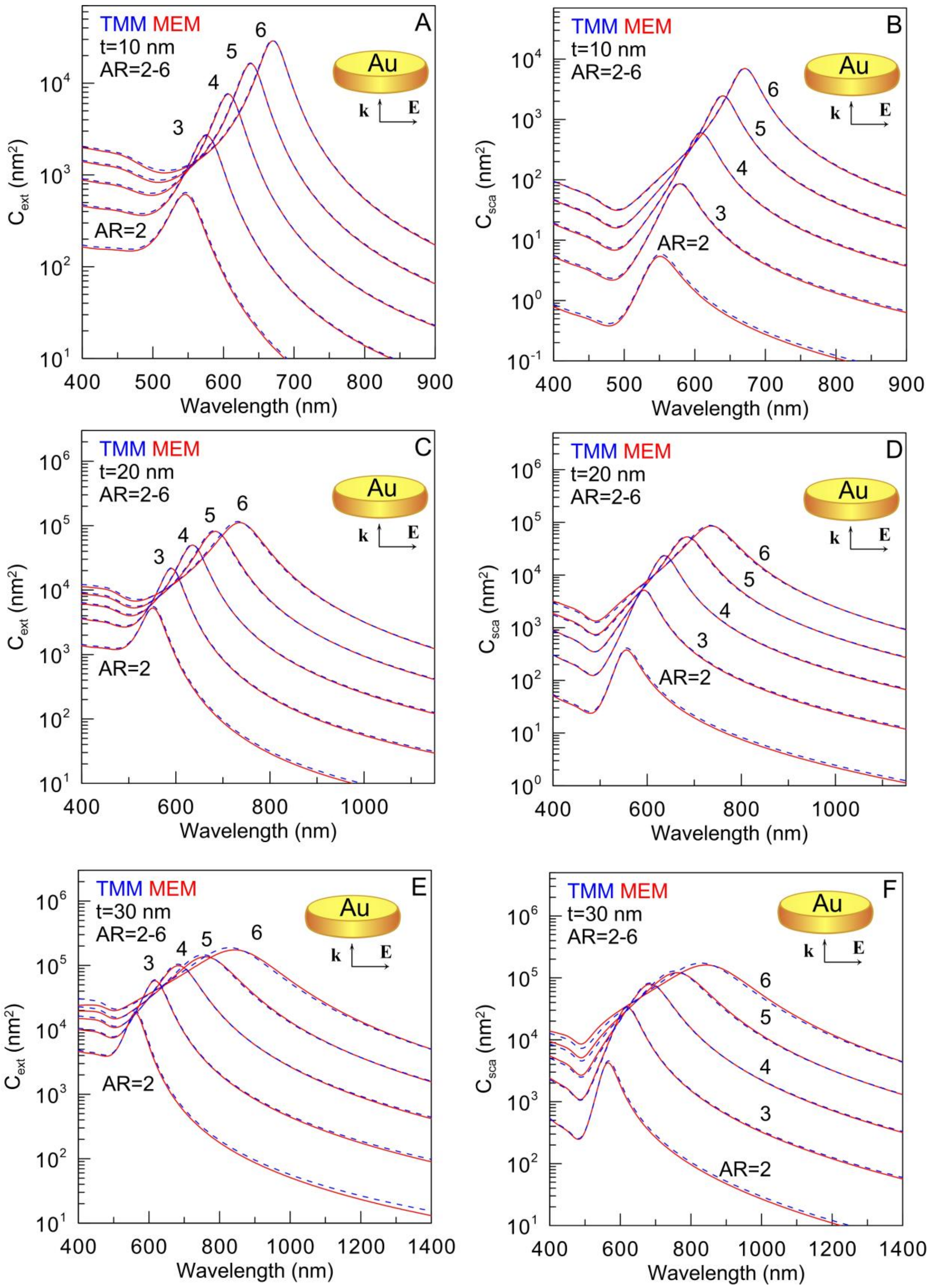

**Figure S11**. Comparison of extinction and scattering spectra calculated by the T-matrix method (TMM, blue) and MEM (red) for excitation along nanodisk diameter $D$. The disk thickness $t$ is 10 (A, B), 20 (C, D), and 30 nm (E, F); the aspect ratio $AR = D/t$ varies from 2 to 6. There is excellent agreement between TMM and MEM spectra for disks of 10 and 20 nm thicknesses and some minor differences for thicknesses of 30 nm.

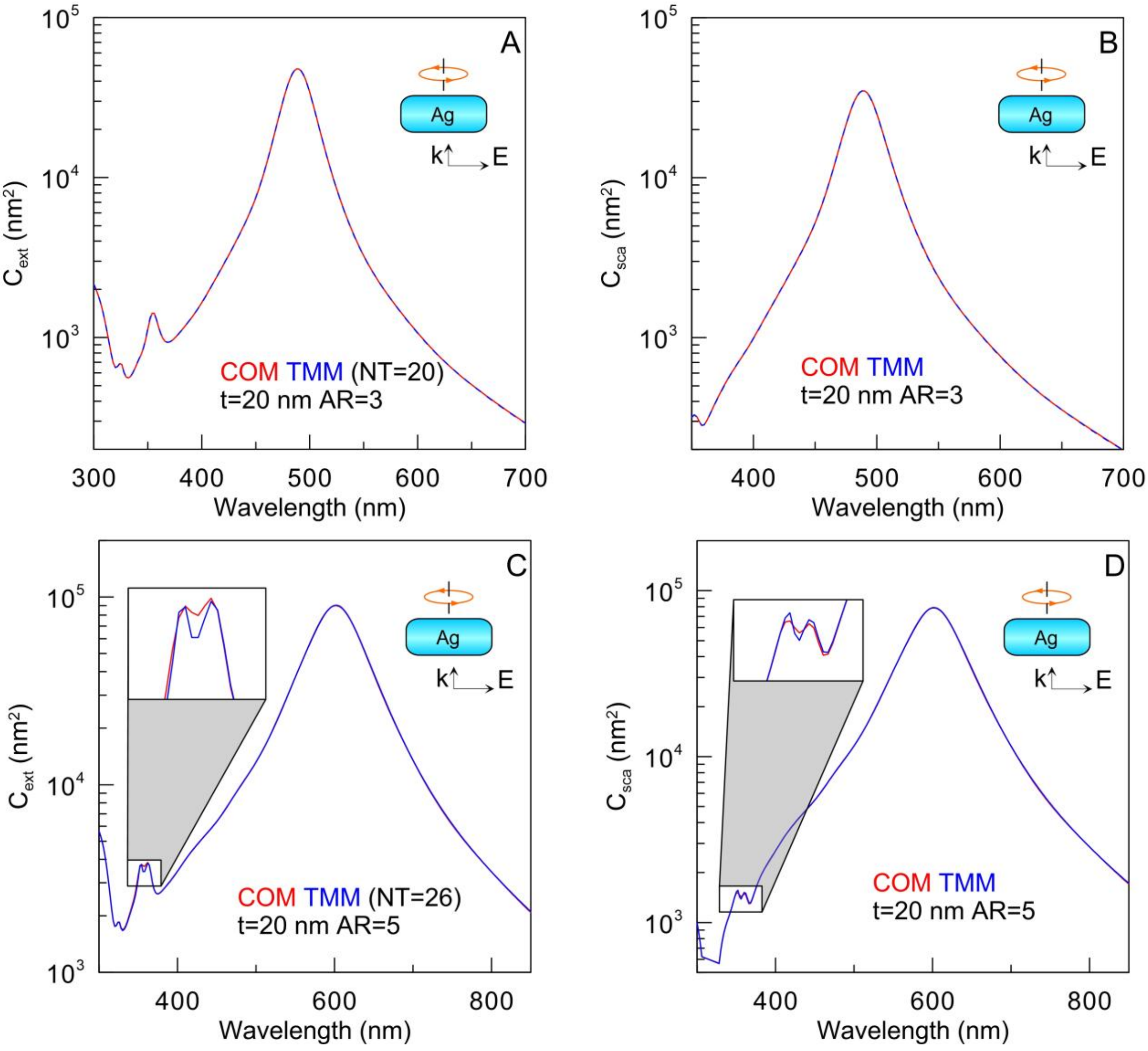

**Figure S12**. Comparison of extinction and scattering spectra calculated by COMSOL (2.5D, red curve COM) and T-matrix method (TMM, blue) for an in-plane excitation of an AgNR with a thickness of 20 nm and aspect ratio of 3 (A, B) and 5 (C, D). Both spectra have an excellent agreement for AR=3 except for minor deviations within the 340-380 nm spectral band for $AR = 5$.

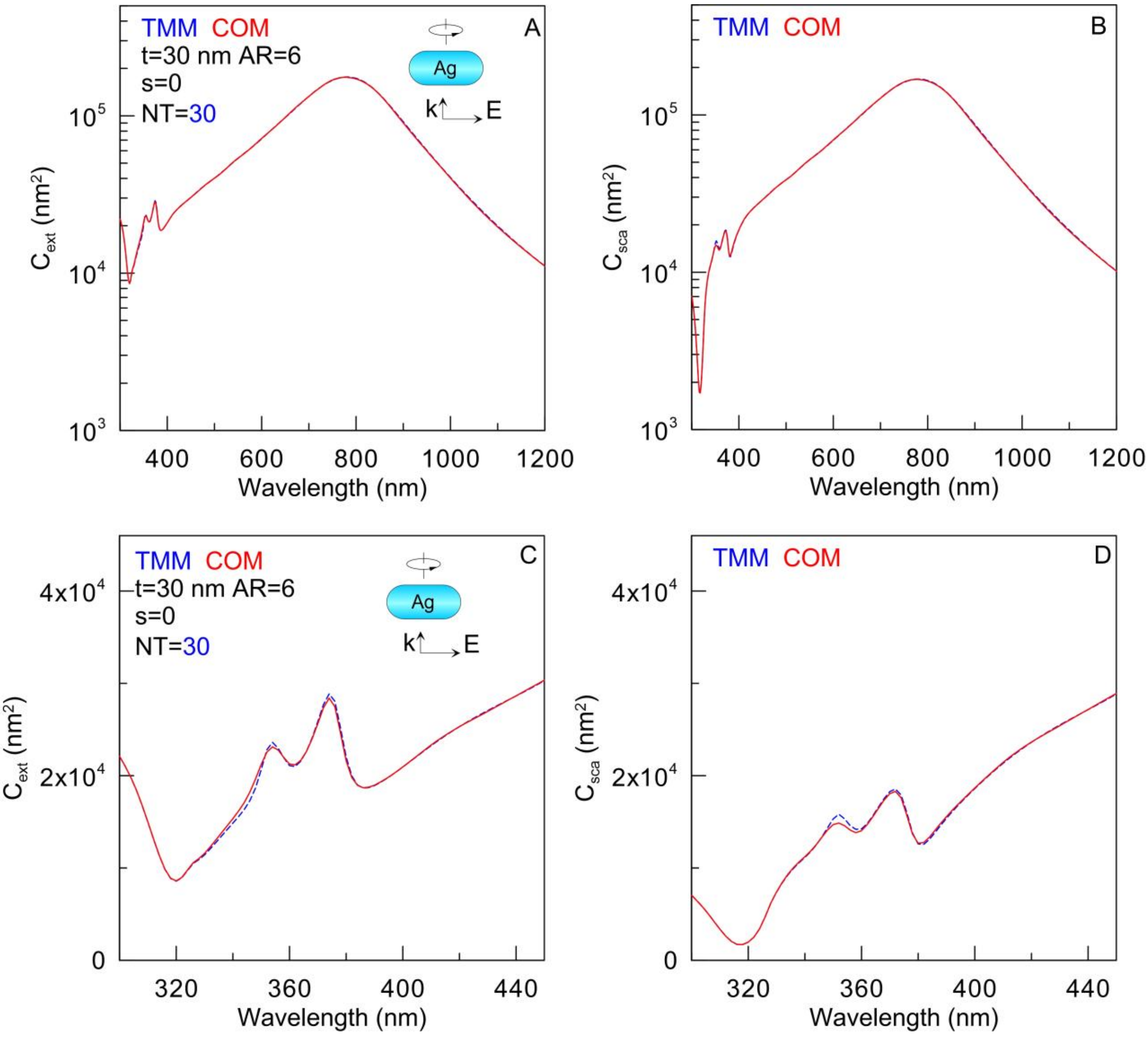

**Figure S13**. Extinction (A, C) and scattering (B, D)spectra for a thick $30\times180$ nm Ag disk calculated by TMM and COM. Panels C and D show minor differences between TMM and COM spectra for a 300-450 nm spectral band.

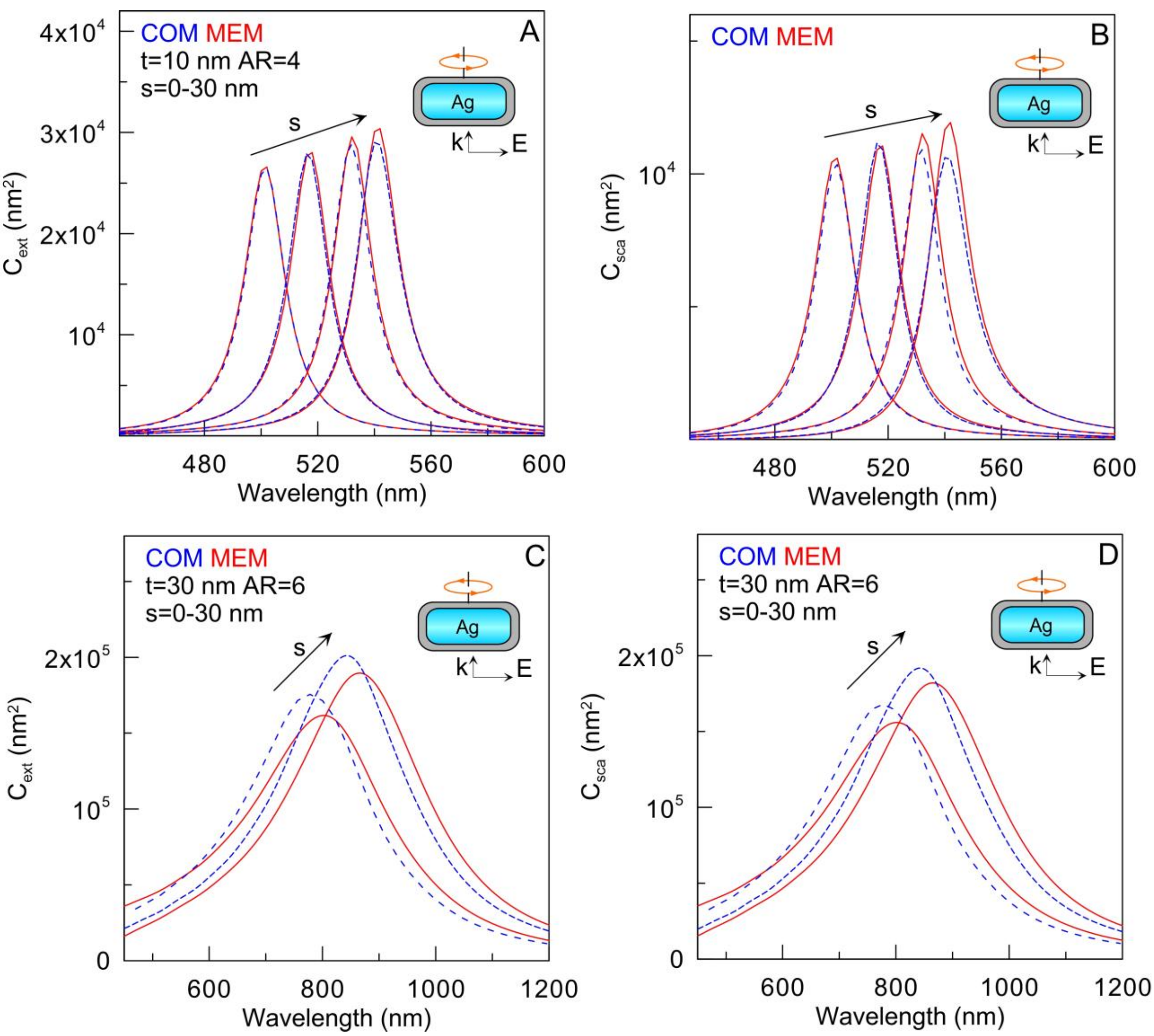

**Figure S14**. Extinction (A, C) and scattering (B, D) spectra of silver bare and coated nanodisks were calculated using MEM (red lines) and COMSOL (blue dashed lines). The particle thickness is $L = 10$ nm (A, B) and 30 nm (C, D); their aspect ratios are 4 (A, B) and 6 (C, D). The coating thicknesses are 0, 3, 10 and 30 nm (A, B) and 0, 30 nm (C, D); the coating refractive index is 1.5. The wave vector is directed along the disk axis.

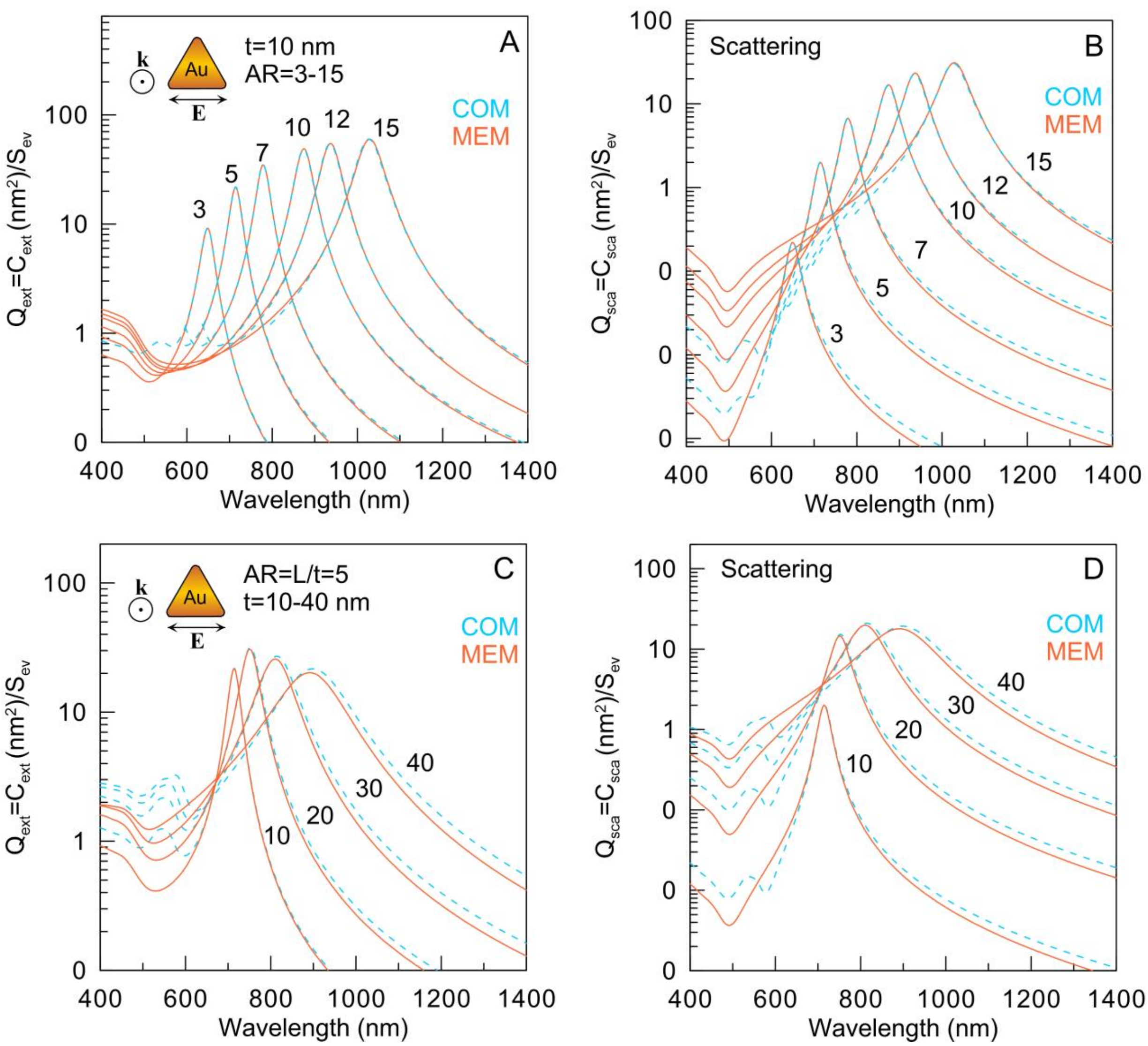

**Figure S15**. Comparison of extinction (A, C) and scattering (B, D) spectra calculated by COM (blue) and MEM (orange) for Au nanoprisms with a thickness of 10 nm and aspect ratios from 3 to 15 (A, B) and the aspect ratio of 5 and thicknesses from 10 to 40 nm (C, D). The electric field is directed along the major side (in-plane excitation). Note that MEM does not reproduce multipole peaks but approximates major plasmonic peaks well.

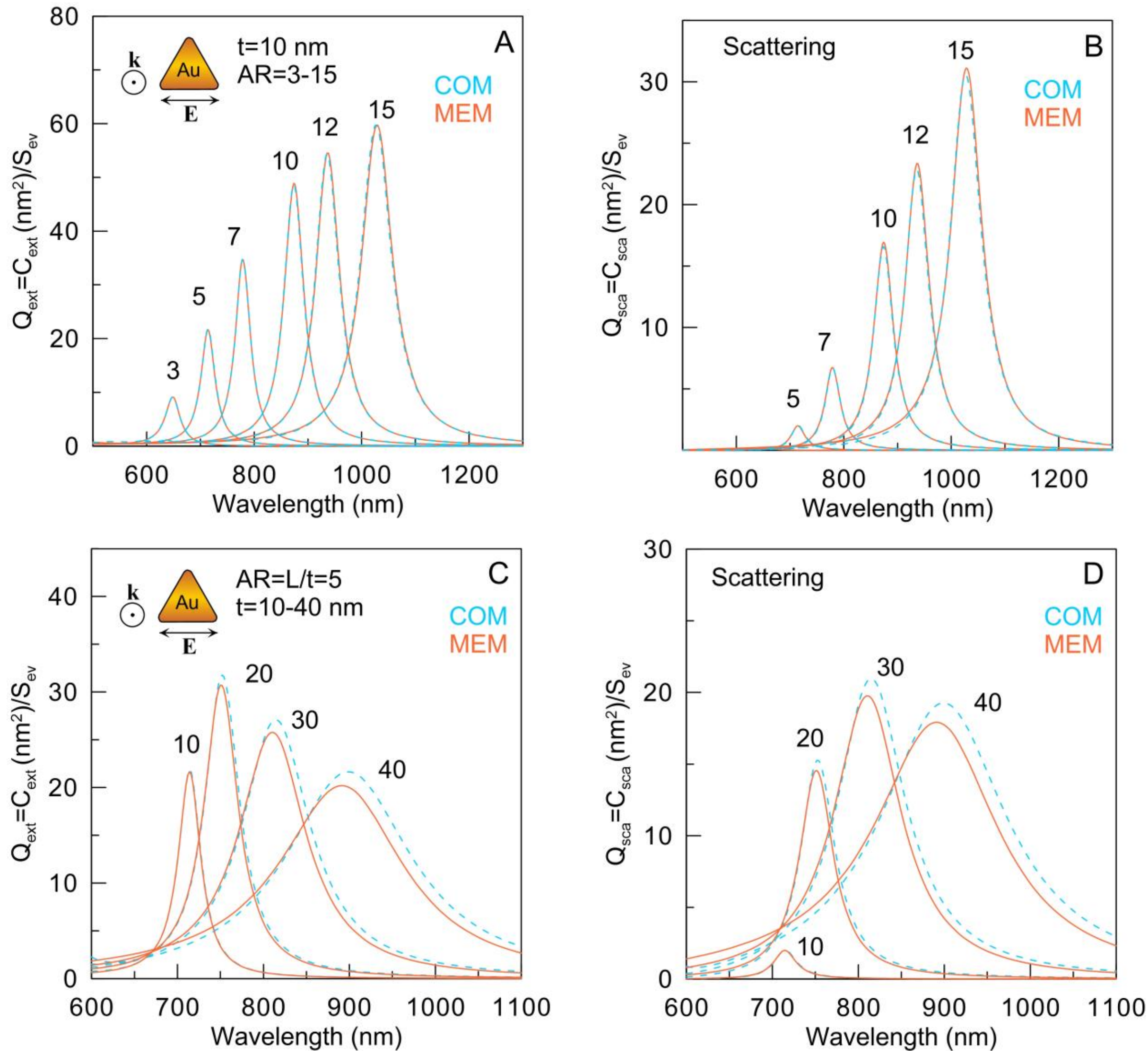

**Figure S16**. Same as in Figure S9 except for the linear ordinate scale. Note excellent agreement between COM and MEM spectra for thin nanoprisms (A, B) and some deviations for thicker ones (C, D).

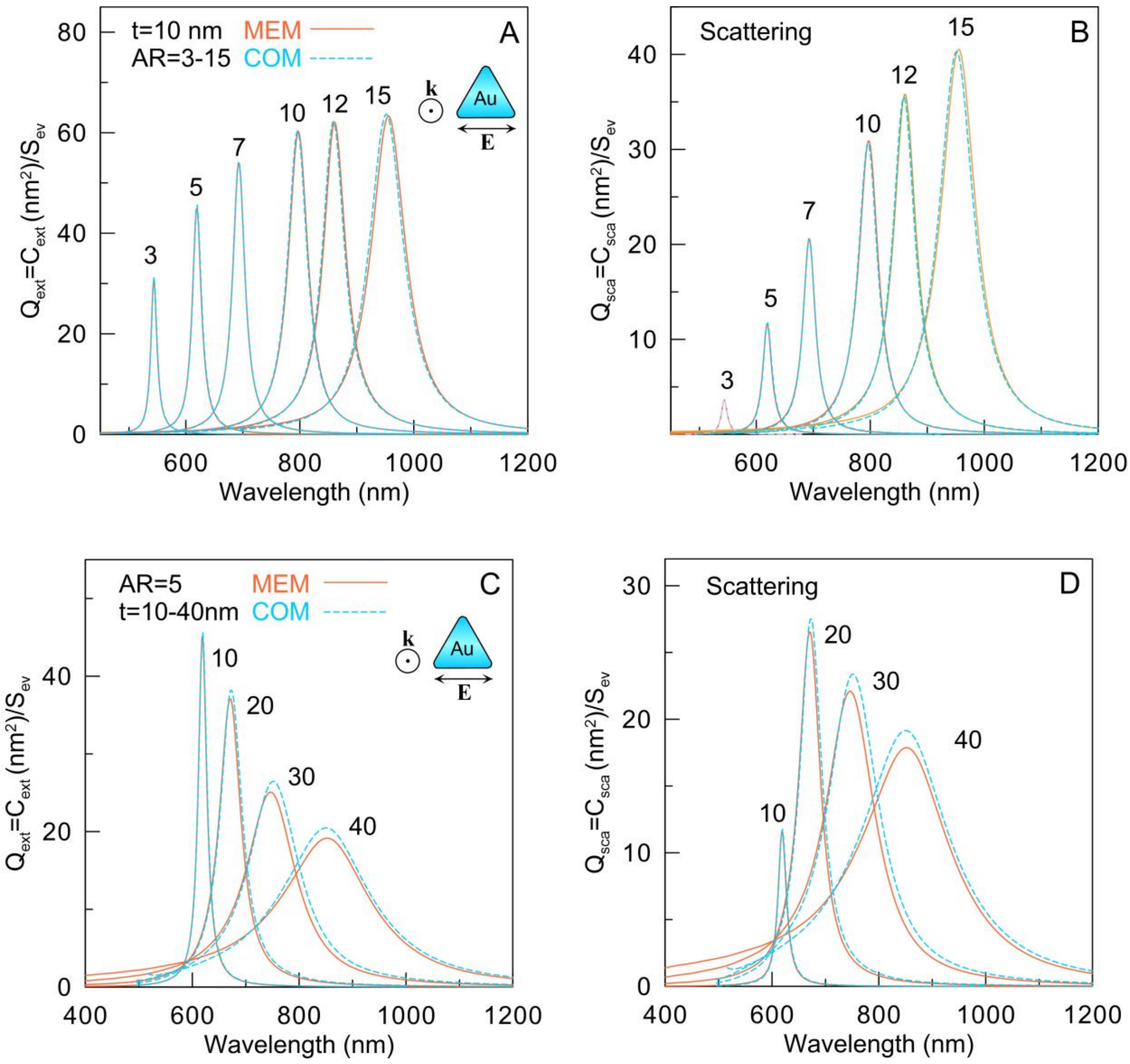

**Figure S17**. Comparison of extinction (A, C) and scattering (B, D) spectra calculated by COM (blue) and MEM (orange) for Ag nanoprisms with a thickness of 10 nm and aspect ratios from 3 to 15 (A, B) and the aspect ratio of 5 and thicknesses from 10 to 40 nm (C, D). The electric field is directed along the major side (in-plane excitation). Note that MEM does not reproduce multipole peaks (COM, not shown) but approximates major plasmonic peaks well.

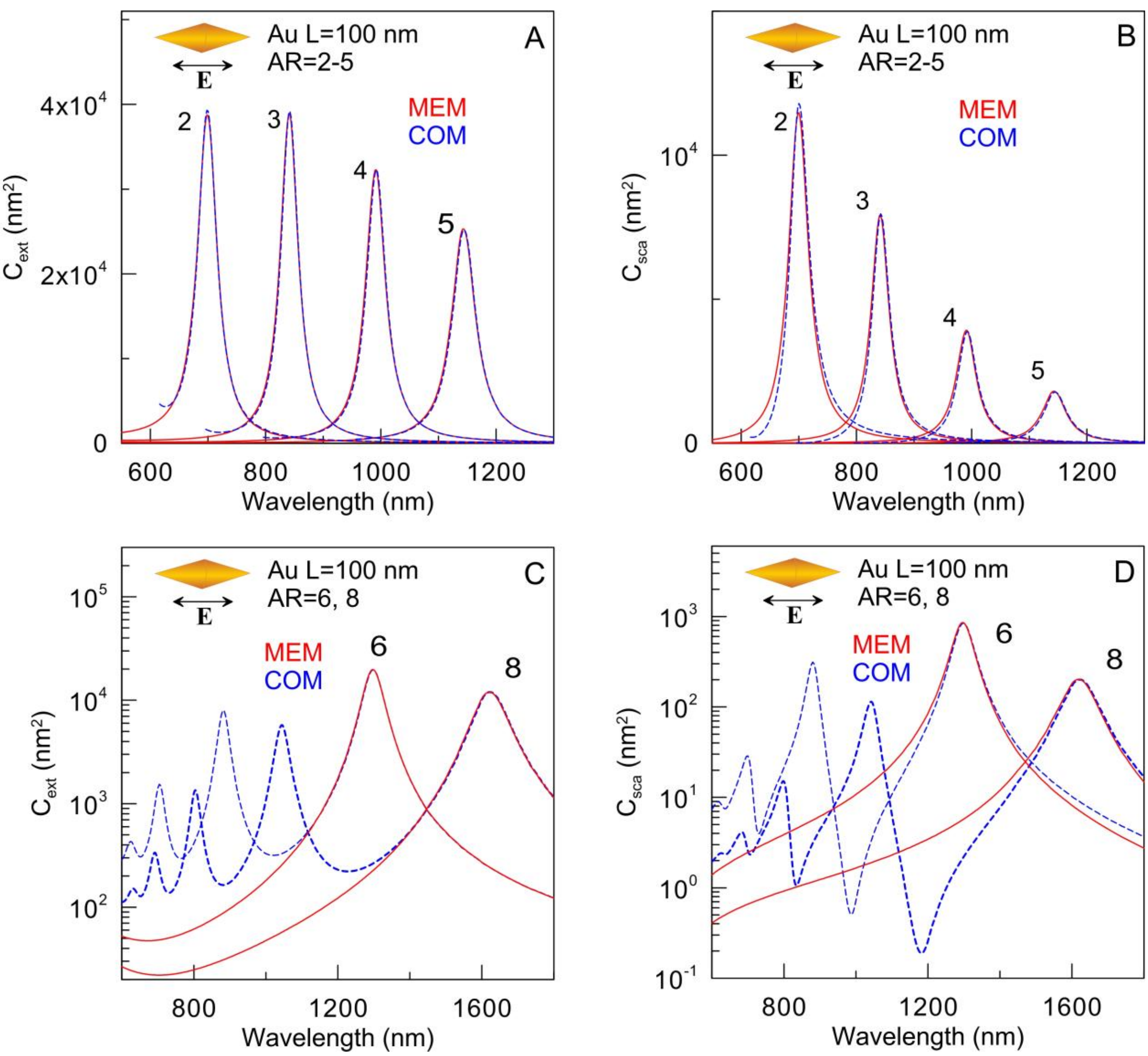

**Figure S18**. Comparison of extinction (A, C) and scattering (B, D) spectra calculated by COMSOL (blue) and MEM (red) for gold bicones of a fixed length of 100 nm and different aspect ratios from 2 to 8. Note that MEM approximation does not reproduce multipole peaks but reproduces the main dipolar resonance.

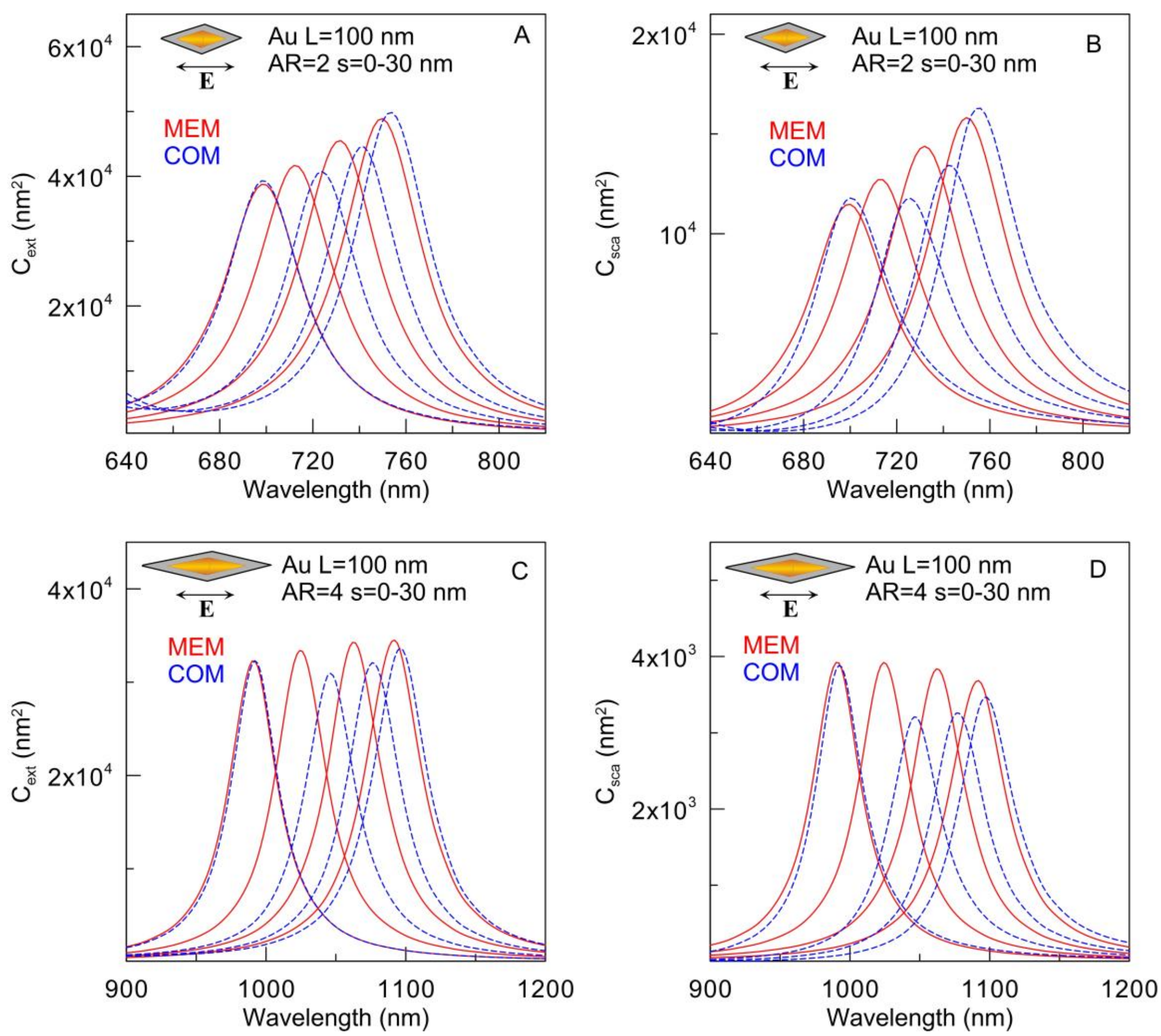

**Figure S19**. Comparison COMSOL and MEM extinction(A, C) and scattering (B, D) spectra calculated for coated gold bicones with a fixed core length of 100 nm, the aspect ratio of 2 (A, B) and 4 (C, D) and the coating shell thickness of 0, 3, 10, and 30 nm. Note the excellent agreement of spectra for the gold core and the significant difference between COMSOL and MEM spectra calculated for a thin 3 nm shell.

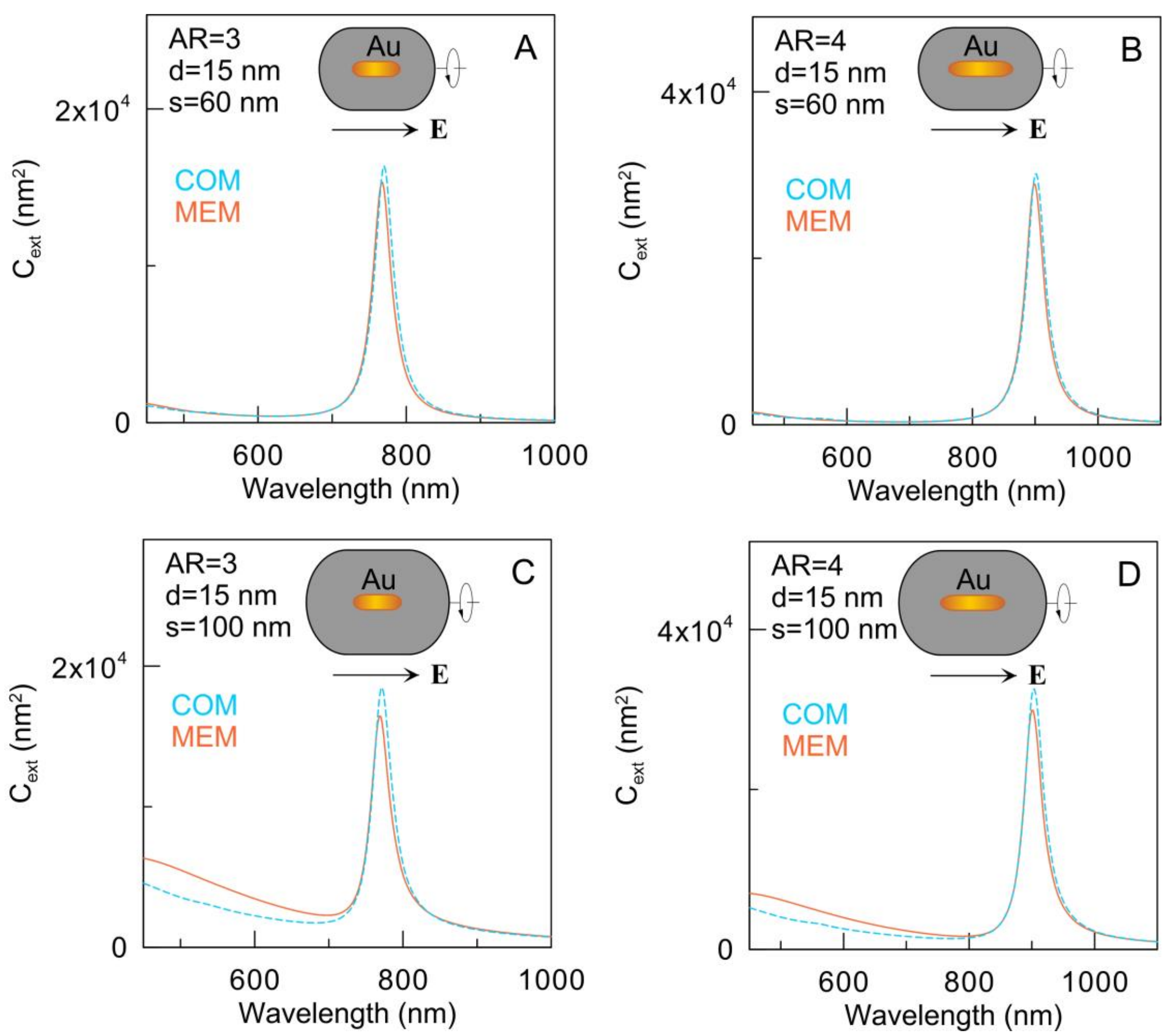

**Figure S20**. Extinction spectra of AuNRs with thick 60-nm (A, B) and 100-nm (C, D) coatings. Calculations were performed using COM (blue) and MEM (red). The particle diameter is 15 nm, with aspect ratios of 3 (A, C) and 4 (B, D). Note that the plasmonic shift and peak amplitudes are already saturated for such thick coatings. For a maximal 100-nm shell, scattering increases in the short-wavelength region.